\documentclass[floatfix,aps,prd,abstract,twocolumn,10pt,nofootinbib,preprintnumbers]{revtex4-1}

\usepackage{amsgen,amsmath,amstext,amsbsy,amsopn,amssymb}
\usepackage{graphicx}
\usepackage[usenames]{color}
\usepackage{epsfig}
\usepackage{multirow}
\usepackage{longtable}

\newcommand{\mb}{\mathbf}

\newcommand{\Ddel}{\delta_{\rm D}   }

\newcommand{\MpcOh}{ \,  \mathrm{Mpc}  \, h^{-1} }
\newcommand{\hOMpc}{ \,  \mathrm{Mpc}^{-1}  \, h  }
\newcommand{\nn}{ \nonumber }
\newcommand{\Msun}{ \,   M_{\odot}  {h}^{-1}   }

\newcommand{\beq}{\begin{equation}}
\newcommand{\eeq}{\end{equation}}
\newcommand{\beqa}{\begin{eqnarray}}
\newcommand{\eeqa}{\end{eqnarray}}

\newcommand{\dg}{\delta_{\rm g}}
\newcommand{\tg}{\theta_{\rm g}}
\newcommand{\bv}{ b_{\rm v} }

\newcommand{\eds}{ \epsilon_{1}^* }
\newcommand{\evs}{ \epsilon_{\rm  v}^* }

\newcommand{\comment}[1]{}

\begin{document}


\title{ Halo Profile Evolution and Velocity Bias }

\author{Kwan Chuen Chan} \email{KwanChuen.Chan@unige.ch}
\affiliation{D\'epartement de Physique Th\'eorique and Center for Astroparticle Physics,
Universit\'e de Gen\`eve, 24 quai Ernest Ansermet, CH--1211 Gen\`eve 4,
Switzerland}


\date{\today}

\begin{abstract}
We propose a simple model that elucidates the generation of halo velocity bias.  The fluid equation approximation is often adopted in modelling the evolution of the halo density field. In this approach, halos are often taken to be point particles even though in reality they are finite-sized objects.  In this paper, we generalize the  fluid equation approximation to halos to include the finite extent of halos by taking into account the halo profile.   We compute the perturbation of the halo density and velocity field to second order and find that the profile correction gives rise to $k^2$ correction terms in Fourier space. These corrections are more important for velocity than for density. In particular, the profile correction generates $k^2$ correction term in the velocity bias and the correction terms do not decay away in the long term limit, but it is not constant.   We  model the halo profile evolution using the spherical collapse model. We also measure the evolution of proto-halo profile at various redshifts numerically. We find that the spherical collapse model gives a reasonable description of the numerical profile evolution.  Static halo profile is often adopted in modelling  halos in theories such as the excursion set theory.  Our work highlights the importance of including the profile evolution in the calculations. 

\end{abstract}

\maketitle
\section{Introduction}

Peculiar velocity can be a powerful probe of cosmology.  On one hand, peculiar velocity causes redshift space distortion \cite{Kaiser1987}, thus to get cosmological information from galaxy surveys, one needs to model the peculiar velocity as well. On the other hand, redshift space distortion causes anisotropy, which gives rise to higher order multipoles in the correlation function/power spectrum \cite{Hamilton1992}, and hence a useful signal.  Peculiar velocity can also help distinguish general relativity from modified gravity, e.g.~\cite{SongPercival_2009, TaruyaKoyama_etal2014}.  The redshift space distortion has been used to constrain the growth rate  and testing gravity from some recent galaxy surveys \cite{GuzzoPierleoni_etal2008,BlakeKazin_2011,ReidSamushia_etal2012,BeutlerBlake_etal2012,SamushiaReid_etal2012,delaTorre_etal2013,OkaSaito_etal2014}.  In future the peculiar velocity surveys can also be fruitful \cite{KodaBlake_etal2014}.

As in galaxy surveys, only galaxies are observable, not the dark matter, it is important to understand if the velocity of the galaxy is biased with respect to that of the underlying dark matter or not.  As galaxies are hosted in halos, and halos are simpler than galaxies because they are only governed by the gravitational physics, in this paper, we take studying the halo bias as a step towards understanding the galaxy bias.  Recently there have been some indications from the measurements using halos at low redshifts that velocity bias may be non-negligible, although the quantitative measurement is still hard \cite{EliaLudlowPorciani2012,ChanScoccimarroSheth2012,BaldaufDesjacquesSeljak2014,JenningsBaughHatt2014,ZhengZhangJing2014b}. Velocity measurement is nontrivial because it requires the velocity field of the tracers to be weighted by volume, and it is easy to mistakenly get the density weighted velocity, i.e. momentum, instead of velocity \cite{Scoccimarro2004}. When the number density of the tracers is low, it suffers from numerical sampling artifact, see e.g.~\cite{Juszkiewiczetal1995,ZhangZhengJing2014,ZhengZhangJing2014a, ZhengZhangJing2014b}.  One way out is to use  momentum instead, e.g.~in \cite{OkumuraSeljak_etal2012,BaldaufDesjacquesSeljak2014}. Unlike the velocity field, however, there is an additional complication that in momentum the galaxy density bias is involved as well.


On the theory side, in the usual fluid approximation for dark matter and galaxy, even if the initial velocity field of the galaxy differs from that of the dark matter, i.e.~there is initial velocity bias, large scale gravitational evolution will naturally drive the galaxy velocity field to that of the dark matter \cite{EliaKulkarnietal2011,ChanScoccimarroSheth2012}. On the other hand, the peak model predicts that the velocity bias persists and remains constant at late time \cite{DesjacquesCrocceetal2014}. This result seems to be favoured by the recent simulation results, which suggest that the halo velocity bias at late time is non-negligible at  $k \sim 0.15 \hOMpc $.  Ref.~\cite{BaldaufDesjacquesSeljak2014} argued that the force on the halo has to be ``biased'' in order for the coupled-fluid approach to agree with peak model result, although no further justification was given.  Ref.~\cite{BiagettiDesjacquesKehagiasRiotto2014b} tried to derive the peak theory results using the distribution function approach.  Here we take a different approach. A halo is a composite object consisting of a collection of particles.  The position of the halo is defined by the position of the center of mass (CM) of its constituent particles. Thus the force acting on the CM position of the halo should be averaged over its constituent particles.   In this way, we give a physical origin for the ``biased'' force on the halo.  We will show that the halo profile correction  naturally gives rise to the leading $k^2$ correction to the velocity bias and it does not decay away.

 We note that our approach also has rather different interpretation for the generation of velocity bias from that in \cite{DesjacquesSheth2010,DesjacquesCrocceetal2014, BaldaufDesjacquesSeljak2014,BiagettiDesjacquesKehagiasRiotto2014b}. In peak theory, although the smoothing window is an important ingredient, the window function is usually assumed to be static.   Sometimes, the attention is focused on the discrete  peak ``points'', which have the same velocity as the dark matter locally, it was argued that the velocity bias is a ``statistical''  effect.   In our model,  the velocity bias physically arises from the fact that halos are finite-sized objects, not point particles, and it also highlights the dynamical nature of window.

On the other hand, our approach may not be mutually exclusive with the peak model approach. In the modelling of halos starting from the Lagrangian space, one defines halos with window function and the smoothing scale is fixed to be the Lagrangian szie when they are transformed to the Eulerian space.  Even if our velocity bias contribution is not the dominant one seen in simulations, the profile correction effects should be taken into account in the calculations as well.


 This paper is organized as follows. As we will show that the halo profile gives rise to the velocity bias correction, to set the stage, we will first review the evolution of the halo profile using the spherical collapse model in Sec.~\ref{sec:spherical_collpase} and the numerical halo profile is measured from simulation and compared with the spherical collapse model in Sec.~\ref{sec:Measurements}.  In Sec.~\ref{sec:LinearBiasComputation} we compute the correction to the linear velocity and density bias due to the halo profile, and the  second order corrections are presented in  Sec.~\ref{sec:SecondBiasComputation}.  We conclude in Sec.~\ref{sec:Conclusions}.

\section{ Halo Profile Evolution}
\label{sec:halo_profile}
Halo profile is often used in the context of halo model for modelling the dark matter power spectrum \cite{Seljak2000, PeacockSmith2000,Scoccimarroetal2001,CooraySheth}. In this case, the virialized halo profile, such as the NFW profile \cite{NFW1996} is often used. However, we will follow the proto-halo from its infancy to the final virialized stage in modelling the bias evolution.  To this end, we will first review the evolution of a halo using the spherical collapse (SC) model.  We will then construct proto-halos at various redshifts and measure the profile evolution in numerical simulations. The results are compared with the SC model. To our knowledge, this is the first systematic measurements of the proto-halo profile evolution.

\subsection{Profile evolution from SC model}
\label{sec:spherical_collpase}
A simple analytic model for halo evolution is given by the SC model \cite{GunnGott1972} (see also \cite{Peebles1980,Padmanabhan1993, MoBoschWhite2010}). Suppose that the initial fluctuations are spherically symmetric about some point in position space. To avoid shell crossing, we assume that the  radial profile is non-increasing as the distance from the center increases.  We will consider the matter dominated universe as the resultant equation can be integrated analytically, and also the more realistic $\Lambda$CDM model. Under the Newtonian approximation, the equation of motion for a mass shell at a distance $r$ from the center is given by 
\beq
\label{eq:EoM_sphericalcollapse}
\frac{d^2 r }{ d t^2 } = -\frac{G M(r)}{ r^2 } + \frac{\Lambda }{3  } r ,
\eeq
where both $r$ and $t$ are the \textit{physical} distance and time, and $G$ is the gravitational constant, $ M(r)$ is the total mass inside the mass shell and $\Lambda  $ is the cosmological constant.

 Integrating Eq.~\ref{eq:EoM_sphericalcollapse} once, we obtain the first integral of motion 
\beq
\label{eq:1stintg_sphericalcollapse}
\frac{1 }{2}\Big(  \frac{dr}{ dt } \Big)^2  - \frac{GM}{ r } - \frac{\Lambda r^2 }{  6} = E, 
\eeq
where the total energy $E$ is a constant of integration.

 We will solve  Eq.~\ref{eq:EoM_sphericalcollapse} numerically when $\Lambda \neq 0 $.   When $\Lambda=0$,  Eq.~\ref{eq:1stintg_sphericalcollapse} can be further integrated analytically, and the solution can be expressed in the form of a cycloid solution
\beqa
\label{eq:r_cycloid}
r &=& A (  1 - \cos \theta  ),   \\
\label{eq:t_cycloid}
t + T &=& B ( \theta - \sin \theta ) , \\
\label{eq:AB_cycloid}
A^3  &=& GM B^2,
\eeqa 
where $A$, $B$ and $T$ are constants.  The parameter $ \theta $, also called the development angle, runs from 0 to $ 2 \pi  $.  When $\theta $ is close to 0, the overdensity inside the mass shell is small, the mass shell essentially follows the Hubble expansion. The mass shell reaches maximum $r_{\rm m} $ at $\theta= \pi  $. Beyond that the mass shell overcomes the Hubble expansion and turns around. At $\theta = 2 \pi  $, the shell collapses to a point according to Eq.~\ref{eq:r_cycloid}. However, it was argued that during the rapid infall the potential varies quickly, the particles no longer follow the energy conserving orbits, instead  the energy available to the particles is  widened, and the system reaches virial equilibrium \cite{LyndenBell1967, BindoniSecco2008}. This procoess is called ``violent relaxation'' \cite{LyndenBell1967}. From virial equilibrium, one finds that the  virial size  $r_{\rm v} $ is related to  $r_{\rm m} $ as
\beq
\label{eq:rvir_rm_half}
r_{\rm v} =  \frac{ r_{\rm m} }{2}. 
\eeq

 The virial size is often given in terms of the virial density, $\Delta_{\rm v}$, as 
\beq
\label{eq:rv_Deltav}
r_{\rm v} =  \Big( \frac{ M }{ \frac{4 \pi}{3}  \bar{\rho}_{\rm m} \Delta_{\rm v} } \Big)^{ \frac{1}{3} },
\eeq
where $M$ is the mass of the halo and $\bar{\rho}_{\rm m}  $ is the comoving density of matter.  For EdS universe, $ \Delta_{\rm v}  $ is equal to 178.  When $\Lambda \neq 0 $, a  fitting formula for the virial density, $\Delta_{\rm v}$, is given in \cite{BryanNorman1998}. For the flat $\Lambda  $CDM with $\Omega_{\rm m} =0.25  $ adopted in this paper,  $\Delta_{\rm v}$ for a halo virailzed at $z=0 $ is 380.  In practice other values of  $\Delta_{\rm v}  $ are  often adopted, such as 200 and 500. We will use  $\Delta_{\rm v} = 500$ as we will see later on it gives a good description of our simulation data. Note that for non-EdS universe, in Eq.~\ref{eq:rv_Deltav}, the critical density is often used instead of $ \bar{\rho}_{\rm m} $ to define $r_{\rm v}$.


To solve  Eq.~\ref{eq:EoM_sphericalcollapse}, the initial conditions that the initial overdensity is obtained by extrapolating the collapse threshold from the present time to the initial time using the linear growth factor and zero initial peculiar velocity is often assumed. However, the zero initial peculiar velocity condition excites both the growing mode and the decaying mode. It can be shown  that the linear amplitude of the perturbation is reduced by a factor of $3/5$ \cite{Peebles1980,Padmanabhan1993,Scoccimarro98}. This is equivalent to setting up the initial condition incorrectly and transient effects are induced.  To get the right final amplitude, one quick fix is to increase the initial perturbation by a factor of $5/3 $ to compensate the loss to the decaying mode.  As we set up the initial conditions at not very high redshifts, the transient effects are not negligible. A better approach is to set the initial peculiar velocity such that the decaying mode vanishes. Thus we will use the initial conditions
\beqa
\label{eq:rstar_SC}
r_* & =&  \Big(  \frac{ 3 M }{  4 \pi  \bar{\rho}_{\rm m} } \Big)^{ \frac{1}{ 3 } } \frac{ a_* }{ ( 1 + \bar{\delta}_*)^{\frac{1 }{3 } } },         \\
\label{eq:rdotstar_SC}
\dot{r}_* &=&  H_* r_* \Big( 1 - \frac{1 }{ 3 } \bar{ \delta}_* \Big) , 
\eeqa
In this paper, we  use ``*'' to denote a quantity at some  initial time. Thus $a_*$, $H_*$ and $ \bar{ \delta }_* $ are the scale factor, Hubble parameter and the average density contrast inside the spherical shell at the initial time. 

Using these initial conditions, we can write the coefficents $A$ and $B$ in Eq.~\ref{eq:r_cycloid} and \ref{eq:t_cycloid} as
\beqa
A &=& \frac{r_*  }{ 2} \frac{ \Omega_{\rm m}^* ( 1 + \bar{\delta}_* ) }{ \Omega_{\rm m }^* ( 1 + \bar{ \delta}_* ) - \big(  1 -  \frac{1 }{ 3 } \bar{\delta}_*   \big)^2  } , \\
B &=& \frac{1 }{ 2 }  \frac{  \Omega_{\rm m}^* ( 1 + \bar{\delta}_* )}{  
 H_* \Big[ \Omega_{\rm m}^* ( 1 + \bar{ \delta}_* ) - \big( 1 -  \frac{1 }{3 }  \bar{\delta}_*  \big)^2  \Big]^{ \frac{3}{2 }} },
\eeqa
where $ \Omega_{\rm m}^*   $ is the density parameter of matter at the initial time. We first note that $A$ and hence $r$ is proportional to $r_*$. On the other hand, the collapse history given by $ t $ is independent of $r_*$, and it depends only on the matter inside through $\Omega_{\rm m}^*$ and $\bar{\delta}_*  $.

In Eq.~\ref{eq:EoM_sphericalcollapse}, after dividing by $r_*$, one can easily see that the collapse history is independent of $r_*$ in $\Lambda  $CDM model. Thus given  $\Omega_{\rm m}$ and  $\Omega_{\Lambda}$, the collapse history depends only on  $\bar{\delta}_* $.

\begin{figure}[!htb]
\centering
\includegraphics[width=\linewidth]{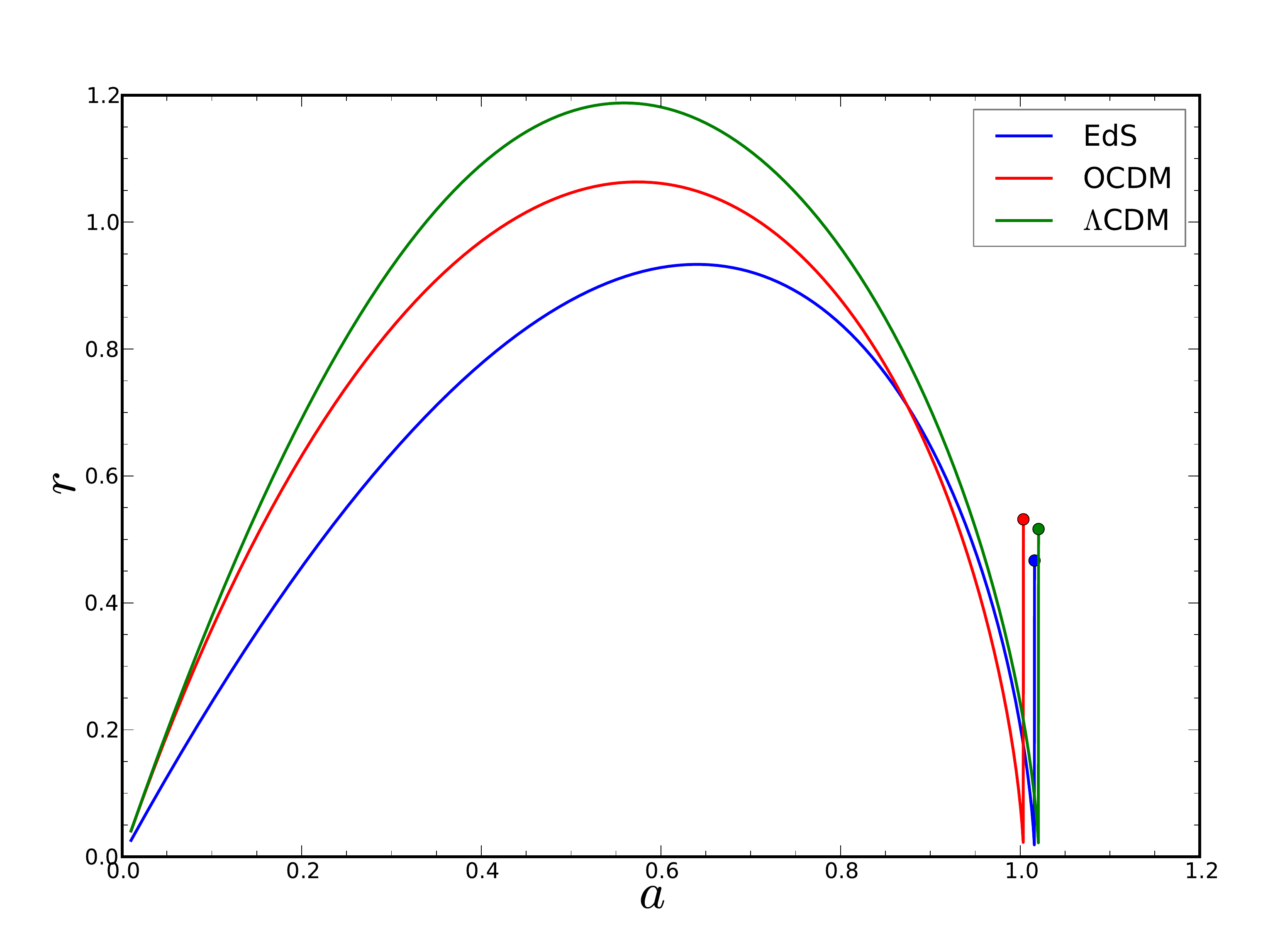}
\caption{ The evolution of the halo profile in three different cosmological models EdS (blue), OCDM with $\Omega_{\rm m}=0.25 $ (red) and $\Lambda$CDM with $\Omega_{\rm m}=0.25 $ and  $\Omega_{\rm m}=0.75 $ (green). The virial  size is indicated as a dot after the collapse.   }
\label{fig:SC_r_EdS_OCDM_LCDM}
\end{figure}

In Fig.~\ref{fig:SC_r_EdS_OCDM_LCDM}, we show the evolution of the profile as a function of $a$ for three different cosmological models: EdS, Open CDM with $\Omega_{\rm m}=0.25 $ and $\Lambda$CDM with $\Omega_{\rm m}=0.25 $ and  $\Omega_{ \Lambda }=0.75 $. Note that in this paper  $\Lambda$CDM always refers to this model. The mass of the halo is chosen to be $ 2 \times 10^{13} \Msun $. The collapse threshold at $z=0$ is set to be $\delta_{\rm c} =1.68  $ and extrapolated to the initial time using  the linear growth factor for the corresponding cosmology. The initial conditions are set using Eq.~\ref{eq:rstar_SC} and \ref{eq:rdotstar_SC}. Nonetheless, we still find that we need to choose $a_*$ to be sufficiently small ($ a_* = 0.01 $ here) to reduce the effects of transients. For example when $a_* =0.02$ is chosen instead, we find that the collapse epochs are increased by a few per cent compared to the ones shown.  We emphasize that this is because we set up the initial conditions using linear theory, and the transients can be further suppressed using higher order perturbation theory. This is analogous to setting up initial conditions in simulations using 2LPT \cite{Scoccimarro98, CroccePeublasetal2006}.  Although these cosmologies are rather different, the final collapse epochs are very similar as long as the correct linear growth factor is used to set $\bar{\delta}_*$. In other words the collapse threshold is insensitive to the cosmological model \cite{EkeColeFrenk1996}. However, the intermediate stages of collapse are quite different among these models. Thus this suggests that we need to solve the model explicitly in order to follow the evolution of the halo profile accurately. We also indicate in Fig.~\ref{fig:SC_r_EdS_OCDM_LCDM} the virial size after the collapse. For EdS and OCDM, it is computed using Eq.~\ref{eq:rvir_rm_half}, while for $\Lambda  $CDM we use Eq.~\ref{eq:rv_Deltav} with  $\Delta_{\rm v} = 500$.

To illustrate the evolution of the profile, we further assume that the initial halo profile is described by a top-hat profile, and in Fourier space, it reads
\beq
W_{\rm TH}( \kappa ) = \frac{ 3  }{ \kappa^3  } ( \sin \kappa - \kappa \cos \kappa ). 
\eeq
Top-hat profile is a good approximation at high redshifts.  In SC with top-hat perturbation, the top-hat shape is preserved during evolution. The only part that changes is the width of the window. On the other hand, the Eulerian virialized spherical halo profile is well described by the NFW profile \cite{NFW1996}. This means that  a top-hat window of perturbation cannot evolve to the NFW-like profile.    We will see in Sec.~\ref{sec:Measurements} that the halo profile measured from simulation goes from one resembling top-hat to an NFW-like profile as the redshift decreases.

\begin{figure}[!htb]
\centering
\includegraphics[width=\linewidth]{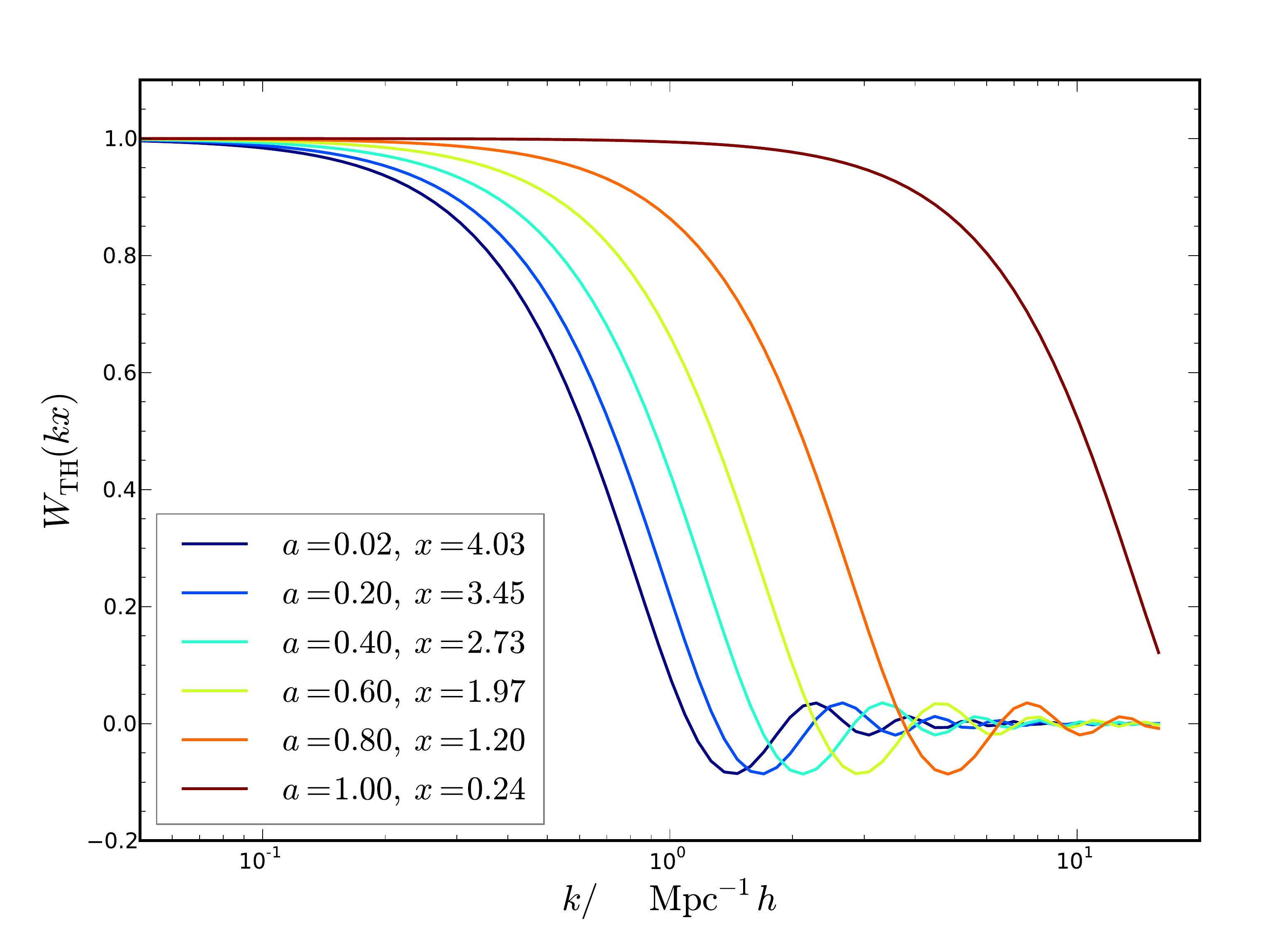}
\caption{ The evolution of the top-hat window, $W_{\rm TH} ( kx )$ for a suite of scale factor  $a$. The comoving size $x$ of the spherical shell is computed using  the SC model.  }
\label{fig:WTH_evolution}
\end{figure}

In Fig.~\ref{fig:WTH_evolution}, we show the evolution of  $W_{\rm TH}(k x) $ for a series of values of the comoving size $x$ at different time $a$. We start with  $x_*=4.07 \MpcOh $ at $z_*=99$, and the size of the spherical shell is then evolved  according to Eq.~\ref{eq:EoM_sphericalcollapse}. We have adopted a flat $\Lambda$CDM model with $\Omega_{\Lambda } = 0.75 $, and $\delta_{\rm c} =1.68  $.  Note that although the physical size $r$ first expands and then collapses as in Fig.~\ref{fig:SC_r_EdS_OCDM_LCDM}, the comoving size $x$ is always decreasing as the effect of expansion is removed. At $a=1$, the shell has not fully collpased yet.  As we will see later on, we are mainly interested in the low $k$ part of the window, thus to a very good approximation, the window is essentially 1 up to $k \sim 3 \hOMpc$ at the present time. In this plot we have not substituted the halo size at $a=1$ with the virial size. Most of time in the paper,  the sudden change during virialization does not matter as it occurs almost instantaneous for our purpose. From now on, the SC model refers the one obtained by evolving an initially top-hat perturbation using Eq.~\ref{eq:EoM_sphericalcollapse}.

\subsection{Measurement of the profile evolution from simulations }
\label{sec:Measurements}

\begin{figure*}[!htb]
\centering
\includegraphics[width=\linewidth]{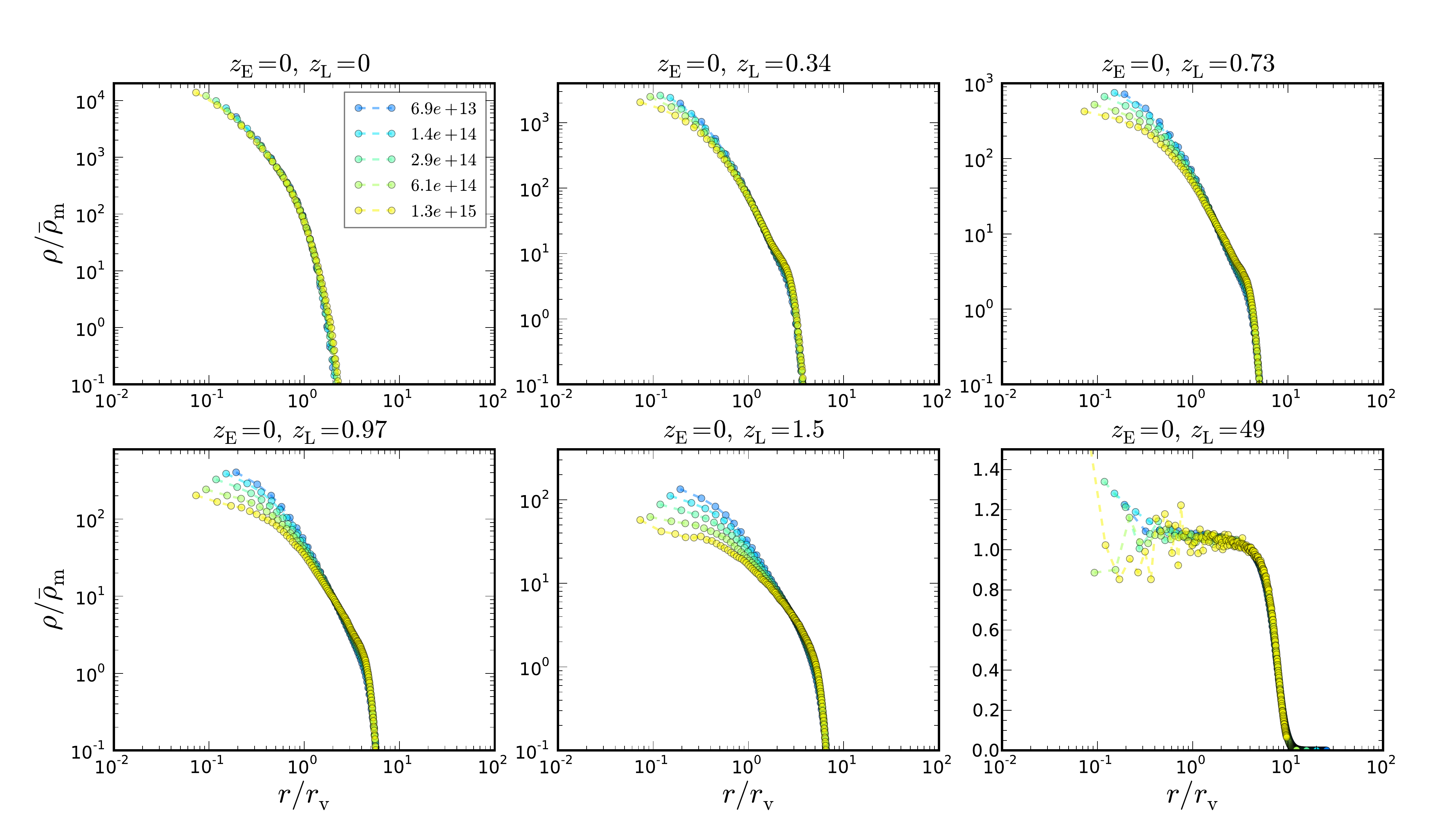}
\caption{ The halo profile at  $z=0$, 0.34, 0.73, 0.97, 1.5, and 49. The  Eulerian profile is at $z=0$, and the proto-halo profiles are obtained by tracing the particles in the Eulerian halo to higher redshifts, $ z_{\rm L}$. Results from halos of mass ranging from $6.9 \times 10^{13} \Msun $ to  $1.3 \times 10^{15} \Msun $ are shown. The density is normalized with respect to the  mean comoving density of matter, $\bar{\rho}_{\rm m} $ and the radial distance  $r$ is normalized with respect to the Eulerian virial size of the halo, $r_{\rm v } $.  The results are from the Oriana simulations.     }
\label{fig:rho_r_zE0_zLMany_bpt156_Oriana}
\end{figure*}

\begin{figure*}[!htb]
\centering
\includegraphics[width=\linewidth]{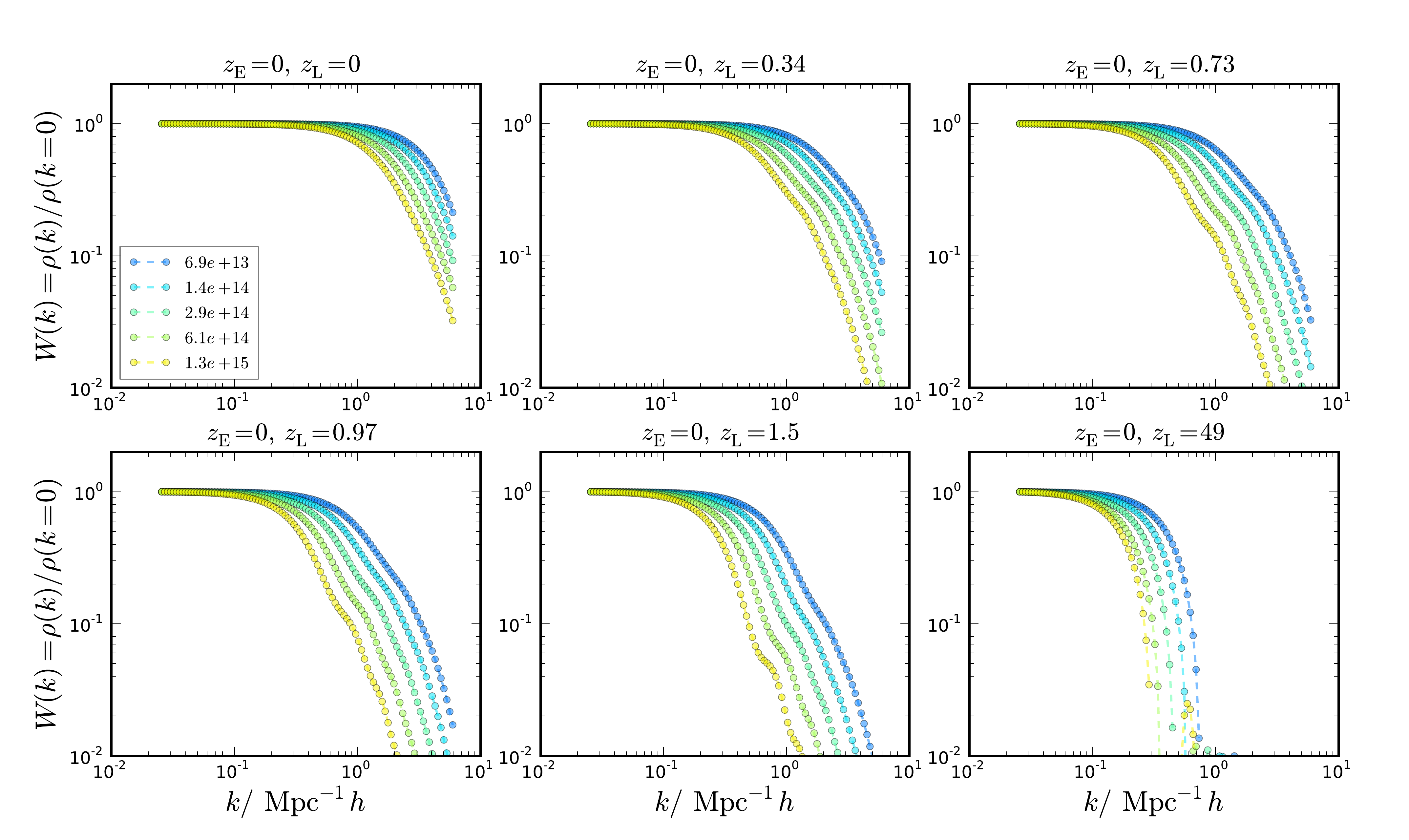}
\caption{  Same as  Fig.~\ref{fig:rho_r_zE0_zLMany_bpt156_Oriana}, except in Fourier space. The halo profile is normalized such that it approaches 1 as $k$ tends to 0.      }
\label{fig:rho_k_norm_zE0_zLMany_bpt156_Oriana}
\end{figure*}

In this section, we will measure the evolution of halo profile from $N$-body simulation.  Starting from Eulerian halos, such as those at $z=0$, we trace the particles in the Eulerian halo   back in time to construct the proto-halos at earlier times.  The position of the proto-halo at redshift $z$ is defined by the CM of its constituent particles at redshift $z$.  We will consider proto-halos at various redshifts.

Before presenting the numerical results we would like to first outline the details of the $N$-body simulation used here. We shall use the Oriana and Carmen simulations in the LasDamas project.  In these simulations, a flat $\Lambda$CDM model with the cosmological parameters, $\Omega_{\rm m} =0.25 $, $\Omega_{\Lambda} =0.75$ and $\sigma_8 =0.8 $ are adopted.  The transfer function is output from CMBFAST \cite{CMBFAST}. The initial conditions are Gaussian with spectral index being 1.   The initial displacement fields are created using 2LPT \cite{CroccePeublasetal2006} at $z=49$. The simulations are evolved using the code Gadget2 \cite{Gadget2}.  In the Oriana simulations, there are $1280^3$ particles in a cubic box of size 2400 $ \MpcOh $, while for Carmen simulation there are $1120^3$ particles in a box of size $1000 \MpcOh$. Thus the particle masses are $4.57 \times 10^{11} $ and $4.94 \times 10^{10} \Msun $ for Oriana and Carmen respectively.  We shall use five realizations for Oriana and seven for Carmen. The halos are obtained using Friend-of-Friend halo finder. For Oriana, the linking length $b=0.156$ is used, while $b=0.2$  for Carmen. To resolve the halo better, we use halos with at least 150 particles. Although the Carmen simulations have better mass resolution than Oriana, we find that their results are quite similar. To avoid redundancy, most of the time, we only show results from Oriana.

In Fig.~\ref{fig:rho_r_zE0_zLMany_bpt156_Oriana}, we show the halo profile at redshifts, $z=0$, 0.34, 0.73, 0.97, 1.5, and 49. The halo profile is obtained by stacking the halos in the same mass bin together and spherically averaged to get the spherically symmetric profile. In this plot, the Eulerian halo is at $z=0$ and the proto-halos at higher redshifts are constructed from the Eulerian ones. We find that when the size of halos of different masses normalized by their corresponding virial size, they coincide well with each other. We computed $r_{\rm v}$ using $\Delta_{\rm v} =500$ although this is immaterial to  our purpose here.

  Note that for redshift $z=0$, the halos are in fact in the Eulerian space. The virialized spherical Eulerian halo profile at low redshift is well fitted by the NFW profile. However, the halos used to construct this profile are carefully selected, see e.g.~\cite{NFW1996,NetoGaoBettetal2007}.  These halos  are constructed using spherical overdensity finder and they are chosen to be spherically symmetric and in a relaxed state without signatures of recent mergers.  Here we use all the halos obtained from the halo finder without further screening.  We find that our Eulerian profile is reasonably well fitted by the NFW profile, but we also find that the profile close to the virial radius  drops faster  than $r^{-3}$, the scaling of the NFW profile near virial radius. Using the NFW profile, one can show that the Eulerian profile is approximately universal in the variable $r/r_{\rm v} $ for different masses because of the fact that the concentration only weakly depends on the mass of the halo \cite{CooraySheth}.

The deviation of the proto-halo profile from the NFW profile increases as the redshift increases.  At $z=49$, the proto-halo profile corresponds  to the one in the initial condition of the simulation. Theoretically, the Lagrangian profile is often assumed to be a top-hat. In \cite{ChanShethScoccimarro2015}, it is found to be in between a Gasussian and a top-hat. More precisely, in Fourier space, the Lagrangian profile is well fitted by a product of a Gaussian and a top-hat window. We note that as $z$ increases, there are large deviations in the profile at small $r$ among different halo masses.

As there is no universal halo profile that can fit the proto-halo profile at various redshifts well, we shall use the numerical profile directly. The profile is Fourier transformed numerically. In the case of the NFW profile, the integration is cut-off at the virial radius  $r_{\rm v }$ \cite{CooraySheth}. In our case, for proto-halos at intermediate redshifts, i.e.~in between $z=0$ and $z=49$, it is not clear what the cut-off size should be. Nonetheless, our profile drops rapidly for $ r$ greater  than a few $r_{\rm  v }$, we can take $r$ to be infinity, and the results are unaffected. As the mass of the proto-halo is conserved, we expect the low-$k$ part of the profile at different redshifts to be the same. Numerically, however, this is not always  achieved.  In each set of simulation, we find that the fractional deviation of the $\rho(k=0) $ across redshifts decreases as the mass of the halo increases. For example, for the lowest mass halo used for Oriana, the fractional deviation of  $\rho(k=0) $ is within 10\%. This is one of the indications that we should use halos with large number of particles.  From now on, we simply normalize the profile so  that it is 1 at low $k$. In Fig.~\ref{fig:rho_k_norm_zE0_zLMany_bpt156_Oriana}, we show the Fourier transform of the halo profile for a selection of halo masses. As the  redshift $z_{\rm L}$ increases, the size of the proto-halo increases, the window in Fourier space decreases and the low-$k$ plateau shrinks. We also note that there are oscillations in the Fourier transform of the window. It is more visible as the mass of the halo increases because the oscillations are pushed to lower $k$. They are also more prominant as $ z_{\rm L} $ increases because the halo profile is more top-hat-like, and hence the wiggles are stronger.

\begin{figure*}[!htb]
\centering
\includegraphics[width=\linewidth]{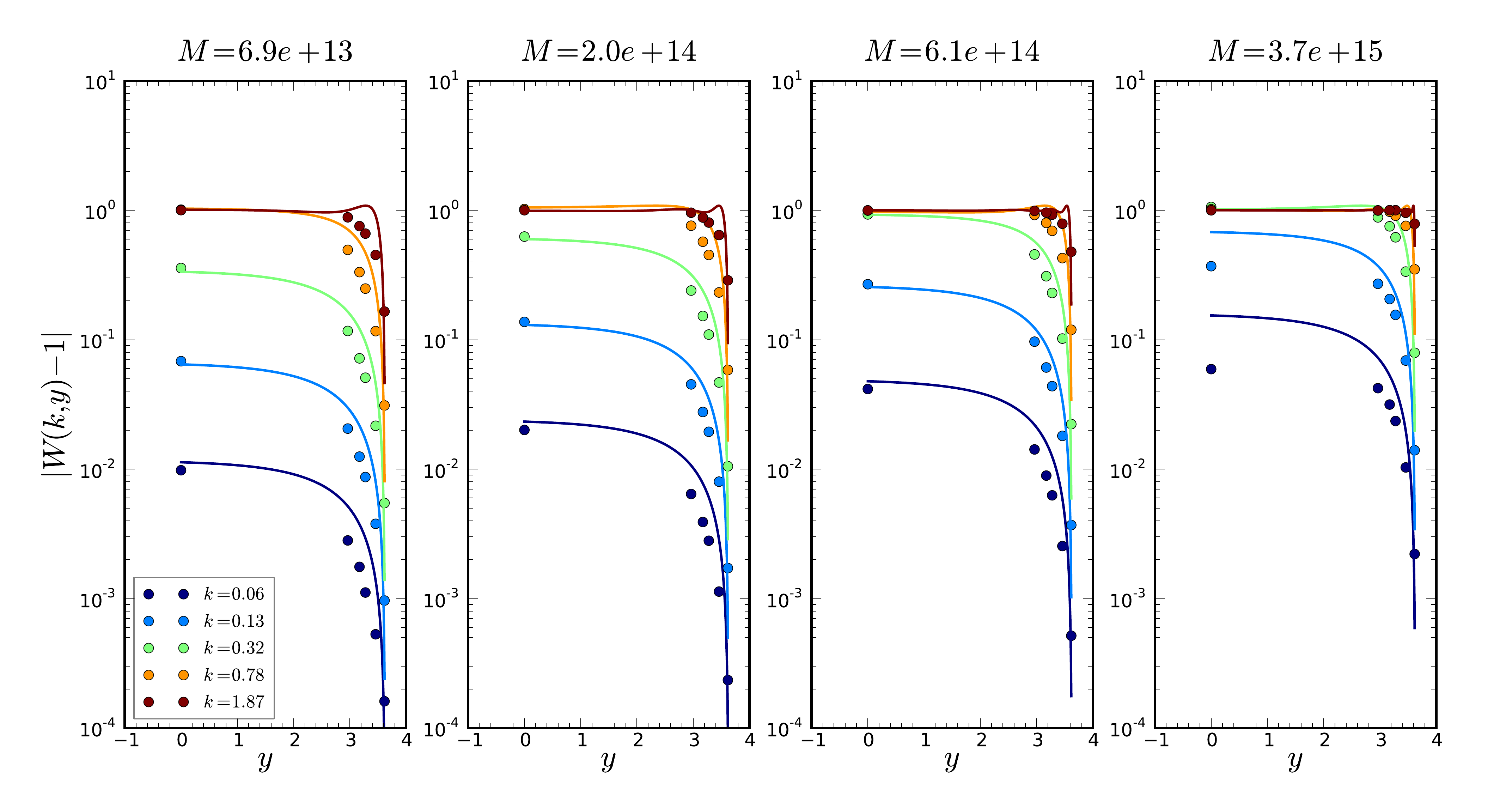}
\caption{  The function $|W-1|$ as a function of $y$, obtained from simulations (filled circles) and SC model (solid line). Results from four mass bins of mass $6.9 \times 10^{13}$, $2.0 \times 10^{14}$, $6.1  \times 10^{14}$, and  $3.7  \times 10^{15} \Msun  $ are shown (from left to right). For each mass bin, $| W(k,y)-1 |$ at six different $k$'s are plotted. The data is from Oriana.       }
\label{fig:W_1_zE0_zLMany_bpt156_Oriana}
\end{figure*}

\begin{figure*}[!htb]
\centering
\includegraphics[width=\linewidth]{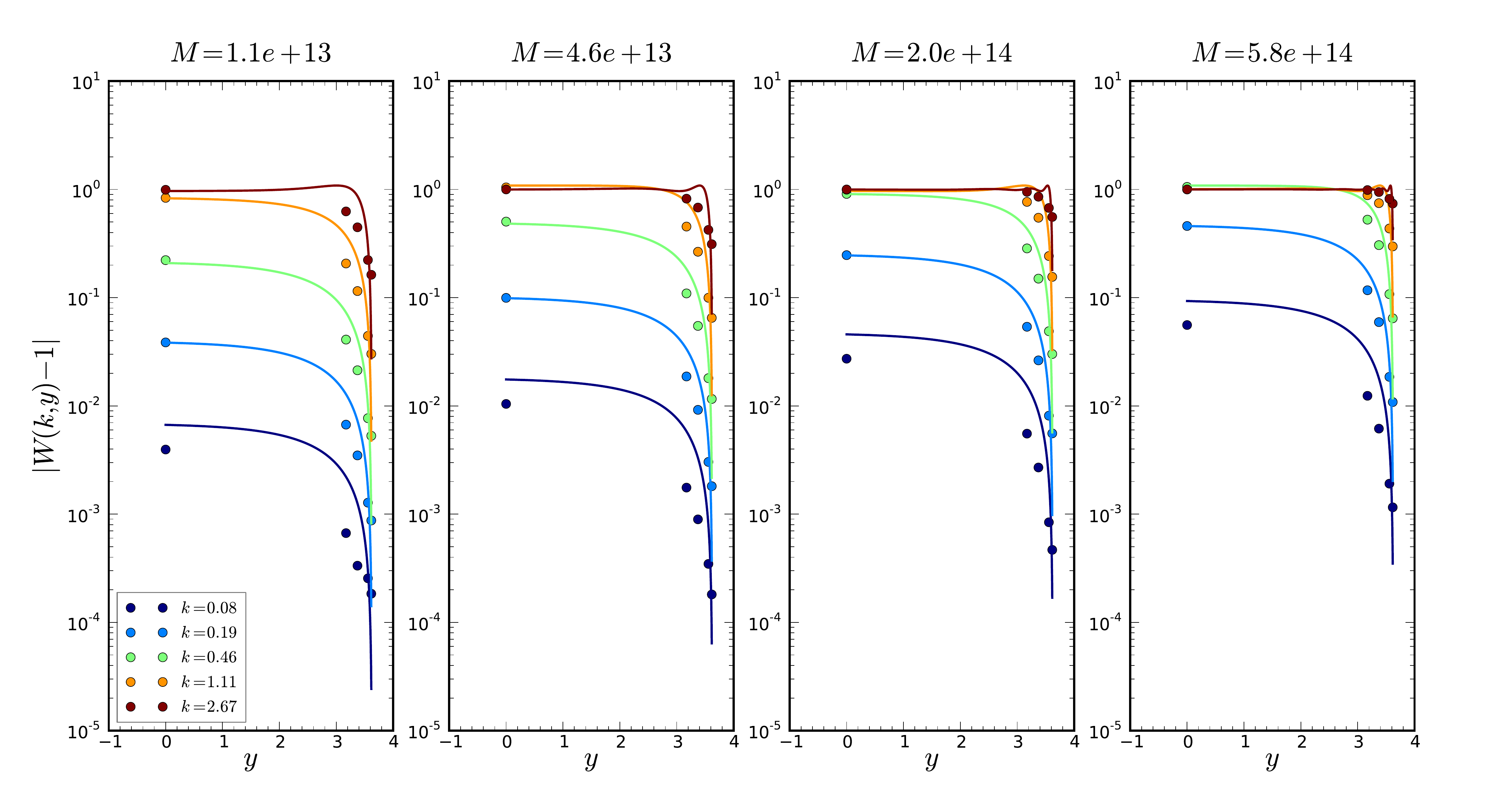}
\caption{ Similar to Fig.~\ref{fig:W_1_zE0_zLMany_bpt156_Oriana}, except for Carmen.      }
\label{fig:W_1_zE0_zLMany_bpt2_Carmen}
\end{figure*}

To show the scale-dependence of the window function, in Fig.~\ref{fig:W_1_zE0_zLMany_bpt156_Oriana}, we plot $ | W - 1 | $ obtained from simulations and the SC model. Absolute value is taken because the $k^2$ correction is negative. We have introduced the time variable 
$y=\ln D $, where $D$ is the linear growth factor (defined by Eq.~\ref{eq:DM_grothfactor}). It is normalized such that  $y=0 $ at $z_*=49$. As we will see in next section, this time variable is convenient.  First at $y=0$, SC model agrees with data well except for the  highest  mass bin shown ($ 3.7 \times 10^{15} \Msun $). Unfortunately, we have no simulation data available in the range $0 \lesssim  y \lesssim 3 $, although we expect that the overdensity is still in the expansion stage ($y \lesssim 3 $), the SC model should work reasonably well. When the region  turns around and collapses, we expect the SC model to fail to describe the simulation data accurately.  In fact, during the turn-around  and collapse  phase, the SC results are  larger than the simulation data. We also note that for various values of $k$ shown, the agreement between the SC model and the data is qualitatively smiliar.   In Fig.~\ref{fig:W_1_zE0_zLMany_bpt2_Carmen} we show the corresponding results obtained using the Carmen simulations. In this plot, the Eulerian halos are at $z=0$ and proto-halos are constructed at  $z=0.13 $, 0.52, 0.97 and 49.   Although Carmen has better mass resolution, the results are quite similar to those obtained from Oriana.  Overall, the agreement between the simulation results and SC model is reasonable. 

 The SC model only works qualitatively at late time. After all, in SC the halo profile shape does not change, but for real halos the halo profile shape does change. In \cite{Engineer_etal2000}, a modified SC model that tries to overcome the jump at the final virialization stage was proposed. The model joins smoothly to the final virial scale at the expense of two additional free parameters. The modified model is valid only when the density is high as it is an expansion in  $1/\delta$. Nonetheless, using these additional parameters, one may get a profile evolution history, especially the part from turn-around to collapse, to agree with the simulation results better.   Another potential way to improve the modelling is to use ellipsoidal model \cite{Peebles1980,BondMyers1996, ShenAbelMoSheth2006,MoBoschWhite2010}. For example, the halo mass function motivated by the ellipsoidal collapse improves the agreement  with simulation  \cite{ShethMoTormen2001} compared to the spherical Press-Schechter one. The halo collapse threshold is also better modelled by the ellipsoidal collapse model \cite{RobertsonKravtsovTinkerZentner2009}.  However,  as the halo profile considered here is spherically averaged, one still need to average over the ellipsoidal profile to get the spherically symmetric one.  On the data side, we hope to get the data to fill the gap in between 0 and 3 in future. As we see the model does not work very well, in practice it will be useful to come up with a parametrized form for the evolution of the profile. Also the paramertized form of halo can be used to improve the halo model. In the standard halo model one assumes that all the matter exists within halos, and the virialized halo profile is used, such as the NFW profile for halos \cite{CooraySheth}. However at higher redshift, virialized halos are rare, and this assumption is not justified. One can improve the halo model using the proto-halo profile instead.

\section{Bias with profile corrections}
\label{sec:BiasComputations}

We shall apply the fluid approximation to model the evolution of the dark matter and the galaxy field. The fluid approximation enable one to derive the nonlocal bias parameters \cite{Fry1996,ChanScoccimarroSheth2012,Baldaufetal2012} which results in better modelling of the halo power spectrum and bispectrum \cite{ChanScoccimarroSheth2012,Baldaufetal2012,SaitoBaldauf_etal2014,BiagettiDesjacquesKehagiasRiotto2014a} and halo 3-point function \cite{BelHoffmannGaztanagn2015}.  In this paper we  use halo and galaxy interchangeably. For dark matter, we will use the standard perturbation theory (SPT) results (see \cite{PTreview} for a review). In this framework, the evolution of the density contrast of the galaxy, $\dg  $,  and its velocity divergence   $\tg  $ are governed by the  continuity equation and the Euler equation
\begin{widetext}
\beqa
\label{eq:continuity_0}
 \frac{\partial \dg  }{ \partial \tau } + \tg  & = & - \int d^3 k_1 d^3 k_2   \Ddel(\mb{k} - \mb{k}_{12} )   \alpha( \mb{k}_1, \mb{k}_2 ) \tg( \mb{k}_1 )   \dg( \mb{k}_2 ) ,  \\
\label{eq:Euler_0}
 \frac{ \partial \tg  }{ \partial \tau } + \mathcal{H}  \tg + \frac{ 3 }{ 2 } \mathcal{H}^2 \Omega_{\rm m} W \delta  &=&  - \int d^3 k_1 d^3 k_2  \Ddel(\mb{k} - \mb{k}_{12} )   \beta( \mb{k}_1, \mb{k}_2 ) \tg( \mb{k}_1 )   \dg( \mb{k}_2 ),  
\eeqa
where $ \tau  $ is the conformal time, $\mathcal{H}$ is the conformal Hubble parameter $d \ln a / d \tau  $, $\mb{k}_{12}$ denotes $ \mb{k}_1 + \mb{k}_2  $, and  $\alpha $ and $\beta $ are the coupling kernels
\beq
\alpha( \mb{k}_1, \mb{k}_2 )  = \frac{ \mb{k}_{12} \cdot \mb{k}_1 }{ k_1^2  }, \quad  \beta( \mb{k}_1, \mb{k}_2 ) = \frac{ k_{12}^2 \mb{k}_1 \cdot \mb{k}_2 }{ 2 k_1^2 k_2^2 }.
\eeq
Here $\Omega_{\rm m} $ is the density parameter of matter.

Eq.~\ref{eq:continuity_0} and \ref{eq:Euler_0} are similar to the fluid equations widely adopted for modeling the evolution of multiple components \cite{SomogyiSmith2010, EliaKulkarnietal2011,BernardeauVandeRijtVernizzi2012,ChanScoccimarroSheth2012}, except with the window function $W$, which is central to the results in this paper. We also note that in \cite{BaldaufDesjacquesSeljak2014}, a similar modification of the Euler equation was proposed, in which the authors argued the forced for halos should be biased. However, the physical origin of this  modification and its form  are  quite different from that in \cite{BaldaufDesjacquesSeljak2014}.  Another important difference from \cite{BaldaufDesjacquesSeljak2014,BiagettiDesjacquesKehagiasRiotto2014b} is that we do not impose the peak constraint in the evolution equations. The proto-halos after initial identification, they simply evolve following Eq.~\ref{eq:continuity_0} and \ref{eq:Euler_0}.

The introduction of $W$ is to model the fact that although $\dg$ denotes the density contrast of the spatial distribution of the CM of the halos, each individual halo consists of a collection of particles. Thus the force on the CM of the halo should be the average force acting on all the individual particles in the halo.  Hence in real space the effective source of the gravitational force for a finite-sized object is $W*\delta$, instead of only $\delta $ at the CM position of the object.   In Fourier space, it is given by the product between $ W $  and $\delta$  thanks to the convolution theorem.  This window function describes the profile of the object. We will use the window function/profile studied in Sec.~\ref{sec:halo_profile}.

To simplify Eq.~\ref{eq:continuity_0} and \ref{eq:Euler_0} further, we introduce the new time variable $y=\ln D $ where $D$ is the linear growth factor for the dark matter satisfying the equation 
\beq
\label{eq:DM_grothfactor}
\frac{d ^2 D  }{ d \tau^2 } + \mathcal{H} \frac{d D }{d \tau } - \frac{3  }{2} \mathcal{H}^2 \Omega_{\rm m} D = 0.
\eeq
We note that $f^2 \approx \Omega_{\rm m}  $, with $f = d \ln D / d \ln a $ is a very good approximation for the  epoch that we are interested in \cite{SCFFHM98}. Using this approximation Eq.~\ref{eq:continuity_0} and \ref{eq:Euler_0} can be written as 
\beqa
\label{eq:continuity_1}
\frac{ \partial \dg  }{ \partial y } -  \tilde{ \tg} & =& \int d^3 k_1  d^3 k_2 \Ddel ( \mb{k}- \mb{k}_{12} )   
\alpha( \mb{k}_1 ,  \mb{k}_2 ) \tilde{ \tg} ( \mb{k}_1 ) \dg ( \mb{k}_2 )  ,  \\ 
\label{eq:Euler_1}
\frac{ \partial \tilde{ \tg}  }{ \partial y } + \frac{ 1 }{2  } \tilde{ \tg}  -  \frac{3  }{2 } W \delta  & = &  \int d^3 k_1  d^3 k_2 \Ddel ( \mb{k}- \mb{k}_{12} )   
 \beta( \mb{k}_1 ,  \mb{k}_2 ) \tilde{ \tg} ( \mb{k}_1 ) \tilde{ \tg} ( \mb{k}_2 ) ,
\eeqa
where  $\tilde{ \tg}$  denotes $ \tg / ( - f \mathcal{H}  )$. In the rest of the paper, we shall abuse the notation and simply use $\theta  $ and  $ \tg $ to denote $ \theta / ( - f \mathcal{H}  )$ and $ \tg / ( - f \mathcal{H}  )$ respectively.

In the following subsections, we will solve Eq.~\ref{eq:continuity_1} and \ref{eq:Euler_1} to linear and second order respectively to reveal the effects of the window function on the bias parameters. In \cite{ChanScoccimarroSheth2012}, the continuity and Euler equation of the galaxy field together with the other two equations for dark matter were written in a concise form, and hence a general perturbative solution was  obtained using the transient formalism \cite{Scoccimarro98}, thanks to the fact that the coefficients of the equations are not explicitly time-dependent. However, $W$ is  time-dependent as we see in Sec.~\ref{sec:halo_profile}. Here we will solve Eq.~\ref{eq:continuity_1} and \ref{eq:Euler_1} directly.

\end{widetext}

\subsection{ Linear biases }
\label{sec:LinearBiasComputation}

 We start from the linearized version of Eq.~\ref{eq:continuity_1} and \ref{eq:Euler_1} 
\beqa
\label{eq:continuity_deltag_linear}
\partial_y \dg^{(1)}  &= &  \tg^{(1)}  , \\
\label{eq:Euler_deltag_linear}
\partial_y \tg^{(1)} + \frac{ 1 }{ 2 } \tg^{(1)} &=& \frac{ 3 }{ 2 } W \delta^{(1)} ,
\eeqa
where the superscript (1) emphasizes that the field is linear.  We have suppressed the explicit $ \mb{k}$-dependence.

\subsubsection{Velocity}

\begin{figure}[!htb]
\centering
\includegraphics[width=\linewidth]{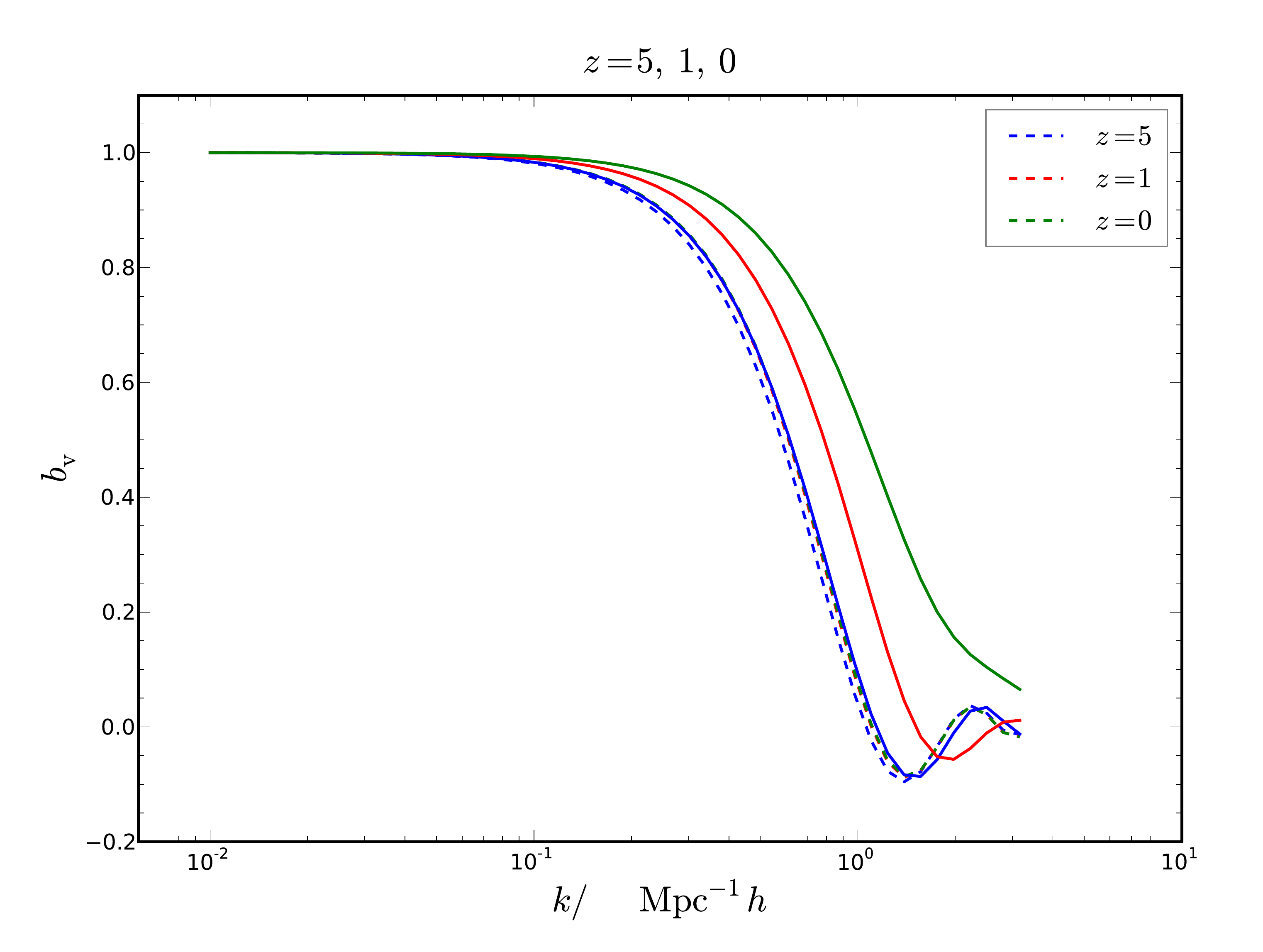}
\caption{ Evolution of the linear velocity bias with  static  window (dashed) and that with window given by SC (solid). Three redshifts are shown $z=5$ (blue), 1 (red) and 0 (green). The case with static window is almost constant (thus some of the curves are covered by the blue curves), while the evolving window case gives decaying, but not negligible  $b_{\rm v} $.   }
\label{fig:bv_tot_evolve}
\end{figure}

\begin{figure*}[!htb]
\centering
\includegraphics[width=\linewidth]{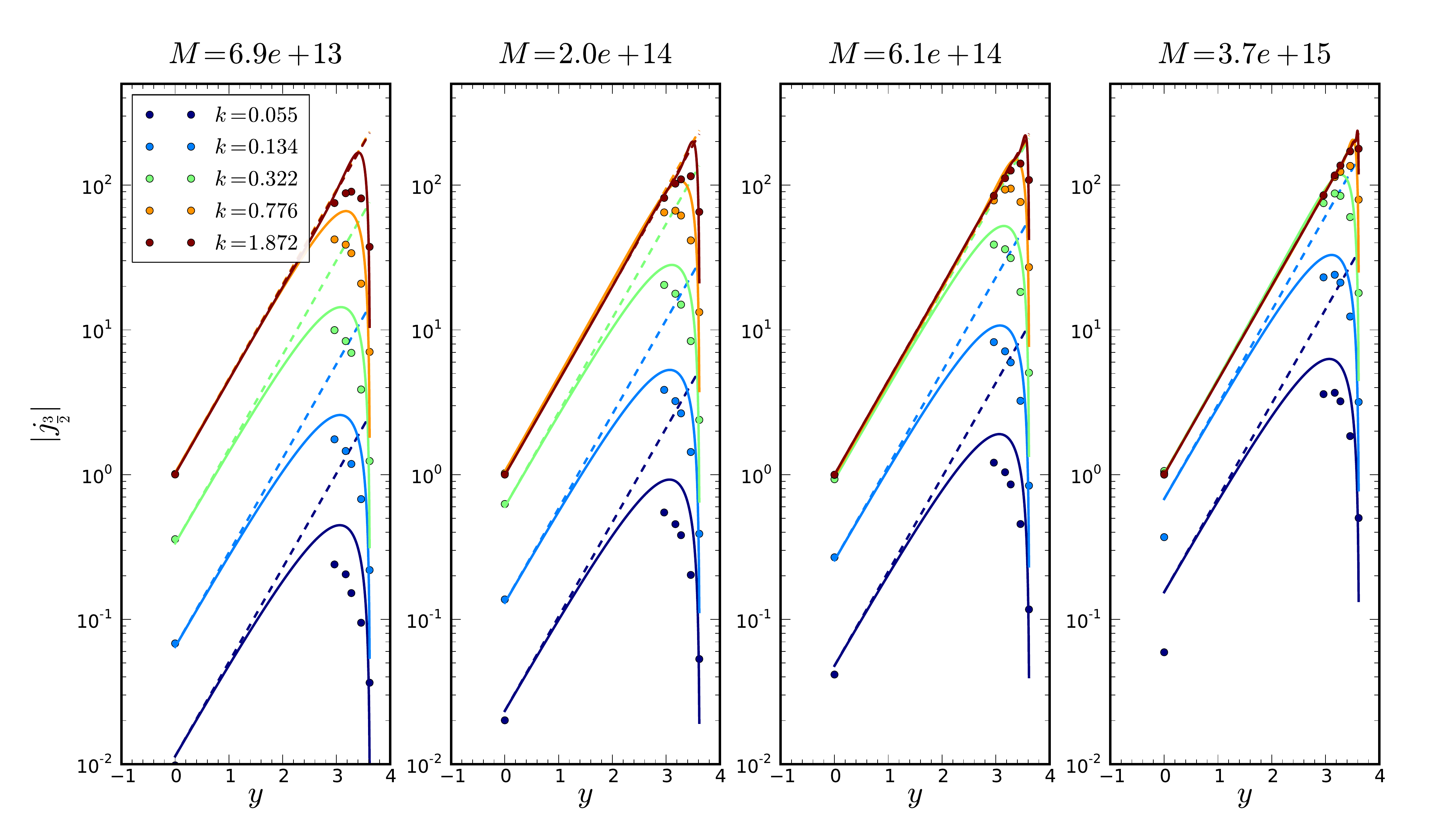}
\caption{ The integrand $ | j_{\frac{3 }{2 }} | $ as a function of $y$ for various values of $k$, obtained from simulation data (filled circles).   The results from the SC (solid) and the static window (dashed) are also shown.  The simulation data is from Oriana.  }
\label{fig:j3o2_zE0_zLMany_bpt156_Oriana}
\end{figure*}

\begin{figure*}[!htb]
\centering
\includegraphics[width=\linewidth]{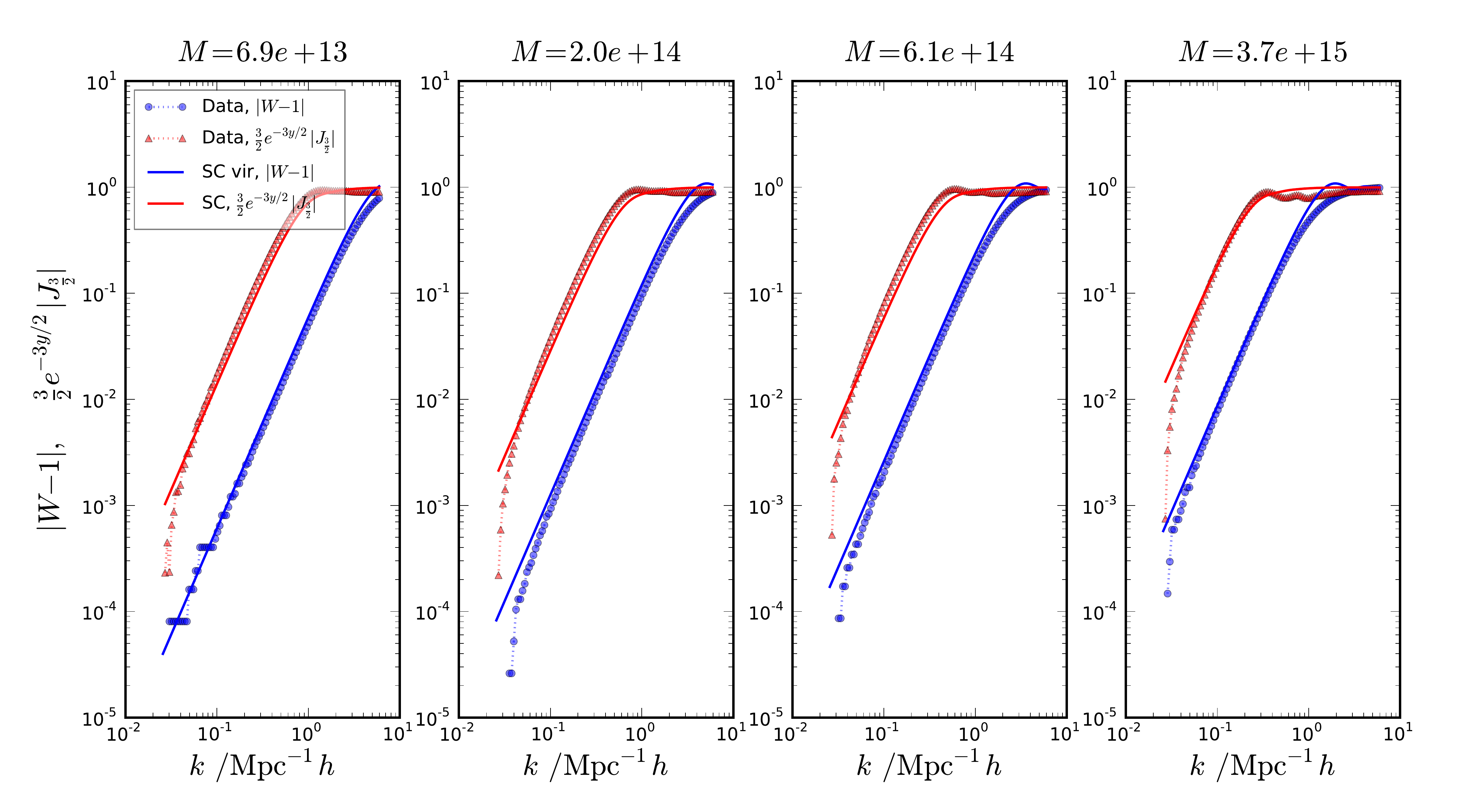}
\caption{ The comparison of the contribution to $k^2$ from $|W-1| $  (blue) and $\frac{3 }{ 2} e^{- \frac{3 y}{ 2} } | J_{ \frac{3}{2} } |$ (red), which is the main contribution to $k^2$ correction in the velocity bias in Eq.~\ref{eq:bw_in_w-1}.  The results from simulations (symbol) and SC (solid line) are shown. The simulation data is from Oriana. For $W-1$ from SC, we have used $r_{\rm v} $ computed with $\Delta_{\rm v}=500  $.   The quantities are evaluated are $z=0$.      }
\label{fig:W1vsU_zE0_zLMany_bpt156_Oriana}
\end{figure*}

We will work on  the Euler equation first because it  depends only on $ \tg^{(1)}  $.  Integrating Eq.~\ref{eq:Euler_deltag_linear} in terms of $W$ and $\delta^{(1)} $, we get 
\beq
\label{eq:theta_g_lin}
\tg^{(1)} (y) = \tg^{*(1)} e^{- \frac{y  }{ 2 } } + \frac{ 3 }{ 2 } \int_0^y  dy' W(y') \delta^{(1)} (y') e^{ -\frac{ 1 }{ 2 }( y - y' ) }.  
\eeq
 As we mentioned before, in this paper, we use star ``*'' to denote a quantity at some initial time. Thus $ \tg^{*(1)} $ is the velocity divergence of the galaxy at the initial time.  For convenience, we define the function $I_n$ as 
\beq
I_n(y)  =  \int_0^y d y' W(y') e^{ n y' }, 
\eeq
thus we have
\beq
\label{eq:thetag_1}
\tg^{(1)}(y) = \tg^{*(1)}  e^{- \frac{y  }{ 2 } } + \frac{ 3 }{ 2 } \delta_*^{(1)} e^{- \frac{ y}{ 2} } I_{\frac{3  }{2} },
\eeq
where we have used the SPT result
\beq
\delta^{(1)}(y) =  \delta_*^{(1)} e^y. 
\eeq
Note that we have normalized the linear growth factor to be 1 at the initial time so that $y_* =0 $.

The linear velocity bias $\bv$ then is given by \footnote{The linear bias parameters  are defined differently from that in peak theory \cite{DesjacquesSheth2010}, where a smoothing window function is divided by. For example, the $b_{\rm v}$ defined here is equal $\tilde{b}_{\rm v}  W_{\rm s} $, where  $W_{\rm s} $ is a smoothing window and only  $\tilde{b}_{\rm v} $ is the called the velocity bias in \cite{DesjacquesSheth2010}.  Our more ``direct'' definition is closer to the standard treatment, where the window is not explicitly written down but its effects will be included in  $b_{\rm v}$. } 
\beqa
\label{eq:bvlin_W}
\bv(y) & \equiv & \frac{ \tg^{(1)}(y)  }{\theta^{(1)} (y)  }   \nn  \\
       &=& \bv^* e^{ - \frac{ 3 y }{ 2 } } + \frac{ 3 }{ 2 }  e^{ - \frac{ 3 }{2 } y } I_{ \frac{ 3 }{ 2} } , 
\eeqa 
where we have used 
\beq
\theta^{(1)}(y) = \theta_*^{(1)} e^y =  \delta_*^{(1)} e^y  . 
\eeq
Correspondingly the initial linear velocity bias $ \bv^* $ is defined as
\beq
\bv^* \equiv \frac{ \tg^{*(1)}  }{ \theta_*^{(1)} }. 
\eeq
In Fourier space, the window function approaches 1 at low $k$, thus it is convenient to express the integral in Eq.~\ref{eq:bvlin_W} in terms of $W-1 $. Hence we have instead 
\beq
\label{eq:bw_in_w-1}
\bv = 1 + ( \bv^* - 1 ) e^{-\frac{ 3 y}{ 2 } } + \frac{ 3 }{ 2 } e^{-\frac{ 3 y}{ 2 } } J_{ \frac{ 3 }{ 2 } }(y),
\eeq 
where $J_n$ denotes the integral 
\beq
J_n (y) = \int_0^y d y' [ W(y') - 1 ] e^{n y' }.
\eeq
The advantage of introducing $ J_n $ is that it gives at least $k^2$ order correction, thus it represents the genuine halo profile correction.  The first two terms in Eq.~\ref{eq:bw_in_w-1} are the velocity bias evolution obtained in \cite{ChanScoccimarroSheth2012}, and the last term is new, which arises from the halo profile. In the limit of large $y$,   the profile correction does not vanish, instead  $\bv $ tends to  $1  +  \frac{ 3 }{ 2 } e^{-\frac{ 3 y}{ 2 } } J_{ \frac{ 3 }{ 2 } }(y) $. If we assume that  the window is static, we get $W$.

In Fig.~\ref{fig:bv_tot_evolve}, we  plot  $\bv$ at $z=5$, 1 and 0.  To set the initial condition,  $b_{\rm v}^* $ at $z_*=49 $, we borrow the results from peak theory \cite{DesjacquesSheth2010}
\beq
\label{eq:bvs_peak}
\bv^* = \Big( 1 -  \frac{ s_0  }{ s_1 }  k^2 \Big) W_{\rm G}( k R_{\rm G}),
\eeq
where $W_{\rm G} $ denotes the Gaussian window function and $s_n $ is the spectral moment defined as 
\beq
s_n = 4 \pi \int dk k^{ 2( n+1) } P(k) W_{\rm G}^2( kR_{\rm G} ) . 
\eeq
Let's clarify the reason that we set the initial conditions using the peak theory even though we dispute about its prediction at late time. Various studies, e.g.~\cite{EliaLudlowPorciani2012,ChanShethScoccimarro2015} show that the density and velocity cross power spectrum between halo and matter in the \textit{Lagrangian} space can be well fitted by the functional form motivated by the peak theory. Thus one may think that here we only use an established fact from simulations. However we argue that the subsequent evolution can be modelled by the simple fluid approximation augmented with the window function.

In the plot, as an example we  will consider halo of mass $2 \times 10^{13} \Msun $.  We map the top-hat window size to the Gaussian window size using the relation $ R_{\rm G} = R_{\rm TH} / \sqrt{5} $. We compare the case when the window is static, with the window size given by the Lagrangian size and the case in which the window size is evolved by the SC model. The difference in treatment is only in $J_{\frac{3}{2} } $ in Eq.~\ref{eq:bw_in_w-1}.   We note that when the window is static, the resultant $b_{\rm v}  $ is almost constant over time. However, when the window is evolved by the SC model,  $b_{\rm v} $ decays over time, but it is not negligible at late time.

Although we have used Eq.~\ref{eq:bvs_peak}, as in the initial condition, the contributions from the initial scale-dependent part is small compared to the ones due to $J_n$, even if we have used  $ \bv^* =1  $, we find that the results are quite similar to that in Fig.~\ref{fig:bv_tot_evolve}. This highlights that the scale dependence of $\bv  $ is mainly driven by the late time halo profile.



In the literature, the velocity bias is often associated with $k^2 $ correction to both the density and velocity biases, e.g. these $k^2 $ corrections can be derived from the peak model \cite{DesjacquesSheth2010}. In \cite{DesjacquesCrocceetal2014}, Zel'dovich approximation was used to displace the peaks to the Eulerian space, and they found that the velocity bias remains constant over time.  The numerical measurement seemed to be in favour of the peak model result \cite{BaldaufDesjacquesSeljak2014}.   Here we show that taking into account that halos are  composite objects there is significant $k^2$-correction to the velocity bias and it does not decay away over time. When the static window is used, we also find that $ b_{\rm v}$ reduces to $W$ in the long term limit.  However, when the evolving SC model is applied, the velocity bias is not constant, as can be seen from Fig.~\ref{fig:bv_tot_evolve}. Even in the static window limit, our result ($W$) is still different from \cite{DesjacquesCrocceetal2014}, which gets $W b_{\rm v}^{\rm pk }$ instead. The reason for this difference is that we do not impose the peak constraint in proto-halo evolution \cite{BiagettiDesjacquesKehagiasRiotto2014b}. These differences can be used to differentiate these two models.

Alternatively we can express the linear bias parameters in terms of the time derivative of the profile. Integrating the integral in Eq.~\ref{eq:bvlin_W} by parts, $ b_{\rm v} $ can be written as
\beqa
\label{eq:bvlin_Wp}
b_{\rm v}( y)  &=&  W(y) +  [ b_{\rm v}^* - W(0) ] e^{  - \frac{ 3 }{ 2 } y }  \nn \\
    &-& \int_0^y d y'  W'(y') e^{ - \frac{ 3 }{ 2 } (y - y') } ,
\eeqa
where $W'$ denotes $ \partial W/  \partial y $. This form shows that there are two contributions to the $k^2$ correction from the profile, one from $W$ and another from $W'$.  The contribution from $W$ is simply the trivial smoothing.  However, numerically performing derivatives on sparse  data can lead to noisy results. Thus we will only use the form in terms of $J_n$, such as  Eq.~\ref{eq:bw_in_w-1}.

As the velocity bias is mainly generated by  $J_{\frac{3}{2}} $ in Eq.~\ref{eq:bw_in_w-1},  to gain insight into which part of the integral of $J_{\frac{3}{2}}  $ contributes most, we plot the integrand of $J_{\frac{3}{2}}$,  $j_{\frac{3}{2}  }$
\beq
j_{\frac{3}{2}} ( y) =  e^{ \frac{3 }{2 }  y  } \Big( W(y) -1 \Big) 
\eeq
 in Fig.~\ref{fig:j3o2_zE0_zLMany_bpt156_Oriana}. Again, in the range $ 0 <y \lesssim 3 $, there are no data available. We also show the prediction from the SC model. SC predicts that the contribution to the results in that range is small, while the contribution around $y\sim 3$ is the largest. However, we note that the SC results often overshoots in this range.   We also show the results obtained with the static window. Static window approximation is good for $y\lesssim 2 $, but it overestimates the results for $y \gtrsim 3 $.

As both the window function $W$ and the time integral $J_{\frac{3}{2} }$ contribute to leading $k^2$ correction, we would like to compare the magnitude of these terms.  In Fig.~\ref{fig:W1vsU_zE0_zLMany_bpt156_Oriana}, we compare $k^2$ contributions from $| W-1 |$ and $\frac{3  }{ 2 } e^{- \frac{3 y}{2} } | J_{ \frac{3}{2}}  | $ using both the numerical results and the SC model. The results are for $z=0$. The SC model gives quite good description of the results from the data. In particular, the value of  $\frac{3  }{ 2 } e^{- \frac{3 y}{2} } | J_{ \frac{3}{2}}  | $ from numerical window and the SC model agrees quite well. 

For the SC result for $ |W-1| $, we have compared a few prescriptions for the size of the window. Although at $z=0$ the size has not yet collapsed exactly to zero, using such a value gives the magnitude of $W-1$ much smaller than the simulation results. We have tried using Eq.~\ref{eq:rv_Deltav} with $\Delta_{\rm v} =200  $, 380 and 500, and  $\Delta_{\rm v} = 500 $ gives the best agreement with simulations.   In Fig.~\ref{fig:W1vsU_zE0_zLMany_bpt156_Oriana} we have shown the results obtained using  $\Delta_{\rm v} = 500 $. In passing, if we simply use Eq.~\ref{eq:rvir_rm_half}, which is strictly only for matter-dominated universe, we get the results very similar to those from $\Delta_{\rm v} = 380 $. We have cross-checked the results using Carmen, and they are consistent with those from Oriana.

We note that recently there are reports of measurements of velocity bias at late time \cite{BaldaufDesjacquesSeljak2014,JenningsBaughHatt2014,ZhengZhangJing2014b}. In \cite{BaldaufDesjacquesSeljak2014}, the momentum was measured, and they found that the model with  $b_{\rm pk} $ arised due to the peak constraint seems to fit the data better at high redshift such as $z=20$ than the evolution model without it. At such high redshifts, the effect of the profile evolution is small as can be seen from  Fig.~\ref{fig:bv_tot_evolve}. Thus if confirmed, this would shows that the profile correction would not be the dominant effect of the velocity bias seen in simulations.  
 In the study of \cite{ZhengZhangJing2014b}, velocity bias was measured with  sampling bias correction applied. They found that the velocity bias at $k\sim 0.08 \hOMpc  $ is slightly positive, with $ \bv \sim 1.01 $. In our model, velocity bias can only be negative in the mildly nonlinear regime.   However, as the number density of halos is low and halos are more inhomogeneously distributed in Eulerian space, thus it is hard to get an accurate volume weighted measurement. It is not clear that these measurements are free of artifacts. Thus we will keep these in mind and hope to report our own numerical comparison in future.





\subsubsection{Density}


\begin{figure}[!htb]
\centering
\includegraphics[width=\linewidth]{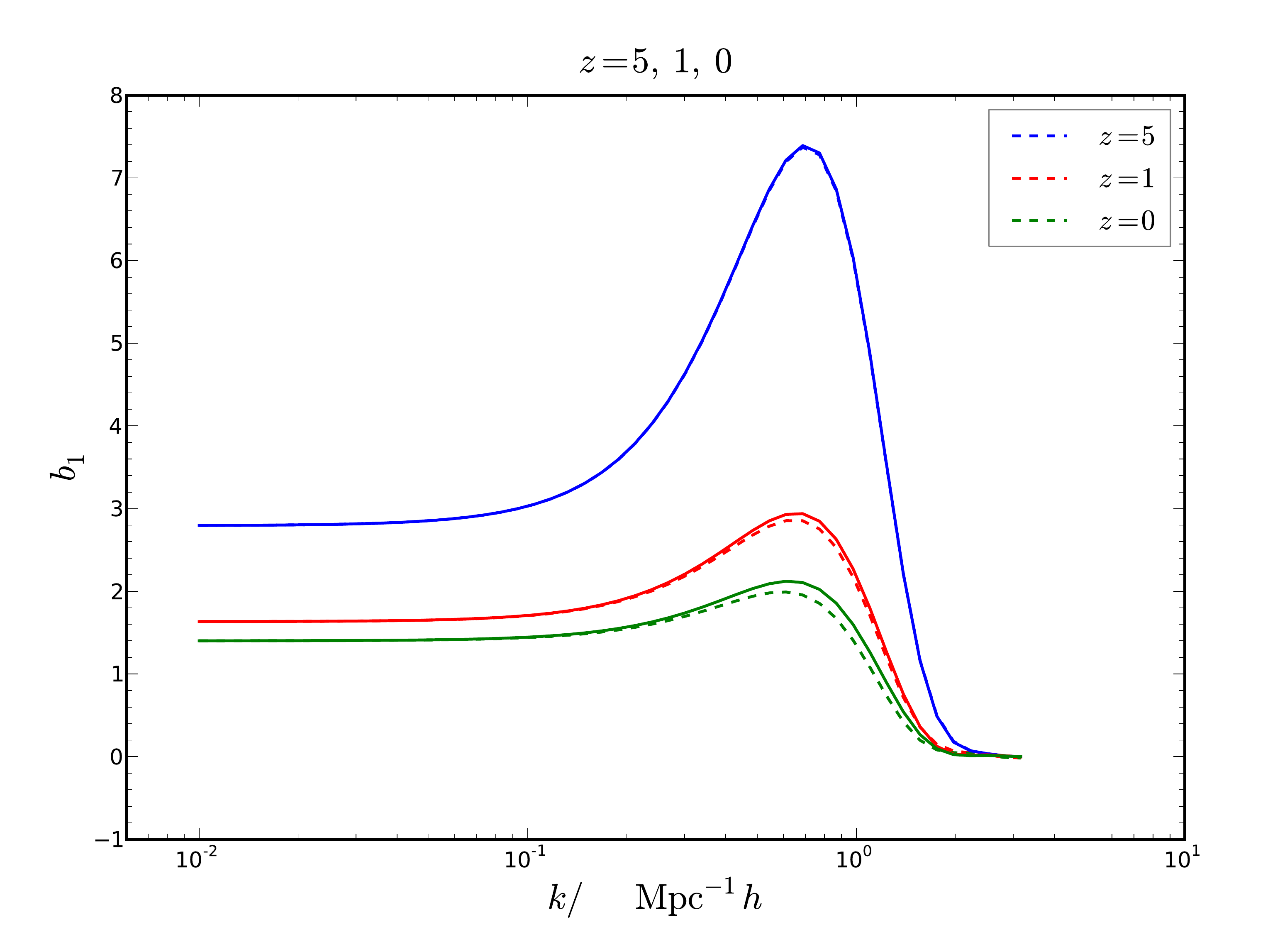}
\caption{ Evolution of the linear density bias with static window (dashed) and evolving window given by the SC (solid).  Three redshifts are shown: $z=5 $ (blue), 1 (red) and 0 (green).     }
\label{fig:b1_tot_evolve}
\end{figure}

\begin{figure}[!htb]
\centering
\includegraphics[width=\linewidth]{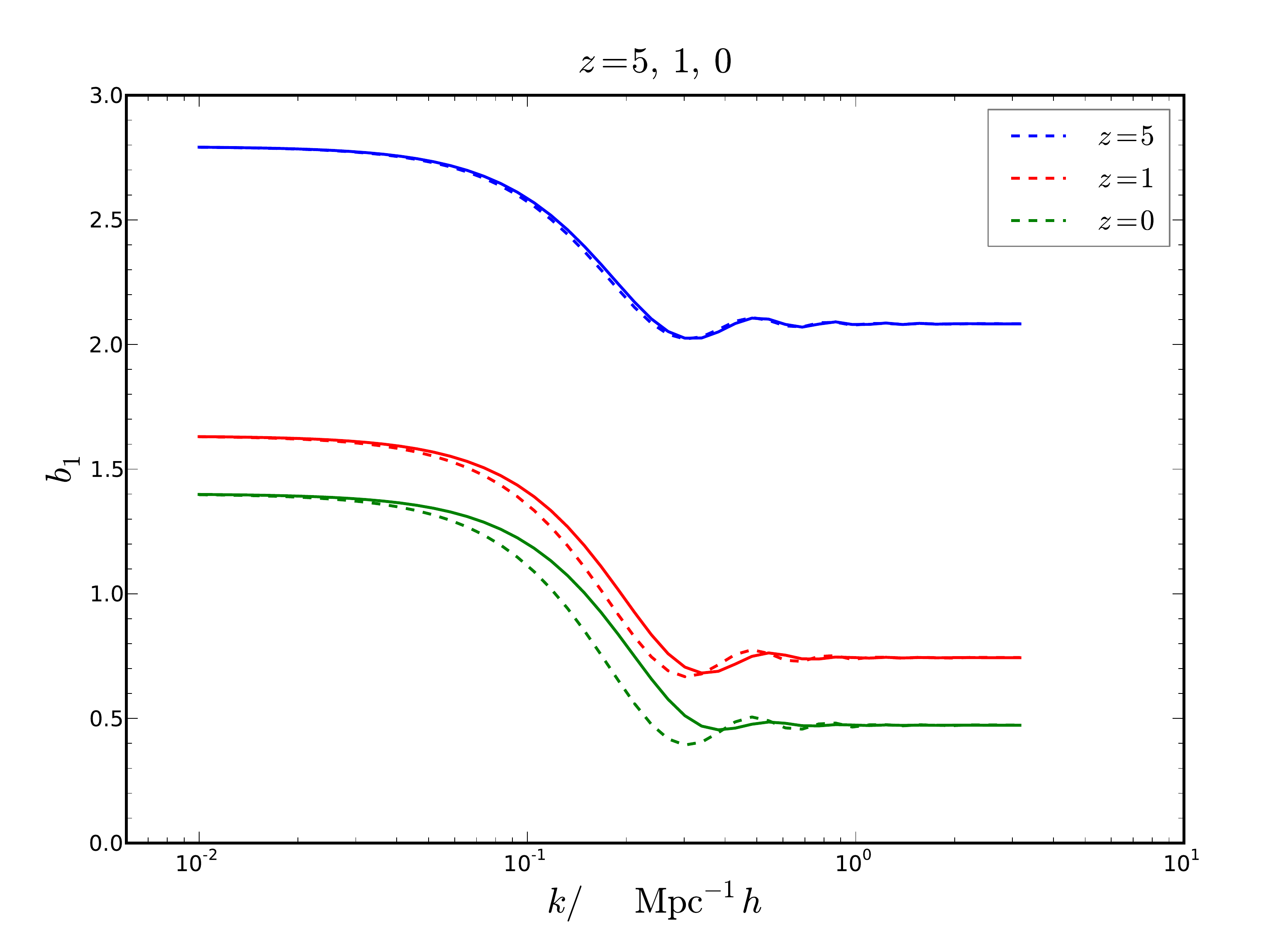}
\caption{ Same as Fig.~\ref{fig:b1_tot_evolve}, except the scale independent initial conditions $b_1^* = b_{\nu} $ and $ \bv^*=1 $ are assumed.    }
\label{fig:b1_tot_evolve_constant_bstar}
\end{figure}


We now turn to the density bias.  Plugging Eq.~\ref{eq:theta_g_lin} into Eq.~\ref{eq:continuity_deltag_linear}, we have 
\beqa
\label{eq:deltag_step1}
\dg^{(1)} (y) &=& \dg^{*(1)}  +  2 \tg^{*(1)} (1 - e^{ - \frac{y }{ 2 } } )  \nn \\
&+& \frac{3 }{ 2 } \delta_*^{(1)} \int_0^y d y' e^{ - \frac{ y' }{ 2 }  }  I_{\frac{ 3 } { 2 }} (y')  .
\eeqa
It is useful to note that 
\beq
\label{eq:int_e_I}
\int_0^y dy' e^{a y'} I_b(y')  = \frac{ 1 }{ a } [ e^{ay} I_b ( y) - I_{a+b} (y)]. 
\eeq
Using Eq.~\ref{eq:int_e_I}, we can simplify Eq.~\ref{eq:deltag_step1} to 
\beq
\label{eq:deltag_1}
\dg^{(1)}(y) = \dg^{*(1)} +  2 \tg^{*(1)} (1 -  e^{ - \frac{y }{ 2 } }  )  - 3 \delta_*^{(1)} [ e^{ - \frac{ y }{ 2 } } I_{\frac{3  }{ 2 }} - I_1   ].
\eeq

Thus the linear density bias is given by 
\beqa
b_1 (y) &\equiv & \frac{  \dg^{(1)} (y) }{ \delta^{(1)}(y) }  \nn \\
      &=&   b_1^* e^{-y} + 2 \bv^* e^{-y} ( 1 -e^{ - \frac{ y }{ 2 }   } )  \nn \\
       & & \quad   -  3 e^{-y} ( e^{ - \frac{ y }{ 2 }  } I_{\frac{ 3 }{ 2 }} - I_1  ) ,
\eeqa
where $b_1^*  $ is defined as
\beq
b_1^* \equiv \frac{ \dg^{*(1)}  }{ \delta_*^{(1)} }. 
\eeq

Or in terms of $J_n$ using $ I_n =  J_n + (e^{ny} -1)/n  $, we have
\beqa
\label{eq:b1_w_1}
b_1(y) &=& 1 + ( b_1^*  + 2 \bv^* - 3 ) e^{-y} + 2 (1 - \bv^* ) e^{ -\frac{3 y}{2} } \nn \\
&+& 3 e^{-y} J_1 - 3 e^{ - \frac{ 3y }{2} } J_{\frac{ 3 }{ 2 } }. 
\eeqa 
The first line in Eq.~\ref{eq:b1_w_1} is the same as the time evolution of linear density bias obtained in \cite{ChanScoccimarroSheth2012}, while the second  line results from the halo profile correction. Unlike the decaying terms in the first line, they do not decay away. In the long term limit,  $b_1$ reduces to $  1 +  3 e^{-y} J_1 - 3 e^{ - \frac{ 3y }{2} } J_{\frac{ 3 }{ 2 } }  $. If we assume that the window is static, we get  $b_1=W$ in the long term limit.

In Fig.~\ref{fig:b1_tot_evolve}, the evolution of the linear density bias is plotted. Again we use the form of the initial condition  motivated by the peak theory \cite{DesjacquesSheth2010}
\beq
\label{eq:b1s_peak}
b_1^* = ( b_{\nu} + b_{ \zeta } k^2 )  W_{\rm G}( k R_{\rm G}).
\eeq
We take $R_{\rm G}  $ corresponding to halo of mass  $2 \times 10^{13} \Msun $.  Instead of using the peak theory results, we take $ b_{\nu}= 15.9$ and $ b_{ \zeta } = 40.0 \, (\MpcOh)^2 $, which are obatined from measurement of the initial cross power spectrum \cite{ChanShethScoccimarro2015}. Both the results from  the  static window  and SC evolving window are shown, however, the differences are very small.

 The initial condition term  $ ( b_1^* + 2 \bv^* - 3 ) e^{- y }  $, especially due to  $b_1^* $, is important at low $k$. In fact, at low $ k$,  it gives the decay of linear bias \cite{Fry1996} . The term due solely to $\bv^*$, $2( 1- \bv^*) e^{-3 y /2} $ is negligible in the whole range of $k$ shown. The sum of the two profile correction terms, $J_1$  and  $J_{\frac{3}{2}} $ gives small overall correction. That is also the reason why the static and evolving window gives almost identical results.  The reason that the profile correction term $J_n$  gives much more significant effect for  $b_{\rm v} $ than for $b_1$ is that unlike the case of $\bv$,  $b_1 $ at late time is still dominated by the reminant effect of $b_1^* $ because the magnitude of $b_1^*   $ is much larger than that of $\bv^* $.  

Unlike the case $\bv$, the magnitude of the scale-dependent part of $b_1^*$ in Eq.~\ref{eq:b1s_peak} is significant compared to other contributions. To highlight its effect, we plot the results when the initial bias is scale-independent, i.e.~$b_1^* =  b_{\nu} $ and $\bv^*=1$   in Fig.~\ref{fig:b1_tot_evolve_constant_bstar}. This plot shows that the bump in Fig.~\ref{fig:b1_tot_evolve} around $k\sim 0.7 \hOMpc $ is due to the large magnitude of the initial $ b_{ \zeta } k^2 $ term. The low $k$ plateau is due to $ b_{\nu}$ from initial condition and the scale-dependent transition comes from $J_1$ and $J_{\frac{3}{2}} $  terms.

\begin{widetext}

\subsection{Second order biases  }
\label{sec:SecondBiasComputation}

To second order,  Eq.~\ref{eq:continuity_1} and \ref{eq:Euler_1} become
\beqa
\label{eq:continuity_g}
\frac{ \partial \dg^{(2)}  }{ \partial y } - \tg^{(2)} & =& \int d^3 k_1  d^3 k_2 \Ddel ( \mb{k}- \mb{k}_{12} )   
\alpha( \mb{k}_1 ,  \mb{k}_2 ) \tg^{(1)} ( \mb{k}_1 ) \dg^{(1)} ( \mb{k}_2 )  ,  \\ 
\label{eq:Euler_g}
\frac{ \partial \tg^{(2)}  }{ \partial y } + \frac{ 1 }{2  } \tg^{(2)}  -  \frac{3  }{2 } W \delta^{ (2)}  & = &  \int d^3 k_1  d^3 k_2 \Ddel ( \mb{k}- \mb{k}_{12} )   
 \beta( \mb{k}_1 ,  \mb{k}_2 ) \tg^{(1)} ( \mb{k}_1 ) \tg^{(1)} ( \mb{k}_2 ) . 
\eeqa
We shall solve Eq.~\ref{eq:continuity_g} and \ref{eq:Euler_g} perturbatively to obtain $\dg^{(2)} $ and $\tg^{(2)}$.

\subsubsection{Velocity}
We will start from Eq.~\ref{eq:Euler_g} to compute $\tg^{(2) }$ first. Using the dark matter SPT result
\beq
\delta^{ (2) } (y ) = e^{2y} \delta^{ (2) }_*. 
\eeq
and Eq.~\ref{eq:thetag_1}, we can integrate Eq.~\ref{eq:Euler_g} to get 
\beqa
\tg^{(2) } ( y ) & = & \tg^{* (2) } e^{- \frac{ y }{ 2 }} + \frac{ 3 }{ 2 } \delta_*^{(2)} e^{-\frac{ y }{ 2 }}  I_{\frac{ 5 }{ 2 }  } (y )  +  \int d^3 k_1  d^3 k_2 \Ddel ( \mb{k}- \mb{k}_{12} ) \beta( \mb{k}_1 ,  \mb{k}_2 )       \nn \\
& \times &  e^{- \frac{ y }{ 2 } }  \Big\{   2 ( 1 - e^{ - \frac{y }{2 }  } ) \tg^{*(1)}( \mb{k}_1 )    \tg^{*(1)} ( \mb{k}_2 )     - \frac{ 3}{2} \big[ (  e^{ - \frac{ y }{ 2 }  } I_{ \frac{ 3 }{ 2 } }( \mb{k}_2 ) - I_1( \mb{k}_2 )  )   \tg^{*(1)} ( \mb{k}_1 ) \delta_*^{(1)} ( \mb{k}_2 ) +  ( \mb{k}_1 \leftrightarrow  \mb{k}_2 )   \big]   \nn \\
&+& \frac{9 }{4 }   \int_0^y d y' e^{ - \frac{ y' }{ 2 } }  I_{\frac{3}{ 2}} (y', \mb{k}_1) I_{\frac{3}{ 2}} (y', \mb{k}_2)    \delta_*^{(1)}(\mb{k}_1 )    \delta_*^{(1)}(\mb{k}_2 )   \Big\}.
\eeqa
Replacing $ \tg^{* (1) }  $ by $ b_{\rm v}^* \delta_*^{(1)} $ and extrapolating $\delta_*^{(1)} $ to the present time, we have 
\beqa
\label{eq:theta_g_2}
\tg^{(2) } ( y ) & = & \tg^{* (2) } e^{- \frac{ y }{ 2 }} + \frac{ 3 }{ 2 } \delta^{(2)}(y)  e^{-\frac{ 5 y }{ 2 }}  I_{\frac{ 5 }{ 2 }  } (y )  +  \int d^3 k_1  d^3 k_2 \Ddel ( \mb{k}- \mb{k}_{12} )  K_{ \theta 2 } ( \mb{k}_1,  \mb{k}_2 ) \delta^{(1)}( \mb{k}_1 )  \delta^{(1)}( \mb{k}_2 ), 
\eeqa
where 
$ K_{ \theta 2 } $ is given by 
\beqa
 K_{ \theta 2 } ( \mb{k}_1,  \mb{k}_2 ) &=&  e^{ - \frac{ 5 y }{ 2 }  } \beta( \mb{k}_1,  \mb{k}_2 ) \Big\{  2( 1 - e^{- \frac{ y}{ 2 } } ) b_{\rm v }^* (\mb{k}_1 )   b_{\rm v }^* (\mb{k}_2 )  
 -  \frac{3}{2} \big[ (  e^{ - \frac{ y }{ 2 }  } I_{ \frac{ 3 }{ 2 } }( \mb{k}_2 ) - I_1( \mb{k_2} )  )  b_{\rm v }^* (\mb{k}_1 ) +      (\mb{k}_1 \leftrightarrow \mb{k}_2 )   \big]   \nn \\
& +&  \frac{9 }{4 }   \int_0^y d y' e^{ - \frac{ y' }{ 2 } } I_{\frac{3}{ 2}} (y', \mb{k}_1)  I_{\frac{3}{ 2}} (y', \mb{k}_2)
\Big\} , 
\eeqa
where $ ( \mb{k}_1  \leftrightarrow \mb{k}_2 ) $ is a shorthand for a similar term obtained with $\mb{k}_1  $ and $\mb{k}_2 $ interchanged.  To be general, we allow the initial linear biases to be scale-dependent. Note that $ K_{ \theta 2 } $ is already symmetric in $\mb{k}_1$ and $\mb{k}_2$, so symmetrization is not required.

In terms of $J_n$, $\tg^{(2)}$ can be expressed as 
\beq
\label{eq:theta_g_2_J_final}
\tg^{(2) } ( y )  =  \tg^{* (2) } e^{- \frac{ y }{ 2 }} +  \int d^3 k_1  d^3 k_2 \Ddel ( \mb{k}- \mb{k}_{12} )  \mathcal{K}_{ \theta 2 } ( \mb{k}_1,  \mb{k}_2 ) \delta^{(1)}( \mb{k}_1 )  \delta^{(1)}( \mb{k}_2 ), 
\eeq
where 
$ \mathcal{K}_{ \theta 2 } $ is given by 
\beq
\mathcal{K}_{ \theta 2 }= T_{F} + T_{G} + T_{\bv^* } + T_{JJ} + T_J . 
\eeq
The five types of terms $ T_{F}$, $T_{G}$,  $ T_{\bv^* }$, $ T_{JJ}$, and  $ T_J$ are given by 
\beqa
T_G &=&  ( 1 - e^{- \frac{ 5 y }{ 2 }  } ) G_2(  \mb{k}_1,  \mb{k}_2 ) ,    \\
T_F &=& \frac{3}{ 2 }  e^{- \frac{ 5 y }{ 2 }  } J_{ \frac{ 5 }{ 2} } (\mb{k}  )F_2(  \mb{k}_1,  \mb{k}_2 ) , \\
T_{\bv^* } &=&  \beta( \mb{k}_1,  \mb{k}_2 ) \Big\{  2  e^{ - \frac{ 5 y }{ 2 }  } ( 1 - e^{- \frac{ y}{ 2 } } ) b_{\rm v }^* (\mb{k}_1 )   b_{\rm v }^* (\mb{k}_2 )  
 +  e^{ - \frac{ 5 y }{ 2 }  }  (   e^{y}   + 2 e^{ - \frac{ y }{ 2 }  }  - 3  )  \big(    b_{\rm v }^* (\mb{k}_1 ) +  b_{\rm v }^* (\mb{k}_2 )      \big)  \nn  \\
 &-&  2 (  e^{ - \frac{3 y}{ 2 } } - 2  e^{ - \frac{ 5y}{ 2 } }  + e^{-3y}  )  
- \frac{3}{2}  e^{ - 3 y  } \Big[  \big( J_{ \frac{ 3 }{2 } }( \mb{k}_2 ) - J_1 ( \mb{k}_2 )  \big)  b_{\rm v }^* (\mb{k}_1 ) +     ( \mb{k}_1 \leftrightarrow \mb{k}_2 )  \Big]  \Big\} ,   \\
T_{JJ} &=&   \beta( \mb{k}_1,  \mb{k}_2 )  e^{ - \frac{ 5 y }{ 2 }  }  \int_0^y d y' e^{ - \frac{ y' }{ 2 } }   \frac{9 }{ 4 }  J_{\frac{3}{ 2}} (y', \mb{k}_1 )  J_{\frac{3}{ 2}} ( y', \mb{k}_2 ),    \\
T_J &=&   \beta( \mb{k}_1,  \mb{k}_2 )  e^{ - \frac{ 5 y }{ 2 }  }   \int_0^y d y' e^{ - \frac{ y' }{ 2 } }    \frac{3}{2} ( e^{ \frac{3y'}{2}  } - 1 ) \big(  J_{\frac{3}{ 2}} (y', \mb{k}_1 ) +  J_{\frac{3}{ 2}} (y', \mb{k}_2 )  \big) ,
\eeqa
where  $\mb{k} = \mb{k}_{12}  $,  and  $F_2$ and $G_2$ represent the coupling kernels
\beq
F_2 ( \mb{k}_1,  \mb{k}_2 ) = \frac{5 }{7}  + \frac{1 }{2} \mu \big( \frac{k_1 }{ k_2 } +   \frac{k_2 }{ k_1 }   \big)   +  \frac{2 }{7} \mu^2,    \quad \quad
G_2 ( \mb{k}_1,  \mb{k}_2 ) = \frac{3 }{7}  + \frac{1 }{2} \mu \big( \frac{k_1 }{ k_2 } +   \frac{k_2 }{ k_1 }   \big)   +  \frac{4 }{7} \mu^2, 
\eeq
with $ \mu =\hat{ \mb{k}}_1 \cdot \hat{ \mb{k}}_2$.

     As a cross-check, we pause to consider the limit $\bv^* =1 $ and $J_n=0$. Then $\mathcal{K}_{\theta 2}$ reduces to  $ (1 -e^{ - \frac{5y}{ 2} }) G_2 $.   Note that in this limit $ \tg^{* (2) } e^{- \frac{ y }{ 2 }} = \theta^{* (2) } e^{- \frac{ y }{ 2 }} = \theta^{ (2) } e^{- \frac{5 y }{ 2 }}  $, thus Eq.~\ref{eq:theta_g_2_J_final} reduces to $\theta^{(2)}  $ because the galaxy field reduces to the dark matter field.   On the other hand, in the long term limit  $y \rightarrow \infty$, the transient terms vanish, in particular those arising from $\bv^* $, and we end up with  $G_2 + T_F + T_{JJ} + T_J  $.

In Fig.~\ref{fig:theta2g_tot_evolve}, we show the evolution of the kernel $\mathcal{K}_{\theta 2 }   $. In this plot, we have used the same parameters as those in the previous section and have set $k_1 = k_2  $ and $\mu = - 1/2$, which corresponds to the equilateral triangle configuration.  We have compared the case with the  static window and the one with SC evolving window, and find that the high $k$ corrections are quite different. In particular, the magnitude of the static one decreases while the evolving one increases over time. When $ \bv^*  $ is assumed to be scale-independent instead, the results are similar to Fig.~\ref{fig:theta2g_tot_evolve}, thus we do not show it here.

We now look at the individual components of  $\mathcal{K}_{\theta 2 }   $ in details in this example.  At low $k$, the only non-vanishing component is $T_G$ and it is (almost) constant for the reshifts shown. The term $T_F$ gives negative $k^2$ correction and its magnitude is large among all the high $k$ correction terms.  The term $T_J$ and $T_{JJ} $ are of opposite signs, but the magnitude of $T_J $ is slightly larger. In particular, as the leading correction from $T_{JJ}$ is of $k^4$,  compared to  $T_{J}$,  it is unimportant for $k\lesssim 0.6 \hOMpc $. $T_{F}$ and $T_J$ are the largest scale-dependent correction terms, they are of similar magnitude but of opposite signs.  The term with $ \bv^* $, $T_{\bv^*} $ gives small negative contribution, which is negligible compared to the other correction terms.   In fact for  $ k \lesssim 0.6 \hOMpc $, the kernel   $\mathcal{K}_{\theta 2 } $ is well captured by the sum $T_F + T_G + T_J $.  This is similar to $\bv  $, for which the term solely due to $\bv^* $ is negligible at late time (even at $z\sim 5$), and the dominant correction term comes from the $J_n$-term.

\subsubsection{Density}

We now compute $\dg^{(2)}  $.   Integrating Eq.~\ref{eq:continuity_g} in terms of $\tg^{(1) }$,  $\tg^{(2) }$, and $\dg^{(1)} $ yields
\beq
\label{eq:deltag_2_general}
\dg^{(2)} = \dg^{*(2)} + \int_0^y d y' \tg^{(2)}(y')  +   \int d^3 k_1  d^3 k_2 \Ddel ( \mb{k}- \mb{k}_{12} ) \alpha( \mb{k}_1 ,  \mb{k}_2 )   \int_0^y dy' \tg^{(1)} ( \mb{k}_1 )  \dg^{(1)} ( \mb{k}_2 ) .  
\eeq
Using Eq.~\ref{eq:theta_g_2}, replacing $ \dg^{* (1) }  $ by $ b_{\rm v}^* \delta_*^{(1)} $ and extrapolating $\delta_*^{(1)} $ to the present time, we have 
\beqa
\label{eq:int_theta2}
\int_0^y dy' \tg^{(2)}(y') & = & 2 \tg^{*(2)} ( 1 - e^{ - \frac{y}{2} } ) - 3 \delta^{(2)} [e^{- \frac{ 5y }{2 } } I_{\frac{ 5 }{2}}  -e^{-2y} I_2   ] +  \int d^3 k_1  d^3 k_2 \Ddel ( \mb{k}- \mb{k}_{12} ) \beta( \mb{k}_1 ,  \mb{k}_2 ) \nn \\
& \times & e^{-2y}  \Big \{ 2( 1 - 2 e^{- \frac{y }{2}  } + e^{-y} ) b_{\rm v}^*( \mb{k}_1 )b_{\rm v}^*( \mb{k}_2 ) + \frac{3}{2} \Big [  \big( e^{-y} I_{\frac{ 3 }{ 2 } }(\mb{k}_2)  - 2 e^{ - \frac{ y }{ 2 }  } I_1( \mb{k}_2 ) + I_{ \frac{ 1 }{ 2 } }( \mb{k}_2)  \big)   b_{\rm v}^*( \mb{k}_1 ) +  ( \mb{k}_1 \leftrightarrow  \mb{k}_2 ) \Big] \nn \\
&-& \frac{ 9 }{2} \int_0^y dy'  ( e^{ -\frac{ y}{ 2} }  -  e^{ -\frac{ y' }{ 2} }   )  e^{ -\frac{ y' }{ 2} }  I_{\frac{ 3 }{ 2} } (y', \mb{k}_1 ) I_{\frac{ 3 }{ 2} } (y', \mb{k}_2 )    \Big\} \delta^{(1)} ( \mb{k}_1 )  \delta^{(1)} ( \mb{k}_2 ) . 
\eeqa

Making use of Eq.~\ref{eq:thetag_1} and \ref{eq:deltag_1}, we can compute the second integral in Eq.~\ref{eq:deltag_2_general} to get  
\beqa
\label{eq:alpha_theta_delta}
&&  \int d^3 k_1  d^3 k_2 \Ddel ( \mb{k}- \mb{k}_{12} )   \alpha( \mb{k}_1, \mb{k}_2 ) \int_0^y dy'  \tg^{(1)} ( y', \mb{k}_1 ) \dg^{(1)} ( y', \mb{k}_2 )          \nn \\
& = &  \int d^3 k_1  d^3 k_2 \Ddel ( \mb{k}- \mb{k}_{12} )    e^{-2y}  \alpha( \mb{k}_1, \mb{k}_2 )    \Big\{ 
 2 b_{\rm v}^* (\mb{k}_1 ) b_{1}^* (\mb{k}_2 ) ( 1 - e^{ - \frac{ y }{2 }  } )   
+  2 b_{\rm v}^* (\mb{k}_1 ) b_{\rm v}^* (\mb{k}_2 )  (1 -2 e^{ - \frac{ y }{ 2 } } +  e^{ -y  }   )  \nn \\
&-&   3 b_1^*( \mb{k}_2 )  \big(  e^{ - \frac{ y }{ 2 }  } I_{\frac{ 3}{ 2 }}( \mb{k}_1 ) -  I_1 ( \mb{k}_1 )   \big) 
+ 3  b_{\rm v}^* ( \mb{k}_1 )  \big( e^{-y} I_{ \frac{ 3 }{ 2 }  } ( \mb{k}_2 ) - 2  e^{- \frac{ y }{ 2 }} I_1 ( \mb{k}_2 ) +   I_{ \frac{ 1 }{ 2 }  }  ( \mb{k}_2 )  \big)
 + 3 b_{\rm v}^* ( \mb{k}_2 ) \big[  (  e^{-y}  - 2 e^{- \frac{ y }{ 2 }} )  I_{\frac{ 3 }{ 2 }}  ( \mb{k}_1 ) \nn\\
& +&  2 I_1 ( \mb{k}_1 )  -   I_{\frac{ 1}{ 2 }}  ( \mb{k}_1 )   \big] 
- \frac{ 9 }{ 2 } \int_0^y d y' e^{- \frac{ y' }{ 2 }  } I_{\frac{3}{ 2}  }(y', \mb{k}_1 )  [ e^{- \frac{ y'}{ 2 } } I_{\frac{3}{2} }(y', \mb{k}_2) - I_1(y' , \mb{k}_2)   ]
\Big\}   \delta^{(1)}( \mb{k}_1 ) \delta^{(1)}( \mb{k}_2 ). 
\eeqa

Therefore, we have
\beq
\dg^{(2)} = \dg^{*(2)} +  2 \tg^{*(2)} ( 1 - e^{ - \frac{y}{2} } ) - 3 \delta^{(2)} [e^{- \frac{ 5y }{2 } } I_{\frac{ 5 }{2}}  -e^{-2y} I_2   ] +  \int d^3 k_1  d^3 k_2 \Ddel ( \mb{k}- \mb{k}_{12} ) K_{\delta2}( \mb{k}_1, \mb{k}_2) \delta^{(1)} (\mb{k}_1 )  \delta^{(1)}(\mb{k}_2 )   , 
\eeq
where $ K_{\delta2} $ is given by

\beqa
 K_{\delta2} ( \mb{k}_1, \mb{k}_2 )
&=&  e^{-2 y} \Big\{ 
2 ( e^{-y} - 2 e^{ - \frac{ y }{ 2 }  } + 1 ) \big( \beta(\mb{k_1}, \mb{k}_2 ) + \alpha (\mb{k_1}, \mb{k}_2 ) \big)  \bv^*( \mb{k}_1 )\bv^*( \mb{k}_2 )  \nn \\  
 &+&   2  ( 1 - e^{ - \frac{ y }{ 2 }  } )  \bv^* ( \mb{k}_1  )  b_1^* ( \mb{k}_2  )  \alpha (\mb{k_1}, \mb{k}_2 )
+ 3 \big[  e^{-y} I_{\frac{ 3 }{ 2 } } ( \mb{k}_2 )  - 2 e^{ - \frac{ y }{ 2 }  } I_1 ( \mb{k}_2 ) + I_{ \frac{ 1 }{ 2 } }  ( \mb{k}_2 )      \big] \bv^* ( \mb{k}_1 )  ( \alpha(\mb{k_1}, \mb{k}_2 ) + \beta(\mb{k_1}, \mb{k}_2 )     )  \nn \\ 
&+&  3 \big[ ( e^{-y} - 2 e^{ - \frac{y }{ 2 } }  ) I_{\frac{3 } { 2 }} ( \mb{k}_1 ) + 2 I_1  ( \mb{k}_1 )  - I_{ \frac{1 }{2 } }  ( \mb{k}_1 )      \big]  \bv^* ( \mb{k}_2 )   \alpha (\mb{k_1}, \mb{k}_2 ) 
-   3  \big[ e^{ - \frac{ y }{2 } } I_{\frac{ 3 }{ 2 } } ( \mb{k}_1 )  - I_1 (\mb{k}_1 )    \big] b_1^* ( \mb{k}_2 ) \alpha (\mb{k_1}, \mb{k}_2 )   \nn \\
&-&  \frac{ 9 }{ 2 } \int_0^y d y' e^{- \frac{ y' }{ 2 } } I_{\frac{ 3 }{ 2 }  }(y', \mb{k}_1 ) \Big[ ( e^{ - \frac{ y }{2 }  } -  e^{ - \frac{ y' }{2 }  } )  I_{\frac{ 3 }{ 2 }  }(y', \mb{k}_2 ) \beta(\mb{k_1}, \mb{k}_2 )  +   \Big(  e^{ - \frac{ y' }{2 }  } I_{\frac{ 3 }{  2 }  }(y', \mb{k}_2) - I_1(y',\mb{k}_2) \Big)  \alpha (\mb{k_1}, \mb{k}_2 ) \Big]
\Big\} .    \nn \\
\eeqa

In terms of $J_n$, $\dg^{(2)}$ can be expressed as
\beq
\label{eq:deltag_final}
\dg^{(2)} = \dg^{*(2)} +  2 \tg^{*(2)} ( 1 - e^{ - \frac{y}{2} } )  
 +  \int d^3 k_1  d^3 k_2 \Ddel ( \mb{k}- \mb{k}_{12} )\mathcal{ K}_{\delta2}( \mb{k}_1, \mb{k}_2) \delta^{(1)}(\mb{k}_1 )  \delta^{(1)}(\mb{k}_2 )  ,
\eeq
where $ K_{\delta2} $ is given by 
\beqa
\label{eq:Kmathcal_final}
\mathcal{K}_{\delta2} ( \mb{k}_1, \mb{k}_2 )
&=&  e^{-2 y} \Big\{ 
[   \frac{ 3 }{ 10 }  ( e^{2y}  + 4 e^{- \frac{ y }{ 2 }  }  - 5  ) - 3  \big(  e^{- \frac{ y }{ 2 }  } J_{\frac{ 5 }{ 2 }  }( \mb{k} )   - J_2 (\mb{k}) \big)         ] F_2( \mb{k}_1, \mb{k}_2 ) \nn \\
&+& 2( e^{-y} - 2 e^{ - \frac{ y}{ 2 }  } + 1 ) \big(  \alpha (\mb{k_1}, \mb{k}_2) +  \beta (\mb{k_1}, \mb{k}_2)     \big)  \bv^*( \mb{k}_1 )   \bv^*( \mb{k}_2 )    
+    2  ( 1 - e^{ - \frac{ y }{ 2 }  } )  \bv^* ( \mb{k}_1  )  b_1^* ( \mb{k}_2  )  \alpha (\mb{k_1}, \mb{k}_2 )
\nn \\
&+&  \big[  2  (   e^{ \frac{ y}{ 2 }  }  - 3 +3 e^{- \frac{ y }{2 }} - e^{-y}  ) 
+  3 \big( e^{-y}J_{\frac{ 3}{2 }} (\mb{k}_2) - 2 e^{- \frac{y}{2} } J_1( \mb{k}_2 )+ J_{\frac{ 1 }{ 2 }}( \mb{k}_2 )   \big)   \big] \bv^* ( \mb{k}_1 )  \big( \alpha(\mb{k_1}, \mb{k}_2 ) + \beta(\mb{k_1}, \mb{k}_2 )  \big)  \nn \\ 
&+&  \big[  2 \big(  e^y - 2 e^{\frac{ y}{ 2 }} + 2 e^{ - \frac{ y }{ 2 }  } -  e^{-y} \big) + 3 \big( (e^{-y} - 2 e^{- \frac{ y }{ 2 } } ) J_{ \frac{ 3 }{ 2 } }( \mb{k}_1 )  + 2 J_1 ( \mb{k}_1 )  - J_{ \frac{1}{2} }  ( \mb{k}_1 )   \big)            \big]  \bv^* ( \mb{k}_2 )   \alpha (\mb{k_1}, \mb{k}_2 )  \nn \\
&+ &    \big[ e^y - 3 + 2 e^{ - \frac{ y }{ 2 } } - 3 \big(  e^{- \frac{ y }{ 2 }  }  J_{ \frac{ 3 }{ 2 }}( \mb{k}_1 ) - J_1 ( \mb{k}_1 )   \big)  \big]  b_1^* ( \mb{k}_2 ) \alpha (\mb{k_1}, \mb{k}_2 ) 
 + A_1 + A_2 +  A_3
\Big\},
\eeqa
where $A_1$, $A_2$, and $A_3$ represent
\beqa
A_1 &= &- \frac{ 9 }{ 2 } \int_0^y dy' e^{ - \frac{ y' }{ 2 } }  J_{ \frac{ 3 }{ 2 } }( y', \mb{k}_1 )  [ ( e^{- \frac{y }{ 2 }} - e^{- \frac{y' }{2 }  } ) J_{\frac{3 }{ 2} } ( y', \mb{k}_2 )  \beta(\mb{k}_1, \mb{k}_2 ) +  \big( e^{- \frac{y' }{2 } } J_{ \frac{ 3 }{ 2 }  } (  y', \mb{k}_2  )    - J_1 ( y', \mb{k}_2 ) \big) \alpha(\mb{k}_1, \mb{k}_2 )  ],   \\
A_2 & =& - 3 \int_0^y dy' e^{ - \frac{ y' }{ 2 } } \Big\{  2  ( e^{- \frac{y }{ 2 }} - e^{- \frac{y' }{2 }  }  )  ( e^{ \frac{ 3y' }{ 2 } } - 1 )  J_{\frac{3}{2}}(y', \mb{k}_1)   \beta( \mb{k}_1, \mb{k}_2  )  \nn \\
&+&  \Big[      \big(  \frac{3}{2} -  \frac{e^{y' } }{ 2 }  -  e^{ - \frac{y' }{ 2}  } \big) J_{\frac{ 3 }{2 } }(y', \mb{k}_1 )  + ( e^{  y' } - e^{ - \frac{ y'  }{ 2 }  } ) J_{\frac{ 3 }{ 2 } } (y', \mb{k}_2 )  - ( e^{ \frac{3 y' }{ 2 } } - 1 ) J_1 ( y', \mb{k}_2 )  \Big]  \alpha( \mb{k}_1, \mb{k}_2 )    \Big\} ,  \\
A_3 &=& \frac{ 1 }{ 10 } e^{-y} ( e^{ \frac{ y}{ 2}  } - 1 )^4 [ 5 ( 2 +  e^{ \frac{ y}{ 2}  } )^2 \alpha( \mb{k}_1, \mb{k}_2 )  + 2  ( e^y +  4 e^{ \frac{ y}{ 2}   } + 10  ) \beta ( \mb{k}_1, \mb{k}_2 )   ] .
\eeqa

When the limit $b_1^* =1 $, $\bv^* = 1 $ and $J_n=0$ are taken, $\dg^{(2)}  $ in Eq.~\ref{eq:deltag_final} reduces to $\delta^{(2)}  $ as the galaxy field becomes the dark matter field. In the large $y$ limit, we have
\beq
\mathcal{K}_{\delta 2 }  = \Big[ \frac{3}{10} -  3 e^{ -2 y } \Big( e^{ -\frac{ y }{ 2 }  } J_{\frac{5}{2}}(\mb{k})  -J_2(\mb{k})  \Big) \Big] F_2 ( \mb{k}_1, \mb{k}_2 ) + e^{ -2 y }  ( A_1 + A_2 + A_3  ). 
\eeq

As only the symmetric part of the kernel $\mathcal{K}_{\delta 2 }  $ contributes to the integral in Eq.~\ref{eq:deltag_final}, we need to symmetrize the kernel as 
\beq
\mathcal{K}_{\delta 2}^{\rm s} (\mb{k}_1, \mb{k}_2 )  = \frac{1}{2} \Big(  \mathcal{K}_{\delta 2} (\mb{k}_1, \mb{k}_2 ) +  \mathcal{K}_{\delta 2} (\mb{k}_2, \mb{k}_1 ) \Big) .
\eeq
However, to reduce the length of the formulas, we do not explicitly symmetrize them.  However, in the final results, we always use the symmetrized kernel.

Similar to that in Ref.~\cite{ChanScoccimarroSheth2012}, we define the terms that deviate from the local biasing prescription as nonlocal terms. Thus at second order, the nonlocal terms are defined as 
\beq
\chi_{\rm nonloc}^{(2)} = \dg^{(2)} - ( b_1 \delta^{(2)} + \frac{b_2 }{2  } (\delta^{(1)} )^2 ).
\eeq
At second order, the nonlocal terms are induced by the initial linear bias, and the initial second order biases do not generate new terms \cite{ChanScoccimarroSheth2012}. For convenience we consider 
\beq
\label{eq:deltag2_nonloc}
 \dg^{(2)} -  b_1 \delta^{(2)} = \dg^{*(2)} +  2 \tg^{*(2)} ( 1 - e^{ - \frac{y}{2} } )  
 +  \int d^3 k_1  d^3 k_2 \Ddel ( \mb{k}- \mb{k}_{12} ) \chi ( \mb{k}_1, \mb{k}_2) \delta(\mb{k}_1 )  \delta(\mb{k}_2 )
\eeq
where $ \chi_{\delta2} $ is given by 
\beqa
\label{eq:chi_delta2}
\chi  ( \mb{k}_1, \mb{k}_2)
&=&  e^{-2 y} \Big\{ 
\Big[  - \frac{ 7 }{ 10 }  e^{2y} - \big ( b_1^*(\mb{k}) + 2 \bv^*( \mb{k}) - 3 \big) e^y - 2 \big(1 -\bv^*( \mb{k} ) \big)  e^{\frac{ y}{ 2}}  \nn \\
&+& \frac{ 6 }{ 5 } e^{ - \frac{ y}{ 2}  } -\frac{ 3 }{ 2 } - 3 e^{ - \frac{ y}{ 2}   } J_{\frac{5 }{2 }}(\mb{k} ) + 3 J_2(\mb{k} )  - 3 e^y J_1(\mb{k})  + 3 e^{\frac{y}{2} } J_{\frac{ 3}{ 2 }} (\mb{k})    \Big]F_2( \mb{k}_1 ,  \mb{k}_2   ) \nn \\
&+& 2( e^{-y} - 2 e^{ - \frac{ y}{ 2 }  } + 1 ) \big(  \alpha (\mb{k_1}, \mb{k}_2) +  \beta (\mb{k_1}, \mb{k}_2)     \big)  \bv^*( \mb{k}_1 )   \bv^*( \mb{k}_2 )    
+    2  ( 1 - e^{ - \frac{ y }{ 2 }  } )  \bv^* ( \mb{k}_1  )  b_1^* ( \mb{k}_2  )  \alpha (\mb{k_1}, \mb{k}_2 )
\nn \\
&+&  \big[  2  (   e^{ \frac{ y}{ 2 }  }  - 3 +3 e^{- \frac{ y }{2 }} - e^{-y}  ) 
+  3 \big( e^{-y}J_{\frac{ 3}{2 }} (\mb{k}_2) - 2 e^{- \frac{y}{2} } J_1( \mb{k}_2 )+ J_{\frac{ 1 }{ 2 }}( \mb{k}_2 )   \big)   \big] \bv^* ( \mb{k}_1 )  \big( \alpha(\mb{k_1}, \mb{k}_2 ) + \beta(\mb{k_1}, \mb{k}_2 )  \big)  \nn \\ 
&+&  \big[  2 \big(  e^y - 2 e^{\frac{ y}{ 2 }} + 2 e^{ - \frac{ y }{ 2 }  } -  e^{-y} \big) + 3 \big( (e^{-y} - 2 e^{- \frac{ y }{ 2 } } ) J_{ \frac{ 3 }{ 2 } }( \mb{k}_1 )  + 2 J_1 ( \mb{k}_1 )  - J_{ \frac{1}{2} }  ( \mb{k}_1 )   \big)            \big]  \bv^* ( \mb{k}_2 )   \alpha (\mb{k_1}, \mb{k}_2 )  \nn \\
&+ &    \big[ e^y - 3 + 2 e^{ - \frac{ y }{ 2 } } - 3 \big(  e^{- \frac{ y }{ 2 }  }  J_{ \frac{ 3 }{ 2 }}( \mb{k}_1 ) - J_1 ( \mb{k}_1 )   \big)  \big]  b_1^* ( \mb{k}_2 ) \alpha (\mb{k_1}, \mb{k}_2 ) 
 + A_1 + A_2 +  A_3
\Big\},
\eeqa
In Eq.~\ref{eq:deltag2_nonloc}, we have used $b_1 $ given by Eq.~\ref{eq:b1_w_1}. The initial second order biases are hidden in $\dg^{*(2)  }  $.

Suppose that the initial conditions are given by
\beq
\dg^{*(2)} = \frac{ b_{2}^*  }{ 2 } ( \delta^{(1)}_*)^2  + b_{1}^* \delta^{(2)}_* , \quad  \bv^* = 1, \quad   \tg^{*(2)} =  \theta^{(2)}_*, 
\eeq
where $ b_{2 }^* $ and  $b_{1}^* $ are scale-independent.  In other words, we suppose that the initial density biases are local in Lagrangian space, and there is no initial velocity bias.  If we also neglect all the $J_n$  terms, then Eq.~\ref{eq:deltag2_nonloc} is simplified substantially and we end up with \cite{ChanScoccimarroSheth2012}
\beq
\dg^{(2)} -  b_1 \delta^{(2)} =  \frac{ b_2  }{ 2 } ( \delta^{(1)})^2  +  \gamma_2 \mathcal{G}_2 ,
\eeq
where $  b_2 $ and  $\gamma_2$ are given by
\beq
b_2 = b_2^* e^{-2y}, \quad \quad   \gamma_2 =    \frac{2}{ 7} ( b_1^* - 1) e^{-2y} ( e^{y} - 1 ),
\eeq
and $\mathcal{G}_2  $ denotes 
\beq
 \mathcal{G}_2 (\mb{k}) = \int  d^3 k_1  d^3 k_2 \Ddel ( \mb{k}- \mb{k}_{12} ) ( \mu^2 -1 ) \delta^{(1)}(\mb{k}_1 )  \delta^{(1)}(\mb{k}_2 ). 
\eeq
In particular, because there is no velocity bias as the dipole term vanishes.

We now consider the correction to these results due to initial scale-dependent biases and the corrections arising from the profile corrections.  For the initial conditions, we will assume that $\bv^* $ and $b_1^*$ are given by Eq.~\ref{eq:bvs_peak} and \ref{eq:b1s_peak} respectively, and for simplicity  $b_2^*$ is a constant and $\tg^{*(2)} = \theta_*^{(2)} $.

 We first define the scale-dependent parameters of the  initial biases
\beq
\eds(k) = b_1^*(k) - b_{\nu} , \quad \quad \evs( k)  = \bv^*(k) - 1 . 
\eeq
Then we have
\beq
\dg^{(2)} - b_1 \delta^{(2)} -  \frac{ b_2 }{ 2 } (\delta^{(1)} )^2  -    \gamma_2 \mathcal{G}_2 =  \int  d^3 k_1  d^3 k_2 \Ddel ( \mb{k}- \mb{k}_{12} ) \psi(\mb{k}_1, \mb{k}_2   ) \delta^{(1)}(\mb{k}_1 )  \delta^{(1)}(\mb{k}_2 ), 
\eeq
where the kernel $ \psi $ is defined as 
\beq
\psi( \mb{k}_1, \mb{k}_2 ) =  T_{\eds } +  T_{\evs } + T_J ,
\eeq
with various terms given by 
\beqa
T_{\eds } &=&  \frac{ 1}{2}  e^{-2 y  } \big[ -  e^{y}  \eds( \mb{k} ) F_2( \mb{k}_1, \mb{k}_2 ) + ( e^y -1 ) \eds( \mb{k}_2 ) \alpha( \mb{k_1}, \mb{k_2} ) \big] + (\mb{k}_1 \leftrightarrow \mb{k}_2  ),  \\
T_{\evs  } &=& \frac{ 1}{2} e^{-2y}  \Big\{ 2 ( e^{-y} - 2 e^{ - \frac{ y }{ 2 }  } + 1 )   ( \alpha( \mb{k}_1, \mb{k}_2 ) + \beta( \mb{k}_1, \mb{k}_2 )   )  \big(  \evs( \mb{k}_1 ) +   \evs( \mb{k}_2 ) +  \evs( \mb{k}_1 )  \evs( \mb{k}_2 )   \big)    \nn \\   
&+&   2( e^{ \frac{y }{ 2 } } - e^{y} )  \evs(\mb{k} )    F_2 (\mb{k}_1 , \mb{k}_2 )     +   2 (1 - e^{ - \frac{ y }{ 2 } } )\evs( \mb{k}_1 ) \big( b_{\nu} + \eds (\mb{k}_2) \big)  \alpha( \mb{k}_1 , \mb{k}_2 )   \nn \\ 
&+& 2( e^{\frac{ y }{ 2 }} - 3 + 3 e^{- \frac{ y }{ 2 }  } - e^{ -y } ) \evs( \mb{k}_1  ) \big(  \alpha( \mb{k}_1 , \mb{k}_2 ) +  \beta( \mb{k}_1 , \mb{k}_2 ) \big)   \nn \\
&+& 2 ( e^y - 2 e^{ \frac{y }{2 } } + 2  e^{ - \frac{y }{2 } } - e^{-y} ) \evs( \mb{k}_2 ) \alpha(  \mb{k}_1 , \mb{k}_2 ) \Big\} + (\mb{k}_1 \leftrightarrow  \mb{k}_2 ), \\ 
T_J &=&  \frac{ 1}{2} e^{-2y}  \Big\{   \big[ -3 e^{ - \frac{ y }{ 2 }} J_{ \frac{ 5 }{ 2 }} ( \mb{k} ) + 3 J_2 ( \mb{k} ) - 3 e^y J_1 ( \mb{k} ) + 3 e^{ \frac{y }{ 2 } } J_{ \frac{ 3 }{ 2 } } ( \mb{k} )   \big] F_2 ( \mb{k}_1, \mb{k}_2 )  \nn \\
 &+&  3  \big[  e^{-y} J_{ \frac{3 }{ 2 } }( \mb{k}_2 ) - 2 e^{- \frac{ y }{2 } } J_1( \mb{k}_2 ) +  J_{\frac{1}{2}} ( \mb{k}_2 )   \big] \bv^*( \mb{k}_1 ) \big( \alpha( \mb{k}_1, \mb{k}_2 )   +  \beta ( \mb{k}_1 , \mb{k}_2 ) \big)   \nn \\
&+ & 3 \big[ ( e^{-y} - 2 e^{- \frac{ y }{ 2}} ) J_{ \frac{ 3 }{2 }} ( \mb{k}_1 ) + 2 J_1( \mb{k}_1 ) - J_{ \frac{ 1}{ 2 }  } ( \mb{k}_1 )    \big]    \bv^* (\mb{k}_2 ) \alpha ( \mb{k}_1 , \mb{k}_2 )   \nn \\
& -&  3 (  e^{ - \frac{ y }{ 2 } } J_{ \frac{ 3}{ 2 }  } ( \mb{k}_1 )  - J_1 ( \mb{k}_1 )  )   b_1^*( \mb{k}_2 ) \alpha( \mb{k}_1, \mb{k}_2 )    + A_1 + A_2 \Big\} +  (\mb{k}_1 \leftrightarrow  \mb{k}_2 )  . 
\eeqa

In Fig.~\ref{fig:delta2g_tot_evolve}, we show the evolution of the kernel $ \psi$ and its components at $ z=5$, 1 and 0 respectively.  The parameters used are the same as those in the previous section and we have set $k_1 = k_2$ and  $\mu= - 1/2$.  Similar to the case of $b_1$, there is no noticeable difference between the case with static window and the evolving one.  All these contributions peaks around $ k \sim 0.7 - 0.9 \hOMpc $. In this case, $T_{ \epsilon_1^* } $ and  $T_{ \epsilon_{\rm v}^* } $ are of similar magnitude but opposite signs, so they roughly cancel each other. Thus the net contribution is mainly given by  $T_{J}$. The overall contribution of $ \psi$ decays over time.


Again to highlight the effects of the initial scale-dependent bias, we show the case when the initial condition is scale-independent, i.e.~$b_1^* = b_{\nu}$ and $\bv^* = 1 $ are assumed in Fig.~\ref{fig:delta2g_tot_evolve_constant_bstar}.   We find that the bump around  $k\sim 1 \hOMpc $ Fig.~\ref{fig:delta2g_tot_evolve} is no longer present, instead there is smooth transition from  $k\sim 0.2$  to $1 \hOMpc $.

The second order kernel will contribtute to the tree level bispectrum. As a quick check of  the importance of the correction term, we compare the kernel $\psi  $ with the nonlocal term kernel  $\gamma_2 ( \mu^2 -1 ) $ in Fig.~\ref{fig:psi_G2_ratio}. The parameters used are the same as the previous ones. From Fig.~\ref{fig:psi_G2_ratio}, we can see the $k^2$ correction at low $k$. However, the $k^2$  correction term starts to surpass the nonlocal term at $k \sim 0.4 \hOMpc $. Hence one should also find the signature of the $k^2$ correction term in the halo bispectrum at $k \gtrsim 0.2 \hOMpc $.

\end{widetext}

\begin{figure}[!htb]
\centering
\includegraphics[width=\linewidth]{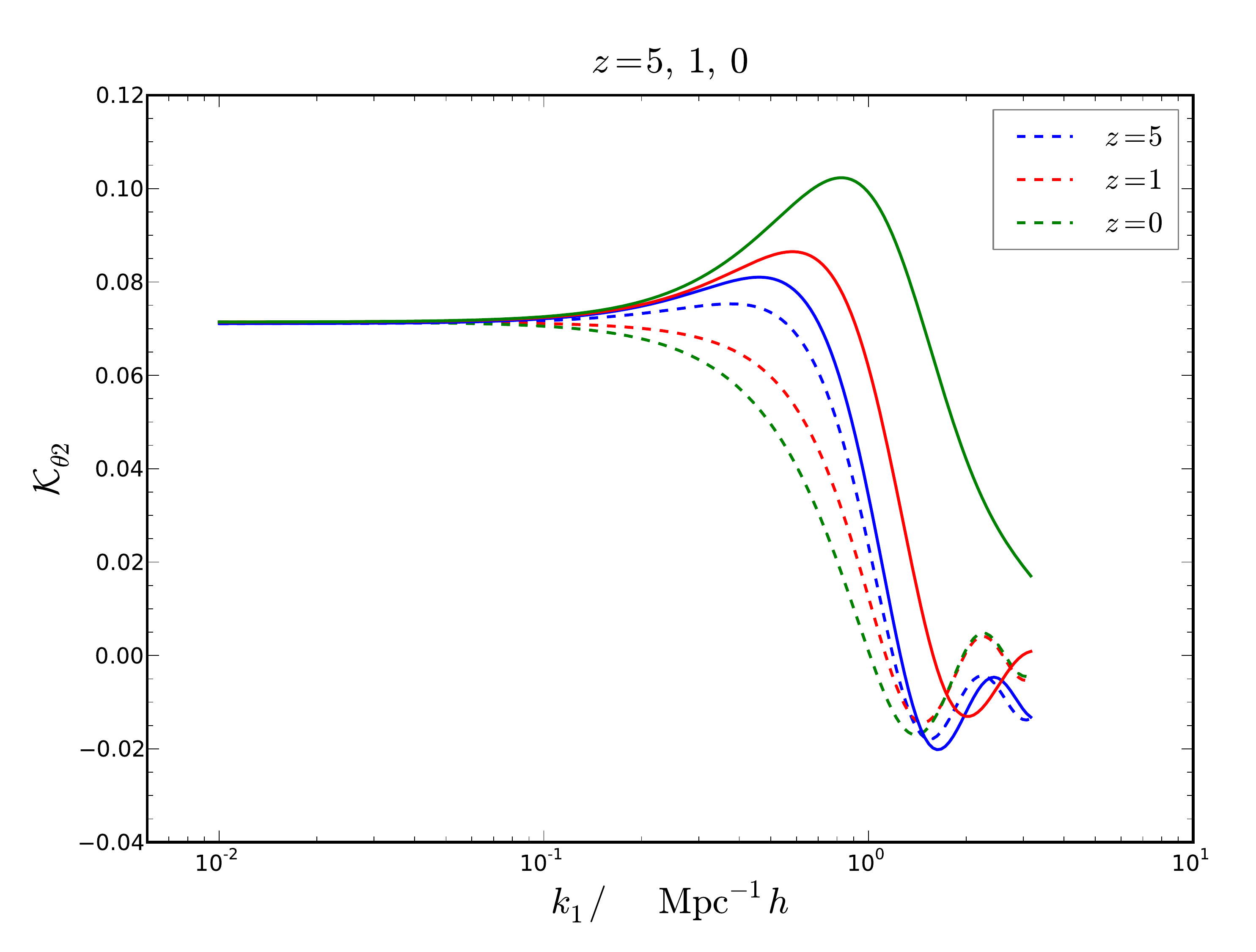}
\caption{ The evolution of the kernel of  $\tg^{(2)} $ for the case with static window (dashed) and the SC evolving one (solid).  The parameters $k_1 = k_2 $ and $\mu =- 1/2$ are used.  Three redshifts are shown, $z=5$ (blue), 1 (red) and 0 (green).    }
\label{fig:theta2g_tot_evolve}
\end{figure}

\begin{figure}[!htb]
\centering
\includegraphics[width=\linewidth]{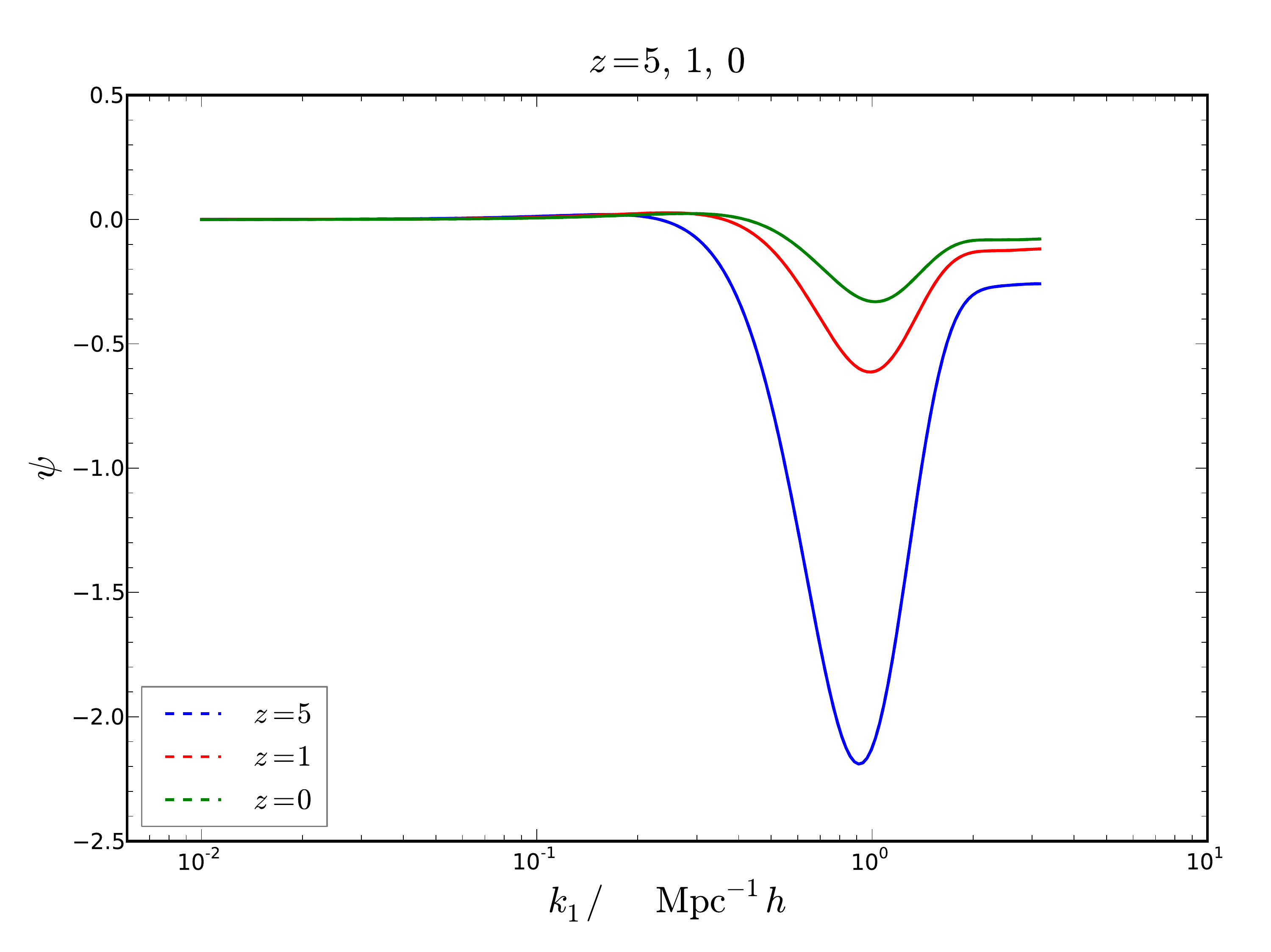}
\caption{ The evolution of the kernel of  $\psi $ for the case with static window (dashed) and the SC evolving one (solid). However, these two cases are indistinguishable.  The parameters are set such that $k_1 = k_2 $ and $\mu = -1/2$.  Three redshifts are shown, $z=5$ (blue), 1 (red) and 0 (green).    }
\label{fig:delta2g_tot_evolve}
\end{figure}

\begin{figure}[!htb]
\centering
\includegraphics[width=\linewidth]{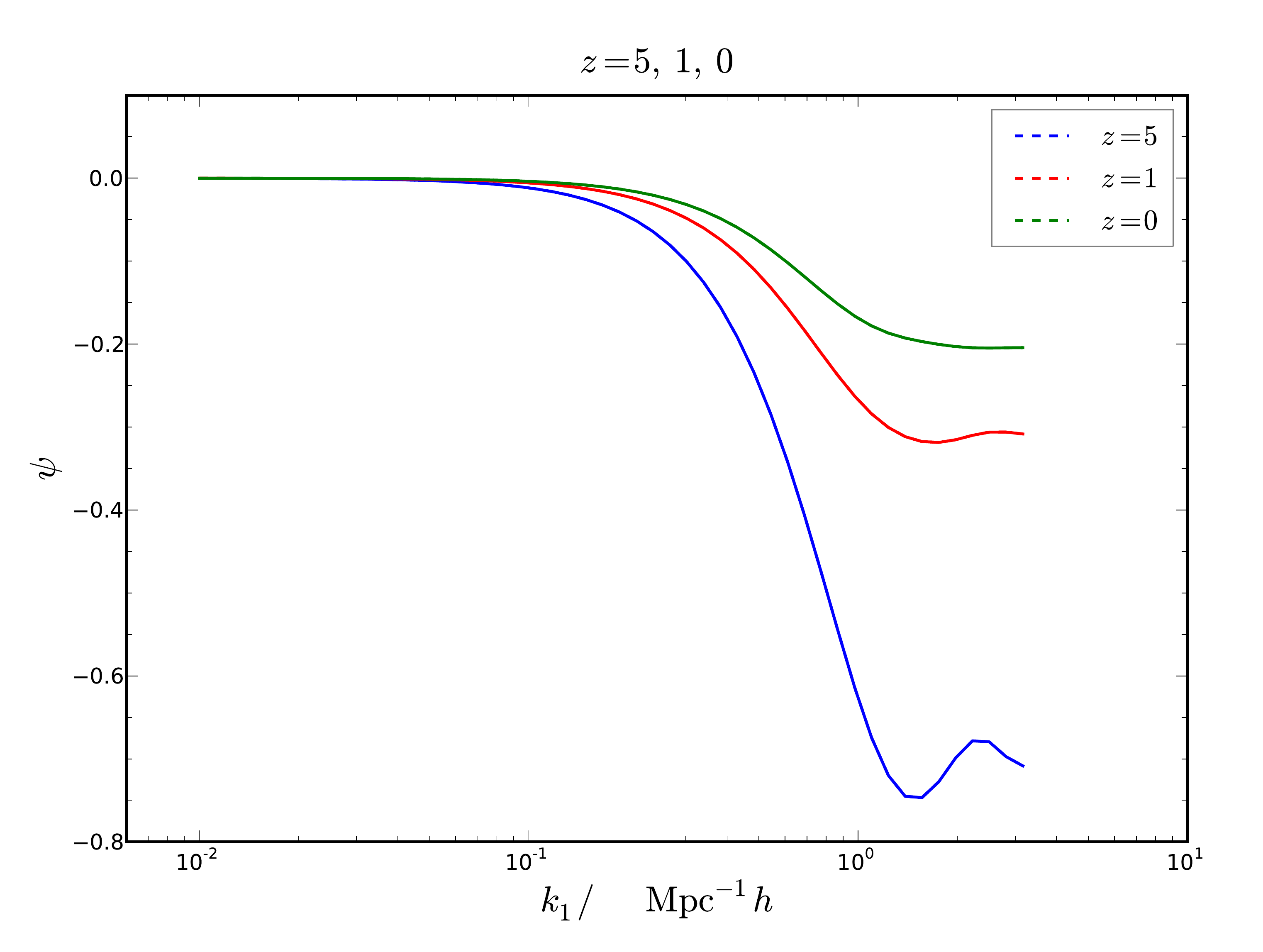}
\caption{ Same as Fig.~\ref{fig:delta2g_tot_evolve}, except with scale-independent initial conditions  $b_1^* = b_{\nu} $ and   $\bv^* =1 $.     }
\label{fig:delta2g_tot_evolve_constant_bstar}
\end{figure}

\begin{figure}[!htb]
\centering
\includegraphics[width=\linewidth]{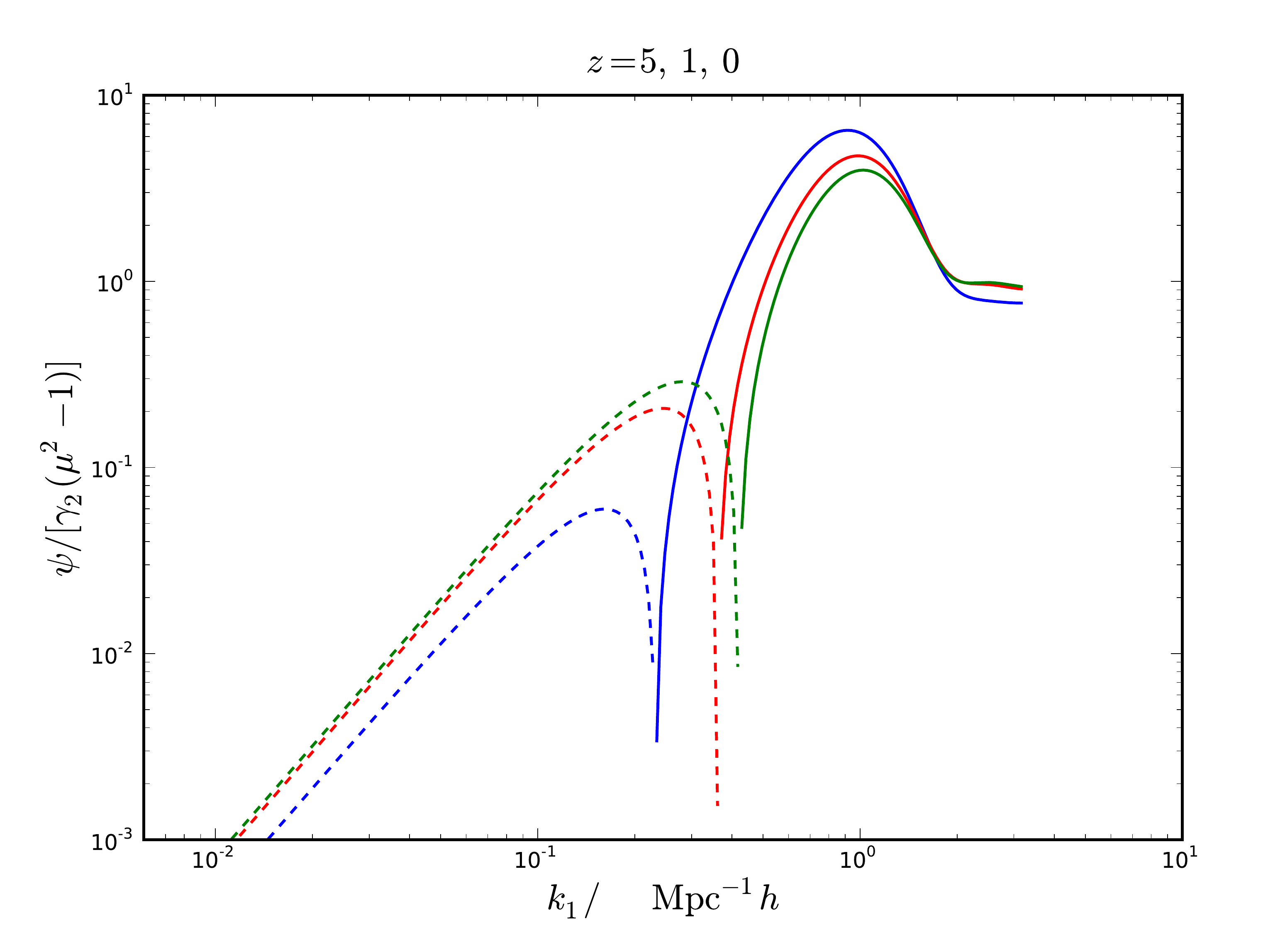}
\caption{  The ratio between the kernel $\psi$ and $\gamma_2 ( \mu^2 -1 )  $, the kernel of the nonlocal term $\mathcal{G}_2$.   The parameters $k_1 = k_2 $ and $\mu =- 1/2$ are used.  Three redshifts are shown, $z=5$ (blue), 1 (red) and 0 (green). Solid line for positive value and dotted line for negative one.   }
\label{fig:psi_G2_ratio}
\end{figure}


\section{Conclusions}
\label{sec:Conclusions}

Recent measurements of the velocity bias suggest that the velocity bias of the halos is non-negligible at the weakly nonlinear regime  $k \sim 0.1 \hOMpc $  at late time. In the time evolution model of the halo field, previously it was shown that it leads to decay of the initial velocity bias using the point particle approximation so that it  becomes negligible at late time. On the other hand, the peak model gives constant velocity bias over time. Thus the measurement seems to be in favour of the peak model result.

It is often assumed that halos are point particles and focus only on their center of mass. Here we argue that as halos consist of a collection of particles, the force acting on its CM of the halo should be the force averaged over its constituent particles instead of only the force at the position of the CM.  To take into account of the halo profile, we introduce a window function in the Euler equation. We find that the window function leads to non-negligible $k^2$ correction to the linear velocity bias.  While the initial $k^2 $ velocity bias decays away, the correction due to profile correction does not.  In contrast,  in the peak model,  the imposition of the peak constraint leads to an extra constant scale-dependent bias.   This difference can be used to distinguish these models.  The profile correction also gives $k^2$ correction to the second order velocity kernel.  For the density bias, the effect  of the profile correction is not important at low $k$ because the magnitude of the initial scale-dependent density bias is large in the peak model. Thus even at low $z$, the magnitude of the terms due to $ b_1^*$ are the most important ones.  Nonetheless, this implies that the $k^2$ correction is non-negligible for $k\gtrsim 0.2\hOMpc$, especially for bispecturm.

Since the window function is dynamical, we model it using the spherical collapse model. We also measure the evolution of the halo  profile  by constructing proto-halos at different redshifts.  To our knowledge this is the first systematic numerical study of the evolution of the proto-halo profile.  We find that the proto-halo profile evolves from a top-hat-like profile to an NFW profile.  We find reasonable agreement between the spherical collapse and  the numerical results.   On the theory side, one may improve the modelling using ellipsoidal collapse model instead. Computationally, it would be useful to come up with a parameterization for the halo profile at various epochs.

Our work has highlighted the importance of halo profile and its evolution on bias.  In theories such as the excursion set theory and peak model,  window function are used to define halos in Lagrangian space.  They are often assumed to be static and the window size given by the Lagrangian size even when they are transformed to the Eulerian space.   The idea of profile evolution can be easily applied to these models as well.

In our model, the effect of the window function correction is most apparent in the velocity bias. Although there are some existing measurements of velocity bias, it is still hard because it is prone to sampling artifacts. We hope to report the comparison of our model with velocity bias measurement in future.

\section*{Acknowledgement} 
I am grateful to Rom\'an Scoccimarro for suggesting that the halo profile evolution as the source of velocity bias and motivating me to look at the effects of halo profile evolution analytically and numerically. He also made numerous valuable suggestions to this paper.  I thank Andreas Berlind, Matteo Biagetti and Ravi Sheth for useful discussions. I also thank  Vincent Desjacques and Ravi Sheth for comments on the draft of the paper.  I thank LasDamas project \footnote{\url{http://lss.phy.vanderbilt.edu/lasdamas}} for the simulations used in the work. The simulations were run using a Teragrid allocation and some RPI and NYU computing resources were also used.  This work is supported by the Swiss National Science Foundation. 

\bibliography{bv_from_profile}

\begin{thebibliography}{59}%
\makeatletter
\providecommand \@ifxundefined [1]{%
 \@ifx{#1\undefined}
}%
\providecommand \@ifnum [1]{%
 \ifnum #1\expandafter \@firstoftwo
 \else \expandafter \@secondoftwo
 \fi
}%
\providecommand \@ifx [1]{%
 \ifx #1\expandafter \@firstoftwo
 \else \expandafter \@secondoftwo
 \fi
}%
\providecommand \natexlab [1]{#1}%
\providecommand \enquote  [1]{``#1''}%
\providecommand \bibnamefont  [1]{#1}%
\providecommand \bibfnamefont [1]{#1}%
\providecommand \citenamefont [1]{#1}%
\providecommand \href@noop [0]{\@secondoftwo}%
\providecommand \href [0]{\begingroup \@sanitize@url \@href}%
\providecommand \@href[1]{\@@startlink{#1}\@@href}%
\providecommand \@@href[1]{\endgroup#1\@@endlink}%
\providecommand \@sanitize@url [0]{\catcode `\\12\catcode `\$12\catcode
  `\&12\catcode `\#12\catcode `\^12\catcode `\_12\catcode `\%12\relax}%
\providecommand \@@startlink[1]{}%
\providecommand \@@endlink[0]{}%
\providecommand \url  [0]{\begingroup\@sanitize@url \@url }%
\providecommand \@url [1]{\endgroup\@href {#1}{\urlprefix }}%
\providecommand \urlprefix  [0]{URL }%
\providecommand \Eprint [0]{\href }%
\providecommand \doibase [0]{http://dx.doi.org/}%
\providecommand \selectlanguage [0]{\@gobble}%
\providecommand \bibinfo  [0]{\@secondoftwo}%
\providecommand \bibfield  [0]{\@secondoftwo}%
\providecommand \translation [1]{[#1]}%
\providecommand \BibitemOpen [0]{}%
\providecommand \bibitemStop [0]{}%
\providecommand \bibitemNoStop [0]{.\EOS\space}%
\providecommand \EOS [0]{\spacefactor3000\relax}%
\providecommand \BibitemShut  [1]{\csname bibitem#1\endcsname}%
\let\auto@bib@innerbib\@empty
\bibitem [{\citenamefont {Kaiser}(1987)}]{Kaiser1987}%
  \BibitemOpen
  \bibfield  {author} {\bibinfo {author} {\bibfnamefont {N.}~\bibnamefont
  {Kaiser}},\ }\href@noop {} {\bibfield  {journal} {\bibinfo  {journal}
  {MNRAS}\ }\textbf {\bibinfo {volume} {227}},\ \bibinfo {pages} {1} (\bibinfo
  {year} {1987})}\BibitemShut {NoStop}%
\bibitem [{\citenamefont {Hamilton}(1992)}]{Hamilton1992}%
  \BibitemOpen
  \bibfield  {author} {\bibinfo {author} {\bibfnamefont {A.~J.~S.}\
  \bibnamefont {Hamilton}},\ }\href@noop {} {\bibfield  {journal} {\bibinfo
  {journal} {ApJ}\ }\textbf {\bibinfo {volume} {385}},\ \bibinfo {pages} {L5}
  (\bibinfo {year} {1992})}\BibitemShut {NoStop}%
\bibitem [{\citenamefont {Song}\ and\ \citenamefont
  {Percival}(2009)}]{SongPercival_2009}%
  \BibitemOpen
  \bibfield  {author} {\bibinfo {author} {\bibfnamefont {Y.-S.}\ \bibnamefont
  {Song}}\ and\ \bibinfo {author} {\bibfnamefont {W.~J.}\ \bibnamefont
  {Percival}},\ }\href@noop {} {\bibfield  {journal} {\bibinfo  {journal}
  {JCAP}\ }\textbf {\bibinfo {volume} {10}},\ \bibinfo {pages} {4} (\bibinfo
  {year} {2009})},\ \Eprint {http://arxiv.org/abs/arXiv:0807.0810}
  {arXiv:0807.0810} \BibitemShut {NoStop}%
\bibitem [{\citenamefont {Taruya}\ \emph {et~al.}(2014)\citenamefont {Taruya},
  \citenamefont {Koyama}, \citenamefont {Hiramatsu},\ and\ \citenamefont
  {Oka}}]{TaruyaKoyama_etal2014}%
  \BibitemOpen
  \bibfield  {author} {\bibinfo {author} {\bibfnamefont {A.}~\bibnamefont
  {Taruya}}, \bibinfo {author} {\bibfnamefont {K.}~\bibnamefont {Koyama}},
  \bibinfo {author} {\bibfnamefont {T.}~\bibnamefont {Hiramatsu}}, \ and\
  \bibinfo {author} {\bibfnamefont {A.}~\bibnamefont {Oka}},\ }\href@noop {}
  {\bibfield  {journal} {\bibinfo  {journal} {Phys. Rev. D}\ }\textbf {\bibinfo
  {volume} {89}},\ \bibinfo {pages} {043509} (\bibinfo {year} {2014})},\
  \Eprint {http://arxiv.org/abs/arXiv:1309.6783} {arXiv:1309.6783} \BibitemShut
  {NoStop}%
\bibitem [{\citenamefont {Guzzo}\ \emph {et~al.}(2008)\citenamefont {Guzzo},
  \citenamefont {Pierleoni}, \citenamefont {Meneux}, \citenamefont {Branchini},
  \citenamefont {Fevre}, \citenamefont {Marinoni} \emph
  {et~al.}}]{GuzzoPierleoni_etal2008}%
  \BibitemOpen
  \bibfield  {author} {\bibinfo {author} {\bibfnamefont {L.}~\bibnamefont
  {Guzzo}}, \bibinfo {author} {\bibfnamefont {M.}~\bibnamefont {Pierleoni}},
  \bibinfo {author} {\bibfnamefont {B.}~\bibnamefont {Meneux}}, \bibinfo
  {author} {\bibfnamefont {E.}~\bibnamefont {Branchini}}, \bibinfo {author}
  {\bibfnamefont {O.~L.}\ \bibnamefont {Fevre}}, \bibinfo {author}
  {\bibfnamefont {C.}~\bibnamefont {Marinoni}},  \emph {et~al.},\ }\href@noop
  {} {\bibfield  {journal} {\bibinfo  {journal} {Nature}\ }\textbf {\bibinfo
  {volume} {451}},\ \bibinfo {pages} {541} (\bibinfo {year} {2008})},\ \Eprint
  {http://arxiv.org/abs/arXiv:0802.1944} {arXiv:0802.1944} \BibitemShut
  {NoStop}%
\bibitem [{\citenamefont {Blake}\ \emph {et~al.}(201)\citenamefont {Blake},
  \citenamefont {Kazin}, \citenamefont {Beutler}, \citenamefont {Davis},
  \citenamefont {Parkinson}, \citenamefont {Brough} \emph
  {et~al.}}]{BlakeKazin_2011}%
  \BibitemOpen
  \bibfield  {author} {\bibinfo {author} {\bibfnamefont {C.}~\bibnamefont
  {Blake}}, \bibinfo {author} {\bibfnamefont {E.~A.}\ \bibnamefont {Kazin}},
  \bibinfo {author} {\bibfnamefont {F.}~\bibnamefont {Beutler}}, \bibinfo
  {author} {\bibfnamefont {T.~M.}\ \bibnamefont {Davis}}, \bibinfo {author}
  {\bibfnamefont {D.}~\bibnamefont {Parkinson}}, \bibinfo {author}
  {\bibfnamefont {S.}~\bibnamefont {Brough}},  \emph {et~al.},\ }\href@noop {}
  {\bibfield  {journal} {\bibinfo  {journal} {MNRAS}\ }\textbf {\bibinfo
  {volume} {418}},\ \bibinfo {pages} {1707} (\bibinfo {year} {201})},\ \Eprint
  {http://arxiv.org/abs/arXiv:1108.2635} {arXiv:1108.2635} \BibitemShut
  {NoStop}%
\bibitem [{\citenamefont {Reid}\ \emph {et~al.}(2012)\citenamefont {Reid},
  \citenamefont {Samushia}, \citenamefont {White}, \citenamefont {Percival},
  \citenamefont {Manera}, \citenamefont {Padmanabhan} \emph
  {et~al.}}]{ReidSamushia_etal2012}%
  \BibitemOpen
  \bibfield  {author} {\bibinfo {author} {\bibfnamefont {B.~A.}\ \bibnamefont
  {Reid}}, \bibinfo {author} {\bibfnamefont {L.}~\bibnamefont {Samushia}},
  \bibinfo {author} {\bibfnamefont {M.}~\bibnamefont {White}}, \bibinfo
  {author} {\bibfnamefont {W.~J.}\ \bibnamefont {Percival}}, \bibinfo {author}
  {\bibfnamefont {M.}~\bibnamefont {Manera}}, \bibinfo {author} {\bibfnamefont
  {N.}~\bibnamefont {Padmanabhan}},  \emph {et~al.},\ }\href@noop {} {\bibfield
   {journal} {\bibinfo  {journal} {MNRAS}\ }\textbf {\bibinfo {volume} {426}},\
  \bibinfo {pages} {2719} (\bibinfo {year} {2012})}\BibitemShut {NoStop}%
\bibitem [{\citenamefont {Beutler}\ \emph {et~al.}(2012)\citenamefont
  {Beutler}, \citenamefont {Blake}, \citenamefont {Colless}, \citenamefont
  {Jones}, \citenamefont {Staveley-Smith}, \citenamefont {Poole} \emph
  {et~al.}}]{BeutlerBlake_etal2012}%
  \BibitemOpen
  \bibfield  {author} {\bibinfo {author} {\bibfnamefont {F.}~\bibnamefont
  {Beutler}}, \bibinfo {author} {\bibfnamefont {C.}~\bibnamefont {Blake}},
  \bibinfo {author} {\bibfnamefont {M.}~\bibnamefont {Colless}}, \bibinfo
  {author} {\bibfnamefont {D.~H.}\ \bibnamefont {Jones}}, \bibinfo {author}
  {\bibfnamefont {L.}~\bibnamefont {Staveley-Smith}}, \bibinfo {author}
  {\bibfnamefont {G.~B.}\ \bibnamefont {Poole}},  \emph {et~al.},\ }\href@noop
  {} {\bibfield  {journal} {\bibinfo  {journal} {MNRAS}\ }\textbf {\bibinfo
  {volume} {423}},\ \bibinfo {pages} {3430} (\bibinfo {year}
  {2012})}\BibitemShut {NoStop}%
\bibitem [{\citenamefont {Samushia}\ \emph {et~al.}(2013)\citenamefont
  {Samushia}, \citenamefont {Reid}, \citenamefont {White}, \citenamefont
  {Percival}, \citenamefont {Cuesta} \emph {et~al.}}]{SamushiaReid_etal2012}%
  \BibitemOpen
  \bibfield  {author} {\bibinfo {author} {\bibfnamefont {L.}~\bibnamefont
  {Samushia}}, \bibinfo {author} {\bibfnamefont {B.~A.}\ \bibnamefont {Reid}},
  \bibinfo {author} {\bibfnamefont {M.}~\bibnamefont {White}}, \bibinfo
  {author} {\bibfnamefont {W.~J.}\ \bibnamefont {Percival}}, \bibinfo {author}
  {\bibfnamefont {A.~J.}\ \bibnamefont {Cuesta}},  \emph {et~al.},\ }\href@noop
  {} {\bibfield  {journal} {\bibinfo  {journal} {MNRAS}\ }\textbf {\bibinfo
  {volume} {429}},\ \bibinfo {pages} {1514} (\bibinfo {year} {2013})},\ \Eprint
  {http://arxiv.org/abs/arXiv:1206.5309} {arXiv:1206.5309} \BibitemShut
  {NoStop}%
\bibitem [{\citenamefont {de~la Torre}\ \emph {et~al.}(2013)\citenamefont
  {de~la Torre}, \citenamefont {Guzzo}, \citenamefont {Peacock}, \citenamefont
  {Branchini}, \citenamefont {Iovino}, \citenamefont {Granett} \emph
  {et~al.}}]{delaTorre_etal2013}%
  \BibitemOpen
  \bibfield  {author} {\bibinfo {author} {\bibfnamefont {S.}~\bibnamefont
  {de~la Torre}}, \bibinfo {author} {\bibfnamefont {L.}~\bibnamefont {Guzzo}},
  \bibinfo {author} {\bibfnamefont {J.~A.}\ \bibnamefont {Peacock}}, \bibinfo
  {author} {\bibfnamefont {E.}~\bibnamefont {Branchini}}, \bibinfo {author}
  {\bibfnamefont {A.}~\bibnamefont {Iovino}}, \bibinfo {author} {\bibfnamefont
  {B.~R.}\ \bibnamefont {Granett}},  \emph {et~al.},\ }\href@noop {} {\bibfield
   {journal} {\bibinfo  {journal} {A\&A}\ }\textbf {\bibinfo {volume} {557}},\
  \bibinfo {pages} {A54} (\bibinfo {year} {2013})},\ \Eprint
  {http://arxiv.org/abs/arXiv:1303.2622} {arXiv:1303.2622} \BibitemShut
  {NoStop}%
\bibitem [{\citenamefont {Oka}\ \emph {et~al.}(2014)\citenamefont {Oka},
  \citenamefont {Saito}, \citenamefont {Nishimichi}, \citenamefont {Taruya},\
  and\ \citenamefont {Yamamoto}}]{OkaSaito_etal2014}%
  \BibitemOpen
  \bibfield  {author} {\bibinfo {author} {\bibfnamefont {A.}~\bibnamefont
  {Oka}}, \bibinfo {author} {\bibfnamefont {S.}~\bibnamefont {Saito}}, \bibinfo
  {author} {\bibfnamefont {T.}~\bibnamefont {Nishimichi}}, \bibinfo {author}
  {\bibfnamefont {A.}~\bibnamefont {Taruya}}, \ and\ \bibinfo {author}
  {\bibfnamefont {K.}~\bibnamefont {Yamamoto}},\ }\href@noop {} {\bibfield
  {journal} {\bibinfo  {journal} {MNRAS}\ }\textbf {\bibinfo {volume} {439}},\
  \bibinfo {pages} {2515} (\bibinfo {year} {2014})},\ \Eprint
  {http://arxiv.org/abs/arXiv:1310.2820} {arXiv:1310.2820} \BibitemShut
  {NoStop}%
\bibitem [{\citenamefont {Koda}\ \emph {et~al.}(2014)\citenamefont {Koda},
  \citenamefont {Blake}, \citenamefont {Davis}, \citenamefont {Magoulas},
  \citenamefont {Springob}, \citenamefont {Scrimgeour} \emph
  {et~al.}}]{KodaBlake_etal2014}%
  \BibitemOpen
  \bibfield  {author} {\bibinfo {author} {\bibfnamefont {J.}~\bibnamefont
  {Koda}}, \bibinfo {author} {\bibfnamefont {C.}~\bibnamefont {Blake}},
  \bibinfo {author} {\bibfnamefont {T.}~\bibnamefont {Davis}}, \bibinfo
  {author} {\bibfnamefont {C.}~\bibnamefont {Magoulas}}, \bibinfo {author}
  {\bibfnamefont {C.~M.}\ \bibnamefont {Springob}}, \bibinfo {author}
  {\bibfnamefont {M.}~\bibnamefont {Scrimgeour}},  \emph {et~al.},\ }\href@noop
  {} {\bibfield  {journal} {\bibinfo  {journal} {MNRAS}\ }\textbf {\bibinfo
  {volume} {445}},\ \bibinfo {pages} {4267} (\bibinfo {year} {2014})},\ \Eprint
  {http://arxiv.org/abs/arXiv:1312.1022} {arXiv:1312.1022} \BibitemShut
  {NoStop}%
\bibitem [{\citenamefont {Elia}\ \emph {et~al.}(2002)\citenamefont {Elia},
  \citenamefont {Ludlow},\ and\ \citenamefont
  {Porciani}}]{EliaLudlowPorciani2012}%
  \BibitemOpen
  \bibfield  {author} {\bibinfo {author} {\bibfnamefont {A.}~\bibnamefont
  {Elia}}, \bibinfo {author} {\bibfnamefont {A.~D.}\ \bibnamefont {Ludlow}}, \
  and\ \bibinfo {author} {\bibfnamefont {C.}~\bibnamefont {Porciani}},\
  }\href@noop {} {\bibfield  {journal} {\bibinfo  {journal} {MNRAS}\ }\textbf
  {\bibinfo {volume} {421}},\ \bibinfo {pages} {3472} (\bibinfo {year}
  {2002})},\ \Eprint {http://arxiv.org/abs/arXiv:1111.4211} {arXiv:1111.4211}
  \BibitemShut {NoStop}%
\bibitem [{\citenamefont {Chan}\ \emph {et~al.}(2012)\citenamefont {Chan},
  \citenamefont {Scoccimarro},\ and\ \citenamefont
  {Sheth}}]{ChanScoccimarroSheth2012}%
  \BibitemOpen
  \bibfield  {author} {\bibinfo {author} {\bibfnamefont {K.~C.}\ \bibnamefont
  {Chan}}, \bibinfo {author} {\bibfnamefont {R.}~\bibnamefont {Scoccimarro}}, \
  and\ \bibinfo {author} {\bibfnamefont {R.~K.}\ \bibnamefont {Sheth}},\
  }\href@noop {} {\bibfield  {journal} {\bibinfo  {journal} {Phys. Rev. D}\
  }\textbf {\bibinfo {volume} {85}},\ \bibinfo {pages} {083509} (\bibinfo
  {year} {2012})},\ \Eprint {http://arxiv.org/abs/arXiv:1201.3614}
  {arXiv:1201.3614} \BibitemShut {NoStop}%
\bibitem [{\citenamefont {Baldauf}\ \emph {et~al.}(2014)\citenamefont
  {Baldauf}, \citenamefont {Desjacques},\ and\ \citenamefont
  {Seljak}}]{BaldaufDesjacquesSeljak2014}%
  \BibitemOpen
  \bibfield  {author} {\bibinfo {author} {\bibfnamefont {T.}~\bibnamefont
  {Baldauf}}, \bibinfo {author} {\bibfnamefont {V.}~\bibnamefont {Desjacques}},
  \ and\ \bibinfo {author} {\bibfnamefont {U.}~\bibnamefont {Seljak}},\
  }\href@noop {} {} (\bibinfo {year} {2014}),\ \Eprint
  {http://arxiv.org/abs/arXiv:1405.5885} {arXiv:1405.5885} \BibitemShut
  {NoStop}%
\bibitem [{\citenamefont {Jennings}\ \emph {et~al.}(2014)\citenamefont
  {Jennings}, \citenamefont {Baugh},\ and\ \citenamefont
  {Hatt}}]{JenningsBaughHatt2014}%
  \BibitemOpen
  \bibfield  {author} {\bibinfo {author} {\bibfnamefont {E.}~\bibnamefont
  {Jennings}}, \bibinfo {author} {\bibfnamefont {C.}~\bibnamefont {Baugh}}, \
  and\ \bibinfo {author} {\bibfnamefont {D.}~\bibnamefont {Hatt}},\ }\href@noop
  {} {} (\bibinfo {year} {2014}),\ \Eprint
  {http://arxiv.org/abs/arXiv:1407.7296} {arXiv:1407.7296} \BibitemShut
  {NoStop}%
\bibitem [{\citenamefont {Zheng}\ \emph
  {et~al.}(2014{\natexlab{a}})\citenamefont {Zheng}, \citenamefont {Zhang},\
  and\ \citenamefont {Jing}}]{ZhengZhangJing2014b}%
  \BibitemOpen
  \bibfield  {author} {\bibinfo {author} {\bibfnamefont {Y.}~\bibnamefont
  {Zheng}}, \bibinfo {author} {\bibfnamefont {P.}~\bibnamefont {Zhang}}, \ and\
  \bibinfo {author} {\bibfnamefont {Y.}~\bibnamefont {Jing}},\ }\href@noop {}
  {} (\bibinfo {year} {2014}{\natexlab{a}}),\ \Eprint
  {http://arxiv.org/abs/arXiv:1410.1256} {arXiv:1410.1256} \BibitemShut
  {NoStop}%
\bibitem [{\citenamefont {Scoccimarro}(2004)}]{Scoccimarro2004}%
  \BibitemOpen
  \bibfield  {author} {\bibinfo {author} {\bibfnamefont {R.}~\bibnamefont
  {Scoccimarro}},\ }\href@noop {} {\bibfield  {journal} {\bibinfo  {journal}
  {Phys. Rev. D}\ }\textbf {\bibinfo {volume} {70}},\ \bibinfo {pages} {083007}
  (\bibinfo {year} {2004})}\BibitemShut {NoStop}%
\bibitem [{\citenamefont {Juszkiewicz}\ \emph {et~al.}(1995)\citenamefont
  {Juszkiewicz}, \citenamefont {Weinberg}, \citenamefont {Amsterdamski},
  \citenamefont {Chodorowski},\ and\ \citenamefont
  {Bouchet}}]{Juszkiewiczetal1995}%
  \BibitemOpen
  \bibfield  {author} {\bibinfo {author} {\bibfnamefont {R.}~\bibnamefont
  {Juszkiewicz}}, \bibinfo {author} {\bibfnamefont {D.~H.}\ \bibnamefont
  {Weinberg}}, \bibinfo {author} {\bibfnamefont {P.}~\bibnamefont
  {Amsterdamski}}, \bibinfo {author} {\bibfnamefont {M.}~\bibnamefont
  {Chodorowski}}, \ and\ \bibinfo {author} {\bibfnamefont {F.}~\bibnamefont
  {Bouchet}},\ }\href@noop {} {\bibfield  {journal} {\bibinfo  {journal} {ApJ}\
  }\textbf {\bibinfo {volume} {442}},\ \bibinfo {pages} {39} (\bibinfo {year}
  {1995})}\BibitemShut {NoStop}%
\bibitem [{\citenamefont {Zhang}\ \emph {et~al.}(2014)\citenamefont {Zhang},
  \citenamefont {Zheng},\ and\ \citenamefont {Jing}}]{ZhangZhengJing2014}%
  \BibitemOpen
  \bibfield  {author} {\bibinfo {author} {\bibfnamefont {P.}~\bibnamefont
  {Zhang}}, \bibinfo {author} {\bibfnamefont {Y.}~\bibnamefont {Zheng}}, \ and\
  \bibinfo {author} {\bibfnamefont {Y.}~\bibnamefont {Jing}},\ }\href@noop {}
  {} (\bibinfo {year} {2014}),\ \Eprint {http://arxiv.org/abs/arXiv:1405.7125}
  {arXiv:1405.7125} \BibitemShut {NoStop}%
\bibitem [{\citenamefont {Zheng}\ \emph
  {et~al.}(2014{\natexlab{b}})\citenamefont {Zheng}, \citenamefont {Zhang},\
  and\ \citenamefont {Jing}}]{ZhengZhangJing2014a}%
  \BibitemOpen
  \bibfield  {author} {\bibinfo {author} {\bibfnamefont {Y.}~\bibnamefont
  {Zheng}}, \bibinfo {author} {\bibfnamefont {P.}~\bibnamefont {Zhang}}, \ and\
  \bibinfo {author} {\bibfnamefont {Y.}~\bibnamefont {Jing}},\ }\href@noop {}
  {} (\bibinfo {year} {2014}{\natexlab{b}}),\ \Eprint
  {http://arxiv.org/abs/arXiv:1409.6809} {arXiv:1409.6809} \BibitemShut
  {NoStop}%
\bibitem [{\citenamefont {Okumura}\ \emph {et~al.}(2012)\citenamefont
  {Okumura}, \citenamefont {Seljak}, \citenamefont {McDonald},\ and\
  \citenamefont {Desjacques}}]{OkumuraSeljak_etal2012}%
  \BibitemOpen
  \bibfield  {author} {\bibinfo {author} {\bibfnamefont {T.}~\bibnamefont
  {Okumura}}, \bibinfo {author} {\bibfnamefont {U.}~\bibnamefont {Seljak}},
  \bibinfo {author} {\bibfnamefont {P.}~\bibnamefont {McDonald}}, \ and\
  \bibinfo {author} {\bibfnamefont {V.}~\bibnamefont {Desjacques}},\
  }\href@noop {} {\bibfield  {journal} {\bibinfo  {journal} {JCAP}\ }\textbf
  {\bibinfo {volume} {02}},\ \bibinfo {pages} {010} (\bibinfo {year} {2012})},\
  \Eprint {http://arxiv.org/abs/arXiv:1109.1609} {arXiv:1109.1609} \BibitemShut
  {NoStop}%
\bibitem [{\citenamefont {Elia}\ \emph {et~al.}(2011)\citenamefont {Elia},
  \citenamefont {Kulkarni}, \citenamefont {Porciani}, \citenamefont
  {Pietroni},\ and\ \citenamefont {Matarrese}}]{EliaKulkarnietal2011}%
  \BibitemOpen
  \bibfield  {author} {\bibinfo {author} {\bibfnamefont {A.}~\bibnamefont
  {Elia}}, \bibinfo {author} {\bibfnamefont {S.}~\bibnamefont {Kulkarni}},
  \bibinfo {author} {\bibfnamefont {C.}~\bibnamefont {Porciani}}, \bibinfo
  {author} {\bibfnamefont {M.}~\bibnamefont {Pietroni}}, \ and\ \bibinfo
  {author} {\bibfnamefont {S.}~\bibnamefont {Matarrese}},\ }\href@noop {}
  {\bibfield  {journal} {\bibinfo  {journal} {MNRAS}\ }\textbf {\bibinfo
  {volume} {416}},\ \bibinfo {pages} {1703} (\bibinfo {year} {2011})},\ \Eprint
  {http://arxiv.org/abs/arXiv:1012.4833} {arXiv:1012.4833} \BibitemShut
  {NoStop}%
\bibitem [{\citenamefont {Desjacques}\ \emph {et~al.}(2010)\citenamefont
  {Desjacques}, \citenamefont {Crocce}, \citenamefont {Scoccimarro},\ and\
  \citenamefont {Sheth}}]{DesjacquesCrocceetal2014}%
  \BibitemOpen
  \bibfield  {author} {\bibinfo {author} {\bibfnamefont {V.}~\bibnamefont
  {Desjacques}}, \bibinfo {author} {\bibfnamefont {M.}~\bibnamefont {Crocce}},
  \bibinfo {author} {\bibfnamefont {R.}~\bibnamefont {Scoccimarro}}, \ and\
  \bibinfo {author} {\bibfnamefont {R.~K.}\ \bibnamefont {Sheth}},\ }\href@noop
  {} {\bibfield  {journal} {\bibinfo  {journal} {Phys. Rev. D}\ }\textbf
  {\bibinfo {volume} {82}},\ \bibinfo {pages} {103529} (\bibinfo {year}
  {2010})},\ \Eprint {http://arxiv.org/abs/arXiv:1009.3449} {arXiv:1009.3449}
  \BibitemShut {NoStop}%
\bibitem [{\citenamefont {Biagetti}\ \emph
  {et~al.}(2014{\natexlab{a}})\citenamefont {Biagetti}, \citenamefont
  {Desjacques}, \citenamefont {Kehagias},\ and\ \citenamefont
  {Riotto}}]{BiagettiDesjacquesKehagiasRiotto2014b}%
  \BibitemOpen
  \bibfield  {author} {\bibinfo {author} {\bibfnamefont {M.}~\bibnamefont
  {Biagetti}}, \bibinfo {author} {\bibfnamefont {V.}~\bibnamefont
  {Desjacques}}, \bibinfo {author} {\bibfnamefont {A.}~\bibnamefont
  {Kehagias}}, \ and\ \bibinfo {author} {\bibfnamefont {A.}~\bibnamefont
  {Riotto}},\ }\href@noop {} {\bibfield  {journal} {\bibinfo  {journal} {Phys.
  Rev. D}\ }\textbf {\bibinfo {volume} {90}},\ \bibinfo {pages} {103529}
  (\bibinfo {year} {2014}{\natexlab{a}})},\ \Eprint
  {http://arxiv.org/abs/arXiv:1408.0293} {arXiv:1408.0293} \BibitemShut
  {NoStop}%
\bibitem [{\citenamefont {Desjacques}\ and\ \citenamefont
  {Sheth}(2010)}]{DesjacquesSheth2010}%
  \BibitemOpen
  \bibfield  {author} {\bibinfo {author} {\bibfnamefont {V.}~\bibnamefont
  {Desjacques}}\ and\ \bibinfo {author} {\bibfnamefont {R.~K.}\ \bibnamefont
  {Sheth}},\ }\href@noop {} {\bibfield  {journal} {\bibinfo  {journal} {Phys.
  Rev. D}\ }\textbf {\bibinfo {volume} {81}},\ \bibinfo {pages} {023526}
  (\bibinfo {year} {2010})},\ \Eprint {http://arxiv.org/abs/arXiv:0909.4544}
  {arXiv:0909.4544} \BibitemShut {NoStop}%
\bibitem [{\citenamefont {Seljak}(2000)}]{Seljak2000}%
  \BibitemOpen
  \bibfield  {author} {\bibinfo {author} {\bibfnamefont {U.}~\bibnamefont
  {Seljak}},\ }\href@noop {} {\bibfield  {journal} {\bibinfo  {journal}
  {MNRAS}\ }\textbf {\bibinfo {volume} {318}},\ \bibinfo {pages} {203}
  (\bibinfo {year} {2000})}\BibitemShut {NoStop}%
\bibitem [{\citenamefont {Peacock}\ and\ \citenamefont
  {Smith}(2000)}]{PeacockSmith2000}%
  \BibitemOpen
  \bibfield  {author} {\bibinfo {author} {\bibfnamefont {J.~A.}\ \bibnamefont
  {Peacock}}\ and\ \bibinfo {author} {\bibfnamefont {R.~R.}\ \bibnamefont
  {Smith}},\ }\href@noop {} {\bibfield  {journal} {\bibinfo  {journal} {MNRAS}\
  }\textbf {\bibinfo {volume} {318}},\ \bibinfo {pages} {1144} (\bibinfo {year}
  {2000})}\BibitemShut {NoStop}%
\bibitem [{\citenamefont {Scoccimarro}\ \emph {et~al.}(2001)\citenamefont
  {Scoccimarro}, \citenamefont {Sheth}, \citenamefont {Hui},\ and\
  \citenamefont {Jain}}]{Scoccimarroetal2001}%
  \BibitemOpen
  \bibfield  {author} {\bibinfo {author} {\bibfnamefont {R.}~\bibnamefont
  {Scoccimarro}}, \bibinfo {author} {\bibfnamefont {R.~K.}\ \bibnamefont
  {Sheth}}, \bibinfo {author} {\bibfnamefont {L.}~\bibnamefont {Hui}}, \ and\
  \bibinfo {author} {\bibfnamefont {B.}~\bibnamefont {Jain}},\ }\href@noop {}
  {\bibfield  {journal} {\bibinfo  {journal} {ApJ}\ }\textbf {\bibinfo {volume}
  {546}},\ \bibinfo {pages} {20} (\bibinfo {year} {2001})}\BibitemShut
  {NoStop}%
\bibitem [{\citenamefont {Cooray}\ and\ \citenamefont
  {Sheth}(2002)}]{CooraySheth}%
  \BibitemOpen
  \bibfield  {author} {\bibinfo {author} {\bibfnamefont {A.}~\bibnamefont
  {Cooray}}\ and\ \bibinfo {author} {\bibfnamefont {R.}~\bibnamefont {Sheth}},\
  }\href@noop {} {\bibfield  {journal} {\bibinfo  {journal} {Phys. Rep.}\
  }\textbf {\bibinfo {volume} {372}},\ \bibinfo {pages} {1} (\bibinfo {year}
  {2002})},\ \Eprint {http://arxiv.org/abs/arXiv:astro-ph/0206508}
  {arXiv:astro-ph/0206508} \BibitemShut {NoStop}%
\bibitem [{\citenamefont {Navarro}\ \emph {et~al.}(1996)\citenamefont
  {Navarro}, \citenamefont {Frenk},\ and\ \citenamefont {White}}]{NFW1996}%
  \BibitemOpen
  \bibfield  {author} {\bibinfo {author} {\bibfnamefont {J.~F.}\ \bibnamefont
  {Navarro}}, \bibinfo {author} {\bibfnamefont {C.~S.}\ \bibnamefont {Frenk}},
  \ and\ \bibinfo {author} {\bibfnamefont {S.~D.~M.}\ \bibnamefont {White}},\
  }\href@noop {} {\bibfield  {journal} {\bibinfo  {journal} {ApJ}\ }\textbf
  {\bibinfo {volume} {462}},\ \bibinfo {pages} {563} (\bibinfo {year}
  {1996})}\BibitemShut {NoStop}%
\bibitem [{\citenamefont {{Gunn}}\ and\ \citenamefont
  {{Gott}}(1972)}]{GunnGott1972}%
  \BibitemOpen
  \bibfield  {author} {\bibinfo {author} {\bibfnamefont {J.~E.}\ \bibnamefont
  {{Gunn}}}\ and\ \bibinfo {author} {\bibfnamefont {J.~R.}\ \bibnamefont
  {{Gott}}, \bibfnamefont {III}},\ }\href@noop {} {\bibfield  {journal}
  {\bibinfo  {journal} {ApJ}\ }\textbf {\bibinfo {volume} {176}},\ \bibinfo
  {pages} {1} (\bibinfo {year} {1972})}\BibitemShut {NoStop}%
\bibitem [{\citenamefont {Peebles}(1980)}]{Peebles1980}%
  \BibitemOpen
  \bibfield  {author} {\bibinfo {author} {\bibfnamefont {P.~J.~E.}\
  \bibnamefont {Peebles}},\ }\href@noop {} {\emph {\bibinfo {title} {The
  Large-Scale Structure of the Universe}}}\ (\bibinfo  {publisher} {Princeton
  University Press},\ \bibinfo {address} {New Jersey},\ \bibinfo {year}
  {1980})\BibitemShut {NoStop}%
\bibitem [{\citenamefont {Padmanabhan}(1993)}]{Padmanabhan1993}%
  \BibitemOpen
  \bibfield  {author} {\bibinfo {author} {\bibfnamefont {T.}~\bibnamefont
  {Padmanabhan}},\ }\href@noop {} {\emph {\bibinfo {title} {Structure formation
  in the Universe}}}\ (\bibinfo  {publisher} {Cambridge University Press},\
  \bibinfo {address} {Cambridge},\ \bibinfo {year} {1993})\BibitemShut
  {NoStop}%
\bibitem [{\citenamefont {Mo}\ \emph {et~al.}(2010)\citenamefont {Mo},
  \citenamefont {van~den Bosch},\ and\ \citenamefont
  {White}}]{MoBoschWhite2010}%
  \BibitemOpen
  \bibfield  {author} {\bibinfo {author} {\bibfnamefont {H.}~\bibnamefont
  {Mo}}, \bibinfo {author} {\bibfnamefont {F.}~\bibnamefont {van~den Bosch}}, \
  and\ \bibinfo {author} {\bibfnamefont {S.}~\bibnamefont {White}},\
  }\href@noop {} {\emph {\bibinfo {title} {Galaxy Formation and Evolution}}}\
  (\bibinfo  {publisher} {Cambridge University Press},\ \bibinfo {address}
  {Cambridge},\ \bibinfo {year} {2010})\BibitemShut {NoStop}%
\bibitem [{\citenamefont {Lynden-Bell}(1967)}]{LyndenBell1967}%
  \BibitemOpen
  \bibfield  {author} {\bibinfo {author} {\bibfnamefont {D.}~\bibnamefont
  {Lynden-Bell}},\ }\href@noop {} {\bibfield  {journal} {\bibinfo  {journal}
  {MNRAS}\ }\textbf {\bibinfo {volume} {136}},\ \bibinfo {pages} {101}
  (\bibinfo {year} {1967})}\BibitemShut {NoStop}%
\bibitem [{\citenamefont {Bindoni}\ and\ \citenamefont
  {Secco}(2008)}]{BindoniSecco2008}%
  \BibitemOpen
  \bibfield  {author} {\bibinfo {author} {\bibfnamefont {D.}~\bibnamefont
  {Bindoni}}\ and\ \bibinfo {author} {\bibfnamefont {L.}~\bibnamefont
  {Secco}},\ }\href@noop {} {\bibfield  {journal} {\bibinfo  {journal} {New
  Astronomy Reviews}\ }\textbf {\bibinfo {volume} {52}},\ \bibinfo {pages} {1}
  (\bibinfo {year} {2008})}\BibitemShut {NoStop}%
\bibitem [{\citenamefont {Bryan}\ and\ \citenamefont
  {Norman}(1998)}]{BryanNorman1998}%
  \BibitemOpen
  \bibfield  {author} {\bibinfo {author} {\bibfnamefont {G.~L.}\ \bibnamefont
  {Bryan}}\ and\ \bibinfo {author} {\bibfnamefont {M.~L.}\ \bibnamefont
  {Norman}},\ }\href@noop {} {\bibfield  {journal} {\bibinfo  {journal} {ApJ}\
  }\textbf {\bibinfo {volume} {495}},\ \bibinfo {pages} {80} (\bibinfo {year}
  {1998})}\BibitemShut {NoStop}%
\bibitem [{\citenamefont {Scoccimarro}(1998)}]{Scoccimarro98}%
  \BibitemOpen
  \bibfield  {author} {\bibinfo {author} {\bibfnamefont {R.}~\bibnamefont
  {Scoccimarro}},\ }\href@noop {} {\bibfield  {journal} {\bibinfo  {journal}
  {MNRAS}\ }\textbf {\bibinfo {volume} {299}},\ \bibinfo {pages} {1097}
  (\bibinfo {year} {1998})},\ \Eprint
  {http://arxiv.org/abs/arXiv:astro-ph/9711187} {arXiv:astro-ph/9711187}
  \BibitemShut {NoStop}%
\bibitem [{\citenamefont {Crocce}\ \emph {et~al.}(2006)\citenamefont {Crocce},
  \citenamefont {Pueblas},\ and\ \citenamefont
  {Scoccimarro}}]{CroccePeublasetal2006}%
  \BibitemOpen
  \bibfield  {author} {\bibinfo {author} {\bibfnamefont {M.}~\bibnamefont
  {Crocce}}, \bibinfo {author} {\bibfnamefont {S.}~\bibnamefont {Pueblas}}, \
  and\ \bibinfo {author} {\bibfnamefont {R.}~\bibnamefont {Scoccimarro}},\
  }\href@noop {} {\bibfield  {journal} {\bibinfo  {journal} {MNRAS}\ }\textbf
  {\bibinfo {volume} {373}},\ \bibinfo {pages} {369} (\bibinfo {year}
  {2006})},\ \Eprint {http://arxiv.org/abs/arXiv:astro-ph/0606505}
  {arXiv:astro-ph/0606505} \BibitemShut {NoStop}%
\bibitem [{\citenamefont {Eke}\ \emph {et~al.}(1996)\citenamefont {Eke},
  \citenamefont {Cole},\ and\ \citenamefont {Frenk}}]{EkeColeFrenk1996}%
  \BibitemOpen
  \bibfield  {author} {\bibinfo {author} {\bibfnamefont {V.~R.}\ \bibnamefont
  {Eke}}, \bibinfo {author} {\bibfnamefont {S.}~\bibnamefont {Cole}}, \ and\
  \bibinfo {author} {\bibfnamefont {C.~S.}\ \bibnamefont {Frenk}},\ }\href@noop
  {} {\bibfield  {journal} {\bibinfo  {journal} {MNRAS}\ }\textbf {\bibinfo
  {volume} {282}},\ \bibinfo {pages} {263} (\bibinfo {year}
  {1996})}\BibitemShut {NoStop}%
\bibitem [{\citenamefont {Seljak}\ and\ \citenamefont
  {Zaldarriaga}(1996)}]{CMBFAST}%
  \BibitemOpen
  \bibfield  {author} {\bibinfo {author} {\bibfnamefont {U.}~\bibnamefont
  {Seljak}}\ and\ \bibinfo {author} {\bibfnamefont {M.}~\bibnamefont
  {Zaldarriaga}},\ }\href@noop {} {\bibfield  {journal} {\bibinfo  {journal}
  {ApJ}\ }\textbf {\bibinfo {volume} {469}},\ \bibinfo {pages} {437} (\bibinfo
  {year} {1996})},\ \Eprint {http://arxiv.org/abs/arXiv:astro-ph/9603033}
  {arXiv:astro-ph/9603033} \BibitemShut {NoStop}%
\bibitem [{\citenamefont {Springel}(2005)}]{Gadget2}%
  \BibitemOpen
  \bibfield  {author} {\bibinfo {author} {\bibfnamefont {V.}~\bibnamefont
  {Springel}},\ }\href@noop {} {\bibfield  {journal} {\bibinfo  {journal}
  {MNRAS}\ }\textbf {\bibinfo {volume} {364}},\ \bibinfo {pages} {1105}
  (\bibinfo {year} {2005})},\ \Eprint
  {http://arxiv.org/abs/arXiv:astro-ph/0505010} {arXiv:astro-ph/0505010}
  \BibitemShut {NoStop}%
\bibitem [{\citenamefont {Neto}\ \emph {et~al.}(2007)\citenamefont {Neto},
  \citenamefont {Gao}, \citenamefont {Bett}, \citenamefont {Cole},
  \citenamefont {Navarro} \emph {et~al.}}]{NetoGaoBettetal2007}%
  \BibitemOpen
  \bibfield  {author} {\bibinfo {author} {\bibfnamefont {A.~F.}\ \bibnamefont
  {Neto}}, \bibinfo {author} {\bibfnamefont {L.}~\bibnamefont {Gao}}, \bibinfo
  {author} {\bibfnamefont {P.}~\bibnamefont {Bett}}, \bibinfo {author}
  {\bibfnamefont {S.}~\bibnamefont {Cole}}, \bibinfo {author} {\bibfnamefont
  {J.~F.}\ \bibnamefont {Navarro}},  \emph {et~al.},\ }\href@noop {} {\bibfield
   {journal} {\bibinfo  {journal} {MNRAS}\ }\textbf {\bibinfo {volume} {381}},\
  \bibinfo {pages} {1450} (\bibinfo {year} {2007})}\BibitemShut {NoStop}%
\bibitem [{\citenamefont {Chan}\ \emph {et~al.}(2015)\citenamefont {Chan},
  \citenamefont {Sheth},\ and\ \citenamefont
  {Scoccimarro}}]{ChanShethScoccimarro2015}%
  \BibitemOpen
  \bibfield  {author} {\bibinfo {author} {\bibfnamefont {K.~C.}\ \bibnamefont
  {Chan}}, \bibinfo {author} {\bibfnamefont {R.~K.}\ \bibnamefont {Sheth}}, \
  and\ \bibinfo {author} {\bibfnamefont {R.}~\bibnamefont {Scoccimarro}},\
  }\href@noop {} {}\bibinfo {howpublished} {in preparation} (\bibinfo {year}
  {2015})\BibitemShut {NoStop}%
\bibitem [{\citenamefont {Engineer}\ \emph {et~al.}(2000)\citenamefont
  {Engineer}, \citenamefont {Kanekar},\ and\ \citenamefont
  {Padmanabhan}}]{Engineer_etal2000}%
  \BibitemOpen
  \bibfield  {author} {\bibinfo {author} {\bibfnamefont {S.}~\bibnamefont
  {Engineer}}, \bibinfo {author} {\bibfnamefont {N.}~\bibnamefont {Kanekar}}, \
  and\ \bibinfo {author} {\bibfnamefont {T.}~\bibnamefont {Padmanabhan}},\
  }\href@noop {} {\bibfield  {journal} {\bibinfo  {journal} {MNRAS}\ }\textbf
  {\bibinfo {volume} {314}},\ \bibinfo {pages} {279} (\bibinfo {year}
  {2000})},\ \Eprint {http://arxiv.org/abs/arXiv:astro-ph/9812452}
  {arXiv:astro-ph/9812452} \BibitemShut {NoStop}%
\bibitem [{\citenamefont {Bond}\ and\ \citenamefont
  {Myers}(1996)}]{BondMyers1996}%
  \BibitemOpen
  \bibfield  {author} {\bibinfo {author} {\bibfnamefont {J.~R.}\ \bibnamefont
  {Bond}}\ and\ \bibinfo {author} {\bibfnamefont {S.~T.}\ \bibnamefont
  {Myers}},\ }\href@noop {} {\bibfield  {journal} {\bibinfo  {journal} {ApJS}\
  }\textbf {\bibinfo {volume} {103}},\ \bibinfo {pages} {1} (\bibinfo {year}
  {1996})}\BibitemShut {NoStop}%
\bibitem [{\citenamefont {Shen}\ \emph {et~al.}(2006)\citenamefont {Shen},
  \citenamefont {Abel}, \citenamefont {Mo},\ and\ \citenamefont
  {Sheth}}]{ShenAbelMoSheth2006}%
  \BibitemOpen
  \bibfield  {author} {\bibinfo {author} {\bibfnamefont {J.}~\bibnamefont
  {Shen}}, \bibinfo {author} {\bibfnamefont {T.}~\bibnamefont {Abel}}, \bibinfo
  {author} {\bibfnamefont {H.~J.}\ \bibnamefont {Mo}}, \ and\ \bibinfo {author}
  {\bibfnamefont {R.~K.}\ \bibnamefont {Sheth}},\ }\href@noop {} {\bibfield
  {journal} {\bibinfo  {journal} {ApJ}\ }\textbf {\bibinfo {volume} {645}},\
  \bibinfo {pages} {783} (\bibinfo {year} {2006})}\BibitemShut {NoStop}%
\bibitem [{\citenamefont {Sheth}\ \emph {et~al.}(2001)\citenamefont {Sheth},
  \citenamefont {Mo},\ and\ \citenamefont {Tormen}}]{ShethMoTormen2001}%
  \BibitemOpen
  \bibfield  {author} {\bibinfo {author} {\bibfnamefont {R.~K.}\ \bibnamefont
  {Sheth}}, \bibinfo {author} {\bibfnamefont {H.~J.}\ \bibnamefont {Mo}}, \
  and\ \bibinfo {author} {\bibfnamefont {G.}~\bibnamefont {Tormen}},\
  }\href@noop {} {\bibfield  {journal} {\bibinfo  {journal} {MNRAS}\ }\textbf
  {\bibinfo {volume} {323}},\ \bibinfo {pages} {1} (\bibinfo {year}
  {2001})}\BibitemShut {NoStop}%
\bibitem [{\citenamefont {Robertson}\ \emph {et~al.}(2009)\citenamefont
  {Robertson}, \citenamefont {Kravtsov}, \citenamefont {Tinker},\ and\
  \citenamefont {Zentner}}]{RobertsonKravtsovTinkerZentner2009}%
  \BibitemOpen
  \bibfield  {author} {\bibinfo {author} {\bibfnamefont {B.~E.}\ \bibnamefont
  {Robertson}}, \bibinfo {author} {\bibfnamefont {A.~V.}\ \bibnamefont
  {Kravtsov}}, \bibinfo {author} {\bibfnamefont {J.}~\bibnamefont {Tinker}}, \
  and\ \bibinfo {author} {\bibfnamefont {A.~R.}\ \bibnamefont {Zentner}},\
  }\href@noop {} {\bibfield  {journal} {\bibinfo  {journal} {ApJ}\ }\textbf
  {\bibinfo {volume} {696}},\ \bibinfo {pages} {636} (\bibinfo {year}
  {2009})}\BibitemShut {NoStop}%
\bibitem [{\citenamefont {Fry}(1996)}]{Fry1996}%
  \BibitemOpen
  \bibfield  {author} {\bibinfo {author} {\bibfnamefont {J.~N.}\ \bibnamefont
  {Fry}},\ }\href@noop {} {\bibfield  {journal} {\bibinfo  {journal} {ApJL}\
  }\textbf {\bibinfo {volume} {461}},\ \bibinfo {pages} {L65} (\bibinfo {year}
  {1996})}\BibitemShut {NoStop}%
\bibitem [{\citenamefont {Baldauf}\ \emph {et~al.}(2012)\citenamefont
  {Baldauf}, \citenamefont {Seljak}, \citenamefont {Desjacques},\ and\
  \citenamefont {McDonald}}]{Baldaufetal2012}%
  \BibitemOpen
  \bibfield  {author} {\bibinfo {author} {\bibfnamefont {T.}~\bibnamefont
  {Baldauf}}, \bibinfo {author} {\bibfnamefont {U.}~\bibnamefont {Seljak}},
  \bibinfo {author} {\bibfnamefont {V.}~\bibnamefont {Desjacques}}, \ and\
  \bibinfo {author} {\bibfnamefont {P.}~\bibnamefont {McDonald}},\ }\href@noop
  {} {\bibfield  {journal} {\bibinfo  {journal} {Phys. Rev. D}\ }\textbf
  {\bibinfo {volume} {86}},\ \bibinfo {pages} {083540} (\bibinfo {year}
  {2012})},\ \Eprint {http://arxiv.org/abs/arXiv:1201.4827} {arXiv:1201.4827}
  \BibitemShut {NoStop}%
\bibitem [{\citenamefont {Saito}\ \emph {et~al.}(2014)\citenamefont {Saito},
  \citenamefont {Baldauf}, \citenamefont {Vlah}, \citenamefont {Seljak},
  \citenamefont {Okumura},\ and\ \citenamefont
  {McDonald}}]{SaitoBaldauf_etal2014}%
  \BibitemOpen
  \bibfield  {author} {\bibinfo {author} {\bibfnamefont {S.}~\bibnamefont
  {Saito}}, \bibinfo {author} {\bibfnamefont {T.}~\bibnamefont {Baldauf}},
  \bibinfo {author} {\bibfnamefont {Z.}~\bibnamefont {Vlah}}, \bibinfo {author}
  {\bibfnamefont {U.}~\bibnamefont {Seljak}}, \bibinfo {author} {\bibfnamefont
  {T.}~\bibnamefont {Okumura}}, \ and\ \bibinfo {author} {\bibfnamefont
  {P.}~\bibnamefont {McDonald}},\ }\href@noop {} {\bibfield  {journal}
  {\bibinfo  {journal} {Phys. Rev. D}\ }\textbf {\bibinfo {volume} {90}},\
  \bibinfo {pages} {123522} (\bibinfo {year} {2014})},\ \Eprint
  {http://arxiv.org/abs/arXiv:1405.1447} {arXiv:1405.1447} \BibitemShut
  {NoStop}%
\bibitem [{\citenamefont {Biagetti}\ \emph
  {et~al.}(2014{\natexlab{b}})\citenamefont {Biagetti}, \citenamefont
  {Desjacques}, \citenamefont {Kehagias},\ and\ \citenamefont
  {Riotto}}]{BiagettiDesjacquesKehagiasRiotto2014a}%
  \BibitemOpen
  \bibfield  {author} {\bibinfo {author} {\bibfnamefont {M.}~\bibnamefont
  {Biagetti}}, \bibinfo {author} {\bibfnamefont {V.}~\bibnamefont
  {Desjacques}}, \bibinfo {author} {\bibfnamefont {A.}~\bibnamefont
  {Kehagias}}, \ and\ \bibinfo {author} {\bibfnamefont {A.}~\bibnamefont
  {Riotto}},\ }\href@noop {} {\bibfield  {journal} {\bibinfo  {journal} {Phys.
  Rev. D}\ }\textbf {\bibinfo {volume} {90}},\ \bibinfo {pages} {045022}
  (\bibinfo {year} {2014}{\natexlab{b}})},\ \Eprint
  {http://arxiv.org/abs/arXiv:1405.1435} {arXiv:1405.1435} \BibitemShut
  {NoStop}%
\bibitem [{\citenamefont {Bel}\ \emph {et~al.}(2015)\citenamefont {Bel},
  \citenamefont {Hoffmann},\ and\ \citenamefont
  {Gaztanaga}}]{BelHoffmannGaztanagn2015}%
  \BibitemOpen
  \bibfield  {author} {\bibinfo {author} {\bibfnamefont {J.}~\bibnamefont
  {Bel}}, \bibinfo {author} {\bibfnamefont {K.}~\bibnamefont {Hoffmann}}, \
  and\ \bibinfo {author} {\bibfnamefont {E.}~\bibnamefont {Gaztanaga}},\
  }\href@noop {} {} (\bibinfo {year} {2015}),\ \Eprint
  {http://arxiv.org/abs/arXiv:1504.02074} {arXiv:1504.02074} \BibitemShut
  {NoStop}%
\bibitem [{\citenamefont {Bernardeau}\ \emph {et~al.}(2002)\citenamefont
  {Bernardeau}, \citenamefont {Colombi}, \citenamefont {Gazta{\~n}aga},\ and\
  \citenamefont {Scoccimarro}}]{PTreview}%
  \BibitemOpen
  \bibfield  {author} {\bibinfo {author} {\bibfnamefont {F.}~\bibnamefont
  {Bernardeau}}, \bibinfo {author} {\bibfnamefont {S.}~\bibnamefont {Colombi}},
  \bibinfo {author} {\bibfnamefont {E.}~\bibnamefont {Gazta{\~n}aga}}, \ and\
  \bibinfo {author} {\bibfnamefont {R.}~\bibnamefont {Scoccimarro}},\
  }\href@noop {} {\bibfield  {journal} {\bibinfo  {journal} {Phys. Rep.}\
  }\textbf {\bibinfo {volume} {367}},\ \bibinfo {pages} {1} (\bibinfo {year}
  {2002})},\ \Eprint {http://arxiv.org/abs/arXiv:astro-ph/0112551}
  {arXiv:astro-ph/0112551} \BibitemShut {NoStop}%
\bibitem [{\citenamefont {Somogyi}\ and\ \citenamefont
  {Smith}(2010)}]{SomogyiSmith2010}%
  \BibitemOpen
  \bibfield  {author} {\bibinfo {author} {\bibfnamefont {G.}~\bibnamefont
  {Somogyi}}\ and\ \bibinfo {author} {\bibfnamefont {R.~E.}\ \bibnamefont
  {Smith}},\ }\href@noop {} {\bibfield  {journal} {\bibinfo  {journal} {Phys.
  Rev. D}\ }\textbf {\bibinfo {volume} {81}},\ \bibinfo {pages} {023524}
  (\bibinfo {year} {2010})},\ \Eprint {http://arxiv.org/abs/arXiv:0910.5220}
  {arXiv:0910.5220} \BibitemShut {NoStop}%
\bibitem [{\citenamefont {Bernardeau}\ \emph {et~al.}(2012)\citenamefont
  {Bernardeau}, \citenamefont {de~Rijt},\ and\ \citenamefont
  {Vernizzi}}]{BernardeauVandeRijtVernizzi2012}%
  \BibitemOpen
  \bibfield  {author} {\bibinfo {author} {\bibfnamefont {F.}~\bibnamefont
  {Bernardeau}}, \bibinfo {author} {\bibfnamefont {N.~V.}\ \bibnamefont
  {de~Rijt}}, \ and\ \bibinfo {author} {\bibfnamefont {F.}~\bibnamefont
  {Vernizzi}},\ }\href@noop {} {\bibfield  {journal} {\bibinfo  {journal}
  {Phys. Rev. D}\ }\textbf {\bibinfo {volume} {85}},\ \bibinfo {pages} {063509}
  (\bibinfo {year} {2012})},\ \Eprint {http://arxiv.org/abs/arXiv:1109.3400}
  {arXiv:1109.3400} \BibitemShut {NoStop}%
\bibitem [{\citenamefont {Scoccimarro}\ \emph {et~al.}(1998)\citenamefont
  {Scoccimarro}, \citenamefont {Colombi}, \citenamefont {Fry}, \citenamefont
  {Frieman}, \citenamefont {Hivon},\ and\ \citenamefont {Melott}}]{SCFFHM98}%
  \BibitemOpen
  \bibfield  {author} {\bibinfo {author} {\bibfnamefont {R.}~\bibnamefont
  {Scoccimarro}}, \bibinfo {author} {\bibfnamefont {S.}~\bibnamefont
  {Colombi}}, \bibinfo {author} {\bibfnamefont {J.~N.}\ \bibnamefont {Fry}},
  \bibinfo {author} {\bibfnamefont {J.~A.}\ \bibnamefont {Frieman}}, \bibinfo
  {author} {\bibfnamefont {E.}~\bibnamefont {Hivon}}, \ and\ \bibinfo {author}
  {\bibfnamefont {A.}~\bibnamefont {Melott}},\ }\href@noop {} {\bibfield
  {journal} {\bibinfo  {journal} {ApJ}\ }\textbf {\bibinfo {volume} {496}},\
  \bibinfo {pages} {586} (\bibinfo {year} {1998})},\ \Eprint
  {http://arxiv.org/abs/arXiv:astro-ph/9704075} {arXiv:astro-ph/9704075}
  \BibitemShut {NoStop}%
\end{thebibliography}%

\end{document}